\documentclass[aps,prd,reprint,amsmath,amssymb,superscriptaddress,nofootinbib]{revtex4-2}
\AtBeginDocument{%
  \heavyrulewidth=.08em
  \lightrulewidth=.05em
  \cmidrulewidth=.03em
  \belowrulesep=.65ex
  \belowbottomsep=0pt
  \aboverulesep=.4ex
  \abovetopsep=0pt
  \cmidrulesep=\doublerulesep
  \cmidrulekern=.5em
  \defaultaddspace=.5em
}
\usepackage{hyperref}
\usepackage{url}
\usepackage{graphicx} 
\usepackage[table]{xcolor}
\usepackage{array}
\usepackage{tabularx}
\definecolor{vargray}{gray}{0.92}

\usepackage{tikz}
\usepackage{tikz-3dplot}
\usetikzlibrary{patterns}

\usepackage{placeins}
\usepackage{amsmath}
\usepackage{booktabs}
\usepackage{longtable}
\usetikzlibrary{shapes.geometric, arrows, decorations.pathreplacing}
\usetikzlibrary{arrows.meta}
\usetikzlibrary{calc}
\usepackage{tabularx}
\usetikzlibrary{fit}
\date{}
\definecolor{mutedsky}{RGB}{40,110,190}
\definecolor{orbitgray}{gray}{0.35}

\definecolor{lightgray}{gray}{0.9}
\newcolumntype{L}{>{\raggedright\arraybackslash}X}
\newcolumntype{G}{>{\columncolor{lightgray}[0pt][\tabcolsep]\raggedright\arraybackslash}X}

\newcommand{\diff}{\mathop{}\!\mathrm{d}}

\tikzstyle{startstop} = [rectangle, rounded corners, minimum width=4cm, minimum height=1cm, text centered, align=center, draw=black, fill=darkgray!20]

\tikzstyle{process} = [rectangle, rounded corners, minimum width=3cm, minimum height=1cm, text centered, align=center, draw=black, fill=gray!7]

\tikzset{
  stepbox/.style={
    draw,
    line width=1.5pt,
    dashed,
    inner sep=6pt
  }
}

\tikzstyle{arrow} = [thick,->,>=stealth]

\begin{document}

\title{Observational selection effects on radio pulsars are minimal for masses, but significant for orbits and spins}

\author{Lisa V. Drummond}
\affiliation{Department of Physics and TAPIR, California Institute of Technology, Pasadena, CA}
\author{Katerina Chatziioannou}
\affiliation{Department of Physics and TAPIR, California Institute of Technology, Pasadena, CA}
\author{Emmanuel Fonseca}
\affiliation{Department of Physics and Astronomy, West Virginia University, PO Box 6315, Morgantown, WV}
\affiliation{Center for Gravitational Waves and Cosmology, West Virginia University, Chestnut Ridge Research Building, Morgantown, WV 26505, USA}

\begin{abstract}
The masses of Galactic pulsars in binaries measured through radio timing point to structure in the mass distribution with implications for dense matter and astrophysical formation processes: a bimodal shape with peaks at ${\sim} 1.3\,M_\odot$ and ${\sim} 1.6\,M_\odot$, with a cutoff above ${\sim}2\,M_\odot$.  
However, this \emph{observed} population of radio pulsars is shaped not only by intrinsic pulsar properties, such as birth masses, binary evolutionary pathways and dense-matter constraints,  but also by observational effects.
In existing catalogs of pulsar mass measurements, observational selection effects influence (i) which systems are detected in the first place (\textit{detectability}) and (ii) which systems yield well-measured masses (\textit{measurability}).  
In this work, we revisit the radio pulsar population distribution in the context of observational effects associated with pulsar radio timing.
Simulating systems of radio pulsars and timing observations, we track the process by which a given system ends up in an observational dataset and estimate the impact of the binary and pulsar parameters.
We fold this selection function into hierarchical population inference and extract the intrinsic \emph{astrophysical} distribution of pulsars observed with radio timing. 
We jointly infer the populations of pulsar mass, companion mass, orbital eccentricity, orbital period and pulsar spin period.
Selection effects have minimal impact on the mass distributions, but more significantly affect other parameters.
The ratio of circular-to-eccentric systems shifts from an observed $1\!:\!1$ to an astrophysical, selection-corrected $2\!:\!1$ ratio, while the true population is shifted toward longer orbital and spin periods than observed.
\end{abstract}

\makeatletter
\renewcommand{\@date}{}
\makeatother
\maketitle

\section{Introduction}


Measurements of the masses of neutron stars in a range of astrophysical environments provide constraints on the dense-matter equation of state~\cite{Ozel2016,Chatziioannou:2024jsr} and strong-field gravity \cite{Freire2024,Stairs2003}. 
Galactic radio pulsars in binaries (around a not-necessarily compact companion) play a key role in this context: the timing of their pulses allows for precision constraints on the component masses.  
Such measurements have revealed that the majority of pulsars have masses ${\sim} 1.3\,M_{\odot}$ and a fraction of them extends to higher masses before the distribution truncates at ${\gtrsim} 2\,M_\odot$~\cite{Antoniadis:2016hxz,Alsing2018,Farr2020}.
This nontrivial structure of the observed mass distribution is shaped by a number of factors.

As an example, consider the upper truncation of the mass distribution that has implications for dense matter~\cite{Annala:2017llu,Alsing2018,Miller:2021qha,Raaijmakers:2021uju,Legred:2021hdx,Finch2025,Koehn2025}.
At the nuclear physics level, the equation of state imposes a maximum mass for nonrotating neutron stars, $M_{\rm TOV}$, that acts as an absolute (up to rotation~\cite{1994ApJ...424..823C}) upper limit~\cite{PhysRev.55.374,PhysRev.55.364}.
At the astrophysics level, the masses of neutron stars realized in nature are shaped by complex astrophysical processes: supernova physics for the birth mass, binary processes that bring neutron stars together with a diverse set of companions, accretion that keeps them pulsating, etc.
Collectively, these processes allow for radio pulsars in binaries up to a mass $M_{\rm max}$ which is strictly ${\leq}M_{\rm TOV}$, though current data cannot distinguish the two~\cite{Golomb2025}.
At the observational level, selection effects inherent to radio detection and timing of the pulses to determine the binary parameters bias datasets.
For example, current searches are less sensitive to pulsars undergoing rapid orbital acceleration which in turn depends on the binary parameters.
Thus the observed $M_{\rm max}$ may be biased compared to its astrophysical value.  
Minus the role of nuclear physics, similar considerations apply to the bulk of the mass distribution~\cite{Farrow:2019xnc,You:2024bmk,SchiebelbeinZwack2025,Anik2025,Chattopadhyay:2026vfv}.
This study is concerned with the impact of the final level: the observational selection effects.

The mass distribution of Galactic pulsars has been the subject of past investigations~\cite{2011MNRAS.414.1427V,Ozel:2012ax,2013ApJ...778...66K}.
\citet{Antoniadis:2016hxz} considered millisecond pulsars and presented evidence for bimodality and a maximum mass cutoff. 
\citet{Alsing2018} and \citet{Shao:2020bzt} expanded the dataset to mass measurements from radio timing, X-ray, and optical observations and yielded qualitatively similar results. 
\citet{Chattopadhyay:2026vfv} explored correlations between the pulsar mass and other binary parameters, recovering hints of an anticorrelation with the spin period in the recycled population.
Beyond the mass distribution at the time of observation, \citet{You:2024bmk} subtracted an estimate of the accreted mass of recycled pulsars to reach a unimodal distribution with a high-mass tail for the underlying birth mass.
\citet{Hu:2026jkn} also argued that super-Eddington accretion during stable mass transfer can grow recycled pulsars to the high-mass peak.

A key element of such studies is how the neutron star (sub)populations are defined. 
For example, \citet{Antoniadis:2016hxz} restricted to millisecond pulsars,
\citet{Farrow:2019xnc} targeted double neutron star binaries (recycled vs non-recycled binary components), while \citet{Anik2025} and \citet{Chattopadhyay:2026vfv} classified the subpopulations by binary type.
Such approaches are motivated by the fact that the mass distribution depends on the binary type~\cite{Anik2025}, for example double neutron stars have component masses ${\sim}1.4\,M_{\odot}$~\cite{Farrow:2019xnc}.
\citet{Alsing2018} instead argued that the heterogeneity of the dataset can be handled by more complex (multi-component) models rather than \emph{a priori} classifications.
We adopt this viewpoint here and consider systems not by their astrophysical classification but instead by their observational detection.

The Galactic pulsar mass distribution can also be contrasted to other populations.
\citet{SchiebelbeinZwack2025} found agreement between the masses of neutron stars with luminous wide-orbit companions observed with the \textit{Gaia} mission and the masses of the first-borns in double neutron star binaries.
A consistent mass distribution over distinct system types suggests a natal origin with little impact from binary evolution. 
\citet{Landry2021} and \citet{Golomb2025} however showed that this is not the case when compared against neutron star masses inferred from gravitational-wave observations. 
Among double neutron stars, GW170817 is consistent with the Galactic binary neutron star population \cite{GW170817}, but GW190425, with a total mass of ${\sim}3.4\,M_\odot$, is significantly heavier \cite{GW190425}. 
A possible explanation is that GW190425-like systems originate from a fast-merging channel and are thus rare in radio \cite{Romero-Shaw:2020aaj, Safarzadeh2020}.

In this study, we consider \emph{radio timing} observational selection effects that impact the pulsar population distribution \emph{at the time of observation}.
Selection effects are inherent to any observational campaign and depend on the details of the detection and measurement process.
For this reason, we need to focus on a specific observational avenue: detection of pulsars by radio observations and parameter estimation by modeling the pulse times of arrival in the context of a binary with Keplerian and post-Keplerian terms.
We quantify radio selection effects by tracking the process by which a pulsar is detected and its mass is measured, and identifying the system dependence in each step.
Folding this ``selection function'' in hierarchical inference~\cite{10.1063/1.1835214,Mandel:2018mve}, we infer the intrinsic distribution of radio pulsars across binary masses, orbital eccentricity and period, and pulsar spin period that is shaped only by nuclear physics and astrophysical processes. 

Before describing the effects we consider, we first clarify the effects we ignore.
A straightforward mass dependence is well-known~\cite{Tauris2017}: present-day radio pulsars are recycled by accretion which raises their masses~\cite{You:2024bmk}.
Population synthesis simulations of pulsars in neutron star~\cite{Chattopadhyay2020} or black hole~\cite{Chattopadhyay2021} binaries follow such astrophysical effects forward in time and, when further projected through radio telescope and propagation selection, find that recycling shifts the detectable mass distribution slightly higher.
Massive systems might form through a fast-merging channel and thus be rare in radio~\cite{Romero-Shaw:2020aaj, Safarzadeh2020}.
Such \emph{astrophysical} selection effects shape the present-day \emph{intrinsic} distribution.
We therefore treat them as part of the underlying population whose distribution we aim to infer and interpret.
We instead target (and correct for) the \emph{observational} selection effects, i.e., effects that enter through the detection and measurement process.

With this clarification in mind, we return to the radio observation process, which we detail in Sec.~\ref{sec:observations}.
We identify two factors: whether a pulsar is detected, i.e., its \textit{detectability}, and whether its mass is measured and thus is included in mass datasets, i.e., its \textit{measurability}. 
The binary dependence on detectability enters through the orbital motion which causes Doppler smearing of the pulsar signal in the Fourier domain.
We quantify the signal-to-noise ratio reduction~\cite{Johnston1991, Bagchi:2013wga, Pol2021} due to pulsar acceleration and recover a bias against tight, short-period binaries \cite{Tauris2017}. 
At the measurement stage, pulsar masses are inferred from ``post-Keplerian'' effects on the orbit that arise from general-relativistic motion and impact the timing of pulses~\cite{DamourTaylor1992,Lorimer2008}. 
For example, the Shapiro delay is more prominent in systems with high inclinations, while periastron advance requires eccentric orbits and scales with the total binary mass. 
In Sec.~\ref{sec:selectioneffects} we estimate these effects with simulated systems and find detectability to be a weak function of the pulsar mass, while the fraction of measurable systems increases from ${\sim}40\%$ at $1\,M_{\odot}$ to ${\sim}50\%$ at $2.9\,M_{\odot}$.

With the selection function in hand, we infer the astrophysical, present-day distribution of the pulsar and companion masses, the binary eccentricity and orbital period, and the pulsar spin period. 
We describe the dataset and our reconstruction of the multi-dimensional posterior for these parameters in Sec.~\ref{sec:posteriorreconstruction} and the relevant population models in Sec.~\ref{sec:hierarchical-models}.
Results are presented in Sec.~\ref{sec:results}.
Accounting for selection effects modestly shifts the inferred distributions to lower-mass pulsars and companions, while the maximum pulsar mass is essentially unchanged; all shifts are within the statistical uncertainty. 
The effect is more pronounced for the orbital and spin parameters: the selection-corrected population contains a larger fraction of nearly-circular binaries (a near-circular to noncircular ratio of approximately $2\!:\!1$, compared with $1\!:\!1$ without selection effects) and is shifted toward longer orbital periods (approximately $8.1$ days, compared with $2.7$ days) and spin periods (approximately $40$ ms, compared with $7$ ms).

\begin{figure*}
\centering
\resizebox{\textwidth}{!}{%
\begin{tikzpicture}[
    font=\small,
    >=Latex,
    outerrule/.style={line width=1.0pt},
    innerrule/.style={densely dashed, line width=0.8pt},
    labelrule/.style={line width=1.6pt},
    doublerule/.style={line width=0.9pt},
    box/.style={
        draw,
        rounded corners=2pt,
        align=center,
        text width=3.0cm,
        minimum height=1.2cm,
        inner sep=4pt
    },
    note/.style={
        align=center,
        text width=3.1cm,
        minimum height=1.0cm,
        inner sep=4pt
    },
    faintnote/.style={
        draw,
        dashed,
        rounded corners=2pt,
        align=center,
        text width=3.0cm,
        minimum height=1.1cm,
        inner sep=4pt
    },
    arrow/.style={->, thick},
    rowlabel/.style={
        rotate=90,
        anchor=center,
        font=\bfseries
    },
    circnum/.style={
        draw,
        circle,
        inner sep=2pt,
        font=\small
    }
]

\def\xlabel{-0.65}
\def\xlabpos{-0.45}

\def\xA{0}
\def\xB{4.32}
\def\xC{8.64}
\def\xD{12.96}
\def\xE{17.28}
\def\xF{21.6}

\def\xcI{2.16}
\def\xcII{6.48}
\def\xcIII{10.8}
\def\xcIV{15.12}
\def\xcV{19.44}

\def\ytop{4.3}
\def\yone{-0.15}
\def\ytwo{-2.35}
\def\ybot{-4.55}

\draw[outerrule] (\xlabel,\ytop) -- (\xF,\ytop);
\draw[outerrule] (\xlabel,\ybot) -- (\xF,\ybot);

\draw[innerrule] (\xlabel,\yone) -- (\xF,\yone);

\draw[doublerule] (\xlabel,\ytwo+0.08) -- (\xF,\ytwo+0.08);
\draw[doublerule] (\xlabel,\ytwo-0.08) -- (\xF,\ytwo-0.08);

\draw[labelrule] (\xA,\ytop) -- (\xA,\ybot);

\foreach \x in {\xB,\xC,\xD,\xE} {
    \draw[innerrule] (\x,\ytop) -- (\x,\ybot);
}

\draw[outerrule] (\xF,\ytop) -- (\xF,\ybot);

\node[rowlabel] at (\xlabpos,2.08) {Workflow};
\node[rowlabel] at (\xlabpos,-1.25) {Selection};
\node[rowlabel] at (\xlabpos,-3.45) {Our model};

\node[circnum] at (0.35,4.00) {1};
\node[circnum] at (4.67,4.00) {2};
\node[circnum] at (8.99,4.00) {3};
\node[circnum] at (13.31,4.00) {4};
\node[circnum] at (17.63,4.00) {5};


\node at (\xcI,2.55) {
\begin{tikzpicture}[scale=0.48, line cap=round, line join=round]

    \begin{scope}[shift={(-0.75,0)}]

        \draw[thin] (2.00,1.65) -- (1.35,1.25);
        \draw[thin] (1.75,1.95) -- (1.10,1.55);
        \draw[thin] (1.50,2.25) -- (0.85,1.85);

        \draw[thick]
            (-1.45,0.65)
            .. controls (-0.85,-0.25) and (0.85,-0.25) ..
            (1.45,0.65);

        \draw[thick] (-1.45,0.65) -- (1.45,0.65);

        \draw[thin] (-1.05,0.55)
            .. controls (-0.55,0.05) and (0.55,0.05) ..
            (1.05,0.55);
        \draw[thin] (-0.55,0.50)
            .. controls (-0.25,0.25) and (0.25,0.25) ..
            (0.55,0.50);

        \draw[thick] (0,1.15) circle (0.13);
        \draw[thick] (-0.22,1.30) rectangle (0.22,1.50);

        \draw[thick] (-1.25,0.65) -- (0,1.15);
        \draw[thick] (1.25,0.65) -- (0,1.15);
        \draw[thick] (0,0.35) -- (0,1.15);

        \draw[thick] (0,0.35) -- (0,-0.45);
        \draw[thick] (-0.35,-0.45) rectangle (0.35,-0.75);

        \draw[thick] (0,-0.75) -- (0,-1.35);
        \draw[thick] (0,-1.00) -- (-0.75,-1.45);
        \draw[thick] (0,-1.00) -- (0.75,-1.45);
        \draw[thick] (-0.95,-1.45) -- (0.95,-1.45);

    \end{scope}

    \begin{scope}[shift={(2.75,2.5)}, scale=0.8]

        \draw[solid] (0,0) ellipse (0.9cm and 0.5cm);

        \fill[gray] (-0.9,0) circle (0.38cm);
        \fill[black] (0.9,0) circle (0.18cm);
    \end{scope}

\end{tikzpicture}
};

\node[align=center] at (\xcI,0.65)
{Radio data recorded;\\Preserved indefinitely};

\node[align=center] at (\xcII,0.55)
{Fourier stacking search};

\node at (\xcII,2.4) {
\scalebox{0.75}{%
\begin{tikzpicture}[x=0.33cm,y=0.78cm,line cap=round,line join=round]

    \draw[->,thick] (0,0) -- (11.2,0);
    \draw[->,thick] (0,0) -- (0,3.9);

    \node[below] at (5.6,-0.05) {\large Frequency};
    \node[rotate=90] at (-0.85,1.95) {\large Power};

    \foreach \x/\h in {
        0.4/0.55,
        0.9/0.85,
        1.4/0.45,
        1.9/0.70,
        2.4/0.50,
        2.9/0.90,
        3.4/0.60,
        3.9/0.40,
        4.4/0.75,
        4.9/0.55,
        5.4/0.70,
        5.9/3.10,
        6.4/0.80,
        6.9/0.50,
        7.4/0.65,
        7.9/0.45,
        8.4/0.85,
        8.9/0.55,
        9.4/0.70,
        9.9/0.50
    }{
        \draw[thick] (\x,0) -- (\x,\h);
    }

\end{tikzpicture}%
}
};

\node at (\xcIII,2.68) {
\begin{tikzpicture}[scale=0.5, line cap=round, line join=round]

    \begin{scope}[
        shift={(-1.55,0.95)},
        scale=0.68,
        line width=1.15pt,
        line cap=round,
        line join=round
    ]

        \draw
            (-1.35,0.00)
            .. controls (-0.62,0.22) and (0.02,0.62) ..
            (0.57,1.02);
        \draw
            (-1.35,0.00)
            .. controls (-0.62,-0.22) and (0.02,-0.62) ..
            (0.57,-1.02);

        \draw (0.57,1.02) -- (1.18,1.55);
        \draw (0.57,-1.02) -- (1.18,-1.55);

        \draw
            (0.57,1.02)
            .. controls (1.17,0.58) and (1.25,-0.55) ..
            (0.57,-1.02);
        \draw
            (0.57,1.02)
            .. controls (0.12,0.48) and (0.12,-0.48) ..
            (0.57,-1.02);

        \draw
            (0.99,0.52)
            .. controls (0.62,0.18) and (0.62,-0.18) ..
            (0.99,-0.52);

        \draw (1.42,0.56) -- (1.78,0.69);
        \draw (1.48,0.00) -- (1.84,0.00);
        \draw (1.42,-0.56) -- (1.78,-0.69);

    \end{scope}

    \begin{scope}[shift={(0.75,-1.4)}]

        \draw[thick] (0,0) -- (3.55,0);
        \draw[thick] (0,0) -- (0,1.85);

        \draw[thick]
        plot[smooth] coordinates {
            (0.10,0.22)
            (0.30,0.24)
            (0.50,0.21)
            (0.70,0.25)
            (0.90,0.23)
            (1.05,0.22)
            (1.18,0.28)
            (1.28,0.48)
            (1.36,0.95)
            (1.43,1.45)
            (1.49,1.63)
            (1.56,1.38)
            (1.66,0.88)
            (1.80,0.46)
            (1.98,0.30)
            (2.20,0.24)
            (2.45,0.21)
            (2.70,0.23)
            (2.95,0.20)
            (3.20,0.22)
            (3.40,0.21)
        };

    \end{scope}

\end{tikzpicture}
};

\node[align=center] at (\xcIII,0.65)
{Candidates confirmed by\\ human verification};

\node at (\xcIV,2.55) {
\begin{tikzpicture}[scale=0.7, line cap=round, line join=round]

    \foreach \dx in {0,2.65} {
        \begin{scope}[shift={(\dx,0)}]
            \draw[thick] (0,0) -- (1.65,0);
            \draw[thick] (0,0) -- (0,1.12);
            \draw[thick]
                plot[smooth] coordinates {
                    (0.08,0.17)
                    (0.30,0.18)
                    (0.50,0.22)
                    (0.62,0.52)
                    (0.72,1.00)
                    (0.82,0.54)
                    (0.96,0.24)
                    (1.22,0.18)
                    (1.55,0.17)
                };
        \end{scope}
    }

    \draw[thick] (1.91,0.52) -- (2.39,0.52);
    \draw[thick] (1.91,0.72) -- (2.39,0.72);

    \draw[->,thick]
        (0.78,1.30)
        .. controls (1.48,1.88) and (2.82,1.88) ..
        (3.50,1.30);

    \draw[line width=1.2pt]
        (4.02,1.48) -- (4.20,1.28) -- (4.57,1.72);

\end{tikzpicture}
};

\node[align=center] at (\xcIV,0.65)
{Repeat observation;\\Matching $P_s$ and DM};

\node at (\xcV,2.55) {
\begin{tikzpicture}[
    scale=0.7,
    line cap=round,
    line join=round
]

    \begin{scope}
        \draw[-,thick] (0,0) -- (1.75,0)
            node[right, font=\scriptsize] {$t$};
        \draw[-,thick] (0,0) -- (0,1.55)
            node[above, font=\scriptsize] {$f$};

\draw[line width=1.2pt]
    plot[smooth] coordinates {
        (1.62,0.16)
        (1.48,0.24)
        (1.25,0.36)
        (0.94,0.54)
        (0.66,0.78)
        (0.49,1.04)
        (0.41,1.28)
        (0.39,1.42)
    };

        \node[font=\scriptsize] at (0.85,-0.38)
            {Dispersed};
    \end{scope}

    \draw[->,thick] (1.65,0.75) -- (2.55,0.75);

    \begin{scope}[shift={(2.80,0)}]
        \draw[-,thick] (0,0) -- (1.75,0)
            node[right, font=\scriptsize] {$t$};
        \draw[-,thick] (0,0) -- (0,1.55)
            node[above, font=\scriptsize] {$f$};

        \draw[line width=1.2pt]
            (0.82,0.16) -- (0.82,1.36);

        \node[font=\scriptsize] at (0.85,-0.38)
            {De-dispersed};
    \end{scope}

\end{tikzpicture}
};

\node[align=center] at (\xcV,0.65)
{Coherently de-dispersed\\mode followup};


\node[align=center] at (\xcI,-1.25)
{Declination\\limits};

\node[align=center] at (\xcII,-0.8)
{Significance threshold};

\node[align=center] at (\xcII,-1.5)
{Fourier smearing\\due to orbital motion};

\node[align=center] at (\xcIII,-1.25)
{Pulse profile health\\$\rm{DM}\neq0$};

\node[align=center] at (\xcIV,-1.25)
{Consistency check};


\node[note] at (\xcV,-1.25)
{N/A};

\foreach \xl/\xr in {\xA/\xB,\xC/\xD,\xD/\xE,\xE/\xF} {
    \fill[pattern=north east lines, pattern color=gray!70]
        (\xl+0.08,\ybot+0.08) rectangle (\xr-0.08,\ytwo-0.08);
}


\node[note, align=center] at (\xcI,-3.45) {\textbf{Ignore}\\[1.2em] Independent of $m_p$};


\node[align=center] at (\xcII,-3.10)
{Significance threshold};

\node[align=center] at (\xcII,-3.85)
{Coherence factor, Sec.~\ref{sec:detect}};

\node[note, align=center] at (\xcIII,-3.45) {\textbf{Ignore}\\[1.2em] Independent of $m_p$};

\node[note, align=center] at (\xcIV,-3.45) {\textbf{Ignore}\\[1.2em] Independent of $m_p$};

\node[note, align=center] at (\xcV,-3.45) {N/A};

\end{tikzpicture}%
}
\caption{
Workflow for the detection of pulsars through radio observations. Each column corresponds to a different stage of this process. The first row describes the real analysis, the second row corresponds to the selection effect introduced at each stage, while the third row describes how we model it.  
}
\label{fig:selection_workflow_sketch}
\end{figure*}

\section{Radio pulsar binary observations}
\label{sec:observations}

We target the properties of pulsars in binaries, regardless of companion type, that are measured with radio timing of relativistic post-Keplerian effects such as Shapiro delay, periastron advance, and gravitational-wave-driven orbital decay \cite{Stairs2003,Freire2024}. 
This setup excludes: (i) gravitational-wave observations of neutron stars in compact binaries~\cite{GW170817,GW190425,LIGOScientific:2021qlt,LIGOScientific:2024elc}, (ii) X-ray pulse profile observations of pulsars that are isolated or in binaries~\cite{Bogdanov2019,Miller2019,Riley2019}, (iii) optical spectroscopic or photometric observations of the binary companion~\cite{Bhalerao2012,vanKerkwijk2011,Romani2012}, (iv) X-ray spectroscopy of neutron stars in low-mass X-ray binaries (thermonuclear bursts and quiescent thermal emission)~\cite{Ozel2016, Falanga2015}, and (v)
astrometric \textit{Gaia} observations of neutron stars with a luminous companion~\cite{2024OJAp....7E..58E}.

In this section, we describe our dataset and selection effects.
We first summarize the observational set of pulsar mass measurements we consider and describe the workflow by which each system enters the set in Sec.~\ref{sec:data}. 
We then discuss two stages of this workflow that introduce selection effects: Sec.~\ref{sec:detect} describes the radio detection of pulsars and the factors that influence their \emph{detectability}, while Sec.~\ref{sec:measure} outlines how timing measurements constrain pulsar masses and the corresponding effects that influence \emph{measurability}.

\subsection{Observational dataset and analysis workflow}
\label{sec:data}

Understanding what selection effects impact an observational dataset requires tracking the workflow by which a system may appear in the dataset.  
Figure~\ref{fig:selection_workflow_sketch} presents the detection workflow for radio pulsars (top), the corresponding selection effect (middle), and how we model it (bottom).
Starting with the full detection workflow:
\begin{itemize}
\item Panel 1: Radio telescopes carry out untargeted, omnidirectional surveys. Data are recorded in a uniform way (detector calibration, time and frequency sampling, receiver frequency, etc.) and are stored for followup analysis.
\item Panel 2: Pulsars are searched for by Fourier-transforming the dedispersed time-series and identifying peaks in the power spectrum. Harmonically-related peaks that pass a significance threshold are flagged as candidates.
\item Panel 3: Candidates are subject to human verification, e.g., ensuring pulse stability across time and frequency, checking for non-zero dispersion measure (DM), etc.
\item Panel 4: Candidates are followed up on using identical detector and search configuration. A pulsar candidate is deemed verified if the independent followup observations yield a second candidate with consistent spin and DM with the original one.
\item Panel 5: Verified pulsars are observed in coherently-dedispersed mode over multiple years to estimate their parameters.
\end{itemize}

The most direct way to model selection effects is to reproduce the full observational workflow for simulated systems, as done in the context of gravitational-wave searches~\cite{LIGOScientific:2025pvj,Essick:2025zed,LIGOScientific:2026ctl}.
This procedure involves starting from a set of pulsar systems, simulating the full set of radio data (pulsar times-of-arrival for each system), and assessing whether the workflow of Fig.~\ref{fig:selection_workflow_sketch} would deem the system detectable or not.
In practice, such an end-to-end analysis is not straightforward.
Besides being computationally prohibitive, it would require modeling the true data-generation process which is unknown, for example the DM distribution of false triggers.\footnote{In the gravitational-wave context, this is overcome by ``injecting'' simulated signals into off-source data that are generated by the real detectors and do not contain real (detectable) signals.}
Additionally, it would require an element of human followup to carry out general data sanity checks.\footnote{This is akin to the event-validation followup of gravitational-wave triggers.}
Instead, we revisit each step of Fig.~\ref{fig:selection_workflow_sketch} and assess its impact:
\begin{itemize}
\item Panel 1: Since the survey is omnidirectional, the only selection effect is the observing site declination limit. We ignore it as the system sky location is independent of the remaining system parameters.
\item Panel 2: Detection efficiency is tied to the prominence of the Fourier spectral peak, with more luminous/nearby pulsars being more detectable. Additionally, the pulsar orbital motion smears the Fourier power and lowers the peak strength. A variety of searches may account for the pulsar orbital velocity, acceleration, or jerk. This orbit-dependent selection effect is quantified in Sec.~\ref{sec:detect} and the same significance threshold is used to assess detectability.
\item Panel 3: The human followup imposes general sanity checks on the pulse properties and interstellar dispersion. We assume that both are independent of the system properties and ignore them.
\item Panel 4: This step penalizes false detections. We assume that any simulated system that has passed the previous steps would pass this one as well and thus ignore it.
\item Panel 5: All pulsars, isolated or in binaries, are followed up on so there is no selection effect. 
\end{itemize}

The pulse times-of-arrival (TOAs) carry information about the binary orbit. 
These TOAs are fit using a timing model that includes Keplerian and post-Keplerian parameters of the orbit. 
The accuracy with which the various system parameters are measured depends on which post-Keplerian parameters are measurable, which in turn depends on the exact binary configuration.

A detected binary pulsar can reveal several levels of mass information, which control how a given system is included in published datasets such as the tables maintained by Freire \cite{FreireNSMasses}:
\begin{enumerate}
    \item For all systems, the mass function (a combination of the companion mass, pulsar mass, and orbital inclination given in Eq.~\eqref{eq:massfunction}) is measured. With one measured quantity and three unknowns, the pulsar mass is effectively unconstrained.
    \item If a single post-Keplerian parameter is measured, this removes one degree of freedom. For example, if the relativistic periastron advance is measured, Eq.~\eqref{eq:periastroadvance}, then the binary total mass is constrained, which yields an upper limit on the pulsar mass.
    \item If multiple post-Keplerian parameters are measured, then the pulsar mass is constrained with a precision related to that of the post-Keplerian parameters. The first table by Freire \cite{FreireNSMasses} lists the systems for which the pulsar mass is measured with less than 15\% relative uncertainty at the 1-$\sigma$ level.  
\end{enumerate}
In principle, the pulsar population distribution can be inferred using systems from any set.
Given the large uncertainty, the first set is never used in practice, a choice that already introduces a selection.
Measurements from the second set require adopting a distribution for the binary inclination; Refs.~\cite{Antoniadis:2016hxz,Alsing2018,Farrow:2019xnc,Farr2020,Golomb2025} assumed isotropic orientations and marginalized over the inclination to obtain a non-Gaussian, weakly informative posterior for the pulsar mass.
The final set is the most constraining, but also subject to the strongest selection effects.

Since the construction of all datasets includes some selection effects that we need to incorporate, we opt to work with the most informative measurements: systems where the pulsar mass is measured with a fractional uncertainty $\leq 15\%$, corresponding to the first table by Freire \cite{FreireNSMasses,Ozel2016,Freire2024}.\footnote{This choice filters out all isolated pulsars as well, both from the detected set and the simulations that estimate the selection.}
This dataset is subject to a \emph{measurability} selection effect that we quantify in Sec.~\ref{sec:measure}.
Details about the systems we analyze are provided in Sec.~\ref{sec:posteriorreconstruction}.

In summary:
\begin{enumerate}
\item We target the distribution of pulsars in binaries detected as radio pulsars, with masses constrained from pulsar timing and post-Keplerian effects;
\item We restrict to systems that are described by relativistic point-particle motion, excluding systems that require additional astrophysical modeling, for example non-degenerate companions \cite{afm+23}, active mass transfer \cite{asr+09}, or significant mass loss \cite{svf+16};
\item We further restrict to systems with well-constrained pulsar masses, namely a fractional mass uncertainty below $15\%$ at the 1-$\sigma$ level;
\item For systems where information beyond standard radio timing and post-Keplerian parameters is available, we restrict to the timing information, see Sec.~\ref{sec:posteriorreconstruction} for how this is achieved in practice.
\end{enumerate}
An additional selection might occur among systems with well-constrained masses.
The ones deemed most ``interesting,'' for example due to an exceptionally high mass or if multiple post-Keplerian parameters are measured, might receive preferential access to telescope time.
\citet{Essick2026} showed that this selection can be ignored if (i) all systems are included in the population inference, regardless of whether they received preferential followup or not, and (ii) the followup decision was based on the same data we analyze, rather than external information.
Both conditions are met in our analysis and thus no additional selection correction is required beyond the detectability and measurability we detail below.

\subsection{Detection of radio pulsars}
\label{sec:detect}

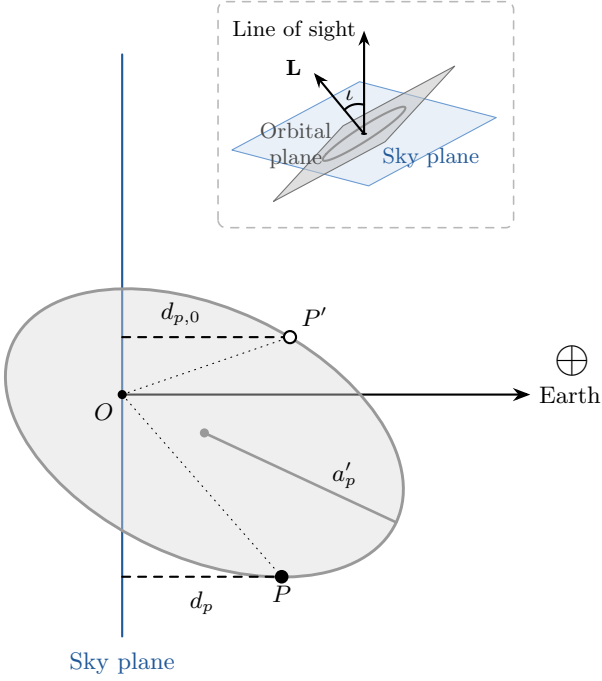
\begin{figure}[t]
\tdplotsetmaincoords{68}{125}

\begin{tikzpicture}[
    >=Stealth,
    line cap=round,
    line join=round
]

\begin{scope}[scale=1]

\coordinate (Omain) at (0,0);

\draw[mutedsky!80!black, thick] (0,-3.2) -- (0,4.5);
\node[mutedsky!80!black, below] at (0,-3.3) {Sky plane};

\draw[thick,->] (Omain) -- (5.4,0);
\node[right] at (5.4,0) {Earth};
\node at (5.95,0.45) {\LARGE $\oplus$};

\begin{scope}[rotate around={-25:(Omain)}]
  \fill[
    gray!30,
    fill opacity=0.4
  ] ($(Omain)+(1.2,0)$)
    ellipse [x radius=2.8, y radius=1.65];

  \draw[
    gray!80,
    line width=1.1pt
  ] ($(Omain)+(1.2,0)$)
    ellipse [x radius=2.8, y radius=1.65];
\end{scope}

\coordinate (Cmain) at ({1.2*cos(-25)},{1.2*sin(-25)});
\fill[gray!80] (Cmain) circle (0.06);

\coordinate (Aone) at ({(1.2+2.8)*cos(-25)},
                       {(1.2+2.8)*sin(-25)});
\draw[gray!80,line width=1pt] (Cmain) -- (Aone);
\node[above=0.5pt] at ($(Cmain)!0.73!(Aone)$) {$a_p'$};

\coordinate (Pp) at (2.22,0.76);

\draw[thick,dashed]
    (Omain |- Pp)
    -- node[pos=0.35,above=1.2pt] {$d_{p,0}$}
    (Pp);

\fill[white] (Pp) circle (0.08);
\draw[thick] (Pp) circle (0.08);
\node[above right=1pt] at (Pp) {$P'$};

\coordinate (P) at (2.11,-2.41);

\draw[thick,dashed]
    (Omain |- P)
    -- node[midway,below=2pt] {$d_p$}
    (P);

\fill (P) circle (0.09);
\node[below] at (P) {$P$};

\fill (Omain) circle (0.06);
\node[below left] at (Omain) {$O$};

\draw[dotted] (Omain) -- ($(Omain)!0.95!(Pp)$);
\draw[dotted] (Omain) -- (P);

\end{scope}


\begin{scope}[
    shift={(3.20,3.45)},
    scale=0.92,
    transform shape
]

\draw[
    gray!55,
    dashed,
    line width=0.6pt,
    rounded corners=3pt
]
(-2.10,-1.35) rectangle (2.15,1.9);

\begin{scope}[
    tdplot_main_coords,
    vector/.style={thick,-{Stealth[length=2.0mm,width=1.5mm]}}
]

\def\inc{50}
\pgfmathsetmacro{\ci}{cos(\inc)}
\pgfmathsetmacro{\si}{sin(\inc)}

\coordinate (Oinc) at (0,0,0);

\filldraw[
    fill=mutedsky!18,
    draw=mutedsky!70,
    fill opacity=0.55
]
(-1.6,-1.2,0) -- (1.6,-1.2,0) -- (1.6,1.2,0) -- (-1.6,1.2,0) -- cycle;

\node[mutedsky!80!black, xshift=-2pt, yshift=8pt]
    at (2.2,2.8,0.7) {Sky plane};

\filldraw[
    fill=orbitgray!35,
    draw=orbitgray!85,
    fill opacity=0.55
]
(-1.4,{-0.95*\ci},{-0.95*\si}) --
( 1.4,{-0.95*\ci},{-0.95*\si}) --
( 1.4,{ 0.95*\ci},{ 0.95*\si}) --
(-1.4,{ 0.95*\ci},{ 0.95*\si}) -- cycle;

\node[orbitgray!85!black, align=center, xshift=-12pt, yshift=-4pt]
    at (1.85,{0.95*\ci},{0.95*\si}) {Orbital\\plane};

\draw[
    gray!85,
    thick,
    variable=\t,
    domain=0:360,
    samples=120
]
plot ({0.95*cos(\t)},{0.45*sin(\t)*\ci},{0.45*sin(\t)*\si});

\draw[vector] (Oinc) -- (0,0,1.6);
\node[left] at (0,0,1.6) {Line of sight};

\draw[vector] (Oinc) -- (0,{-1.15*\si},{1.15*\ci});
\node[left] at (0,{-1.25*\si},{1.25*\ci}) {$\mathbf{L}$};

\draw[
    thick,
    variable=\t,
    domain=0:\inc,
    samples=40
]
plot ({0},{-0.45*sin(\t)},{0.45*cos(\t)});

\node at (0,{-0.6*sin(\inc/2)},{0.6*cos(\inc/2)}) {$\iota$};

\fill (Oinc) circle (0.045);

\end{scope}
\end{scope}

\path[use as bounding box]
    (current bounding box.south west)
    rectangle
    ($(current bounding box.north east)+(1.2,0)$);

\end{tikzpicture}
\caption{
Geometry of the line-of-sight motion of a pulsar in a binary system. 
The vertical line marks the sky plane viewed edge-on, while the plane of the page contains the line of sight to Earth and is perpendicular to the sky plane. 
The ellipse represents the pulsar orbit projected onto the plane of the page.  Point $P'$ denotes the pulsar position at $t=0$ and $P$ denotes its position at a later time $t$; the corresponding pulsar--Earth displacements are $d_{p,0}$ and $d_p$.
 The quantity $a_p' = a_p \sin \iota$ is the line-of-sight projected semi-major axis, where $a_p$ is the true semi-major axis of the pulsar orbit and $\iota$ is the inclination angle. 
 The inset shows $\iota$ as the angle between the line of sight and the orbital angular-momentum vector $\mathbf{L}$; equivalently, $\iota$ is the angle between the orbital plane and the sky plane. 
With this convention, a face-on orbit has $\iota=0^\circ$ and produces no line-of-sight orbital motion, while an edge-on orbit has $\iota=90^\circ$ and gives the maximal line-of-sight projection. The geometrical setup follows Fig.~1 of Ref.~\cite{Bagchi:2013wga}.
}
\label{fig:pulsargeom}
\end{figure}

The first selection effect concerns detecting the pulsar system in the first place, Panel 2 of Fig.~\ref{fig:selection_workflow_sketch}. 
This step involves obtaining radio data for some integration time $t_{\rm int}$ and then searching the data for periodic signals: a Fourier transform reveals excess power above the noise background in bins associated with the pulsar’s spin frequency.
The relevant detection statistic is the signal-to-noise ratio (SNR) of the received signal given by the pulsar radiometer equation \cite{Dewey1985,Lorimer2004}, written in Ref.\ \cite{Pol2021} as 
\begin{equation}
\label{eq:SNR}
  \text{SNR}=  S \times \xi \times f_1 \times \gamma_{im}^2 \,,
\end{equation}
where
\begin{equation}
  \xi = \frac{G \sqrt{B N_p}}{T_{\rm sys}}\,, \qquad
  f_1 = \sqrt{\frac{t_{\rm int}(P_s - w)}{w}}\,.
\end{equation}
Here $S$ is the pulsar flux density, which linearly scales the SNR.
The dependence on the instrument setup is encoded in $\xi$ that collects the telescope gain $G$, usable observing bandwidth $B$, number of summed polarizations $N_p$, and system temperature $T_{\rm sys}$.
For a fixed survey configuration, such as Panel 1 of Fig.~\ref{fig:selection_workflow_sketch}, $\xi$ is independent of the binary parameters. 
The factor $f_1$ encodes the dependence on the pulse properties, where $P_s$ is the pulsar spin period, $w$ is the effective pulse width, and $t_{\rm int}$ is the observation time. 
For fixed $t_{\rm int}$, a smaller duty cycle $w/P_s$ (narrower pulses or larger spin periods) means that the emission is concentrated into a smaller fraction of the pulse period, resulting in a higher Fourier peak and giving greater contrast between the on-pulse and off-pulse parts of the profile. 
Conversely, as $w$ approaches $P_s$ the signal occupies nearly the full rotation period and the contrast with the background is reduced.

The factor $\gamma_{im}\in [0,1]$ encodes the dependence on the binary parameters. 
The index $m$ corresponds to the Fourier harmonic under consideration.
The index $i$ tracks the type of pulsar search and specifically the assumed pulsar line-of-sight motion which affects the recovered Fourier power.
A constant line-of-sight velocity simply Doppler-shifts the spin frequency without affecting the height of the Fourier peak, whereas time-dependent motion causes Fourier smearing, with signal power spread across Fourier bins.
Specifically then, $i=1$ corresponds to a constant-velocity search, $i=2$ is a constant-acceleration search, and $i=3$ is a constant-jerk search.  
The factor $\gamma_{im}^2$ is defined as the fraction of SNR retained after accounting for the pulsar’s line-of-sight motion, with $\gamma_{im}=1$ corresponding to no loss of power.
Despite that, it has been termed the ``degradation factor''~\cite{Johnston1991,Bagchi:2013wga}; we here instead call it the ``coherence factor'' to avoid confusion about its role.

An analytical framework for the impact of the binary motion on the SNR was introduced by \citet{Johnston1991} for circular orbits and extended to eccentric binaries by \citet{Bagchi:2013wga}.  
A pulsar is taken to be detected in the Fourier bin where power is maximized and search degradation is defined in terms of the reduction in peak Fourier power.
The signal emitted by a pulsar with spin period $P_s$ is written as a Fourier series with angular spin frequency $\omega_s = 2\pi / P_s$. 
Taking $d_p$ to be the pulsar displacement toward the Earth,\footnote{This is sometimes identified in the literature as the distance between the pulsar and the Earth. However, under that definition, the retarded time in Eq.~\eqref{eq:sig} would have a minus sign. Though a sign change does not affect the resulting framework, we opt to retain the familiar equation and clarify the definition of $d_p$.} up to a constant phase the signal received by a radio telescope is
\begin{equation}
\label{eq:sig}
S_R(t)=\sum_{m=1}^{\infty}a_m\exp \left[im\omega_s\left(t+\frac{d_p}{c}\right)+i\psi_m\right]\,,
\end{equation}
where $a_m$ is the effective amplitude of the $m$th harmonic and $\psi_m$ is the phase. 
Any overall binary-independent normalization, such as propagation or instrumental factors, is absorbed into $a_m$.
SNR degradation is due to phase modulation induced by orbital motion.

Line-of-sight motion will result in a time-dependent pulsar distance and modulate the Fourier phase.
To obtain some intuition, we Taylor-expand the distance and line-of-sight velocity $v_l$
\begin{align}
d_p &= d_{p,0} + v_{l,0} t + \frac{a_{l,0} t^2}{2!} + \frac{j_{l,0} t^3}{3!} + \dots \, , \label{eq:dist} \\
v_l &= v_{l,0} + a_{l,0}t + \frac{j_{l,0} t^2}{2!} + \dots \,,
\end{align}
where $a_l$ and $j_l$ denote the line-of-sight acceleration and jerk respectively, and the subscript $0$ indicates evaluation at $t=0$. 
For an isolated pulsar, the line-of-sight velocity is constant over a survey. 
In a binary system, however, orbital motion produces a time-dependent line-of-sight velocity, and higher-order terms impact the received signal.
A schematic of the binary geometry is given in Fig.~\ref{fig:pulsargeom}.

The Fourier transform of the $m$th harmonic of the received signal is
\begin{equation}
\mathcal{F}[S^{(m)}_R(t)] = \int_0^{t_{\rm int}} S_R^{(m)}(t)\, \exp({-i\omega t})\, dt \,,
\end{equation}
where $S^{(m)}_R(t)$ is the $m$th harmonic of the signal and $\omega$ is the angular Fourier frequency.
Invoking Eqs.~\eqref{eq:dist} and~\eqref{eq:sig}, the corresponding Fourier power is

\begin{align}
\left| \mathcal{F}\!\left[S_R^{(m)}(t)\right] \right|^2
=&
\left|
\int_0^{t_{\rm int}}
a_m \exp\!\left( \frac{i m \omega_s  d_{p,0}}{c} + i \psi_m \right)\right.\nonumber\\
&\left. \times
\exp\!\left[
i m \omega_s
\left(1 + \frac{v_{l,0}}{c}\right) t
\right]\right.\nonumber\\
&\left. \times
\exp\!\left[
 \frac{i m \omega_s}{c}
\left(
 \frac{a_{l,0} t^2}{2!}
+ \frac{j_{l,0} t^3}{3!}
+ \dots
\right)
\right]\right.\nonumber\\
&\left. \times
\exp(-i \omega t)\, dt
\right|^2 \,.
\end{align}
The constant distance, $d_{p,0}$, in the first term contributes an overall phase factor and does not affect the Fourier spectrum. 
A constant line-of-sight velocity, $v_{l,0}$, in the second term Doppler-shifts the signal frequency and does not affect the height of the Fourier peak.
The third term, however, is a time-dependent phase that causes the signal power to spread across neighboring Fourier bins. 

To compensate for the time-dependent phase, searches fit for additional parameters. 
A standard Fourier search fits a constant line-of-sight velocity parameter, $\alpha_v$. 
An acceleration search adds a constant-acceleration parameter, $\alpha_a$, while an acceleration-jerk search further adds a constant-jerk parameter, $\alpha_j$. 
The corresponding coherence factors, $\gamma_{1m}$, $\gamma_{2m}$ and $\gamma_{3m}$, are the normalized magnitudes of the integral of the complex exponential of the residual phase \cite{Johnston1991,Bagchi:2013wga} that is mismodeled by each type of search.\footnote{The definition in terms of the residual, mismodeled phase motivates our chosen term of ``coherence factor.''} 
If the residual phase is zero, $\gamma_{im}^2=1$ and there is no SNR loss. 
If the residual phase varies, the integral averages out and SNR is lost.
For the standard velocity search
\begin{widetext}
\begin{align}
\label{eq:gamma1m}
\gamma_{1m}(\alpha_v)  &\equiv  \frac{1}{t_{\rm int}}  \Biggl\lvert  \int_{0}^{t_{\rm int}} {\rm exp} \left[ \frac{i m \omega_s}{c} \left( v_{l,0} t + \frac{a_{l,0} t^2}{2!} + \frac{j_{l,0} t^3}{3!} + \ldots - \alpha_v t \right) \right] dt  \Biggr\rvert\nonumber\\
&=\frac{1}{t_{\rm int}}
\left|
\int_0^{t_{\rm int}}
\exp\!\left[
\frac{i m \omega_s}{c}
\left(
\int_0^t v_l(t')\, dt'
- \alpha_v\, t
\right)
\right]
dt
\right| \, ,
\end{align}
where in the second equality we have resummed the Taylor expansion and have used $d_p=d_{p,0}+\int_0^t v_l(t')\,dt'$. 
The search maximizes over $\alpha_v$, and $\gamma_{1m}$ characterizes the sensitivity loss when the pulsar is modeled as having a constant velocity while there is additional binary evolution with acceleration $a_{l,0}$, jerk $j_{l,0}$, and higher order terms.
Similarly, the coherence factor for a search over a constant velocity and a constant acceleration is
\begin{align}
\label{eq:gamma2m}
\gamma_{2m}(\alpha_a, \alpha_v)  &\equiv  
\frac{1}{t_{\rm int}} \Biggl\lvert  \int_{0}^{t_{\rm int}} {\rm exp} \left[ \frac{i m \omega_s}{c}   \left( v_{l,0} t + \frac{a_{l,0} t^2}{2!} + \frac{j_{l,0} t^3}{3!} + \ldots  -\alpha_v t - \alpha_a t^2 \right) \right] dt  \Biggr\rvert\nonumber\\
&=
\frac{1}{t_{\rm int}}
\left|
\int_0^{t_{\rm int}}
\exp\!\left[
\frac{i m \omega_s}{c}
\left(
\int_0^t v_l(t')\, dt'
- \alpha_v\, t - \alpha_a t^2 
\right)
\right]
dt
\right| \, ,
\end{align} 
characterizing the sensitivity loss when the pulsar is modeled as having a constant velocity and acceleration, while there is additional binary evolution with jerk $j_{l,0}$ and higher order terms.
The coherence factor of a constant jerk search can be defined analogously.
\end{widetext}


The binary dependence on the sensitivity loss enters through the line-of-sight velocity in Eq.~\eqref{eq:gamma2m}.
Consider a binary with pulsar mass $m_p$, companion mass $m_c$, orbital period $P_b$, eccentricity $e$, semi-major axis of the relative orbit $a$ (i.e., the ellipse traced by the separation between the pulsar and companion), longitude of periastron $\omega_p$, and inclination $\iota$.
From Kepler's third law
\begin{equation}
a^3 = G(m_p+m_c)\left(\frac{P_b}{2\pi}\right)^2\,.
\end{equation}
However the relevant orbit is not the relative one, but the orbit of the pulsar about the barycenter with semi-major axis $a_p$. 
If $a_c$ denotes the semi-major axis of the companion's orbit about the barycenter, then $a=a_p+a_c$ and the center-of-mass condition, $m_p a_p = m_c a_c$, implies
\begin{equation}
a_p = a\,\frac{m_c}{m_p+m_c}\,.
\end{equation}
The factor $m_c/(m_p+m_c)$ converts the relative orbital scale set by Kepler's law into the size of the pulsar's barycentric orbit; the more massive object lies closer to the barycenter.
Projecting the pulsar orbit along the line of sight gives an orbit with semi-major axis
\begin{equation}
\label{eq:app}
a_p' =a_p\sin \iota= \left( \frac{P_b}{2\pi} \right)^{2/3} G^{1/3} \frac{m_c}{(m_p + m_c)^{2/3}} \sin \iota\,.
\end{equation}
The full line-of-sight velocity is then obtained from Keplerian geometry, see Fig.~\ref{fig:pulsargeom},
\begin{equation}
v_l(t) = \frac{2 \pi}{ P_{b}} \, \frac{a_{p}'}{\sqrt{1-e^2}} \,
\left[ \cos\!\left(f(t)+ \omega_p\right) + e \cos\omega_p \right]\,.
\end{equation}
The orbital time dependence enters through the true anomaly $f(t)$ with Keplerian evolution
\begin{equation}
\frac{df}{dt}
=
\frac{2\pi}{P_b}
\frac{\left(1+e\cos f\right)^2}{\left(1-e^2\right)^{3/2}} \,.
\end{equation}
The line-of-sight acceleration and jerk are obtained by time derivatives:
\begin{align}
a_l &= - \left( \frac{2 \pi}{ P_b} \right)^2 \frac{a_p'}{(1 - e^2)^2} \sin(f + \omega_p) (1 + e \cos f)^2\,,\\
j_l &= - \left( \frac{2\pi}{P_b} \right)^3 \frac{a_p'}{(1 - e^2)^{7/2}} (1 + e \cos f)^3 \nonumber\\
&\qquad\times\left[ \cos(f +  \omega_p) + e \cos  \omega_p - 3 e \sin(f +  \omega_p) \sin f \right].
\label{eq:jerk-def}
\end{align}
For an orbit with given period, inclination, and eccentricity, the line-of-sight velocity and its time derivatives inherit the scaling $a_p' \propto m_c(m_p+m_c)^{-2/3}$. 
Thus, both component masses affect the coherence factors; detectability is not uniform across binary parameters, introducing a selection that biases pulsar surveys toward particular systems.

\begin{figure}[t]
\centering
\includegraphics[width=0.48\textwidth]{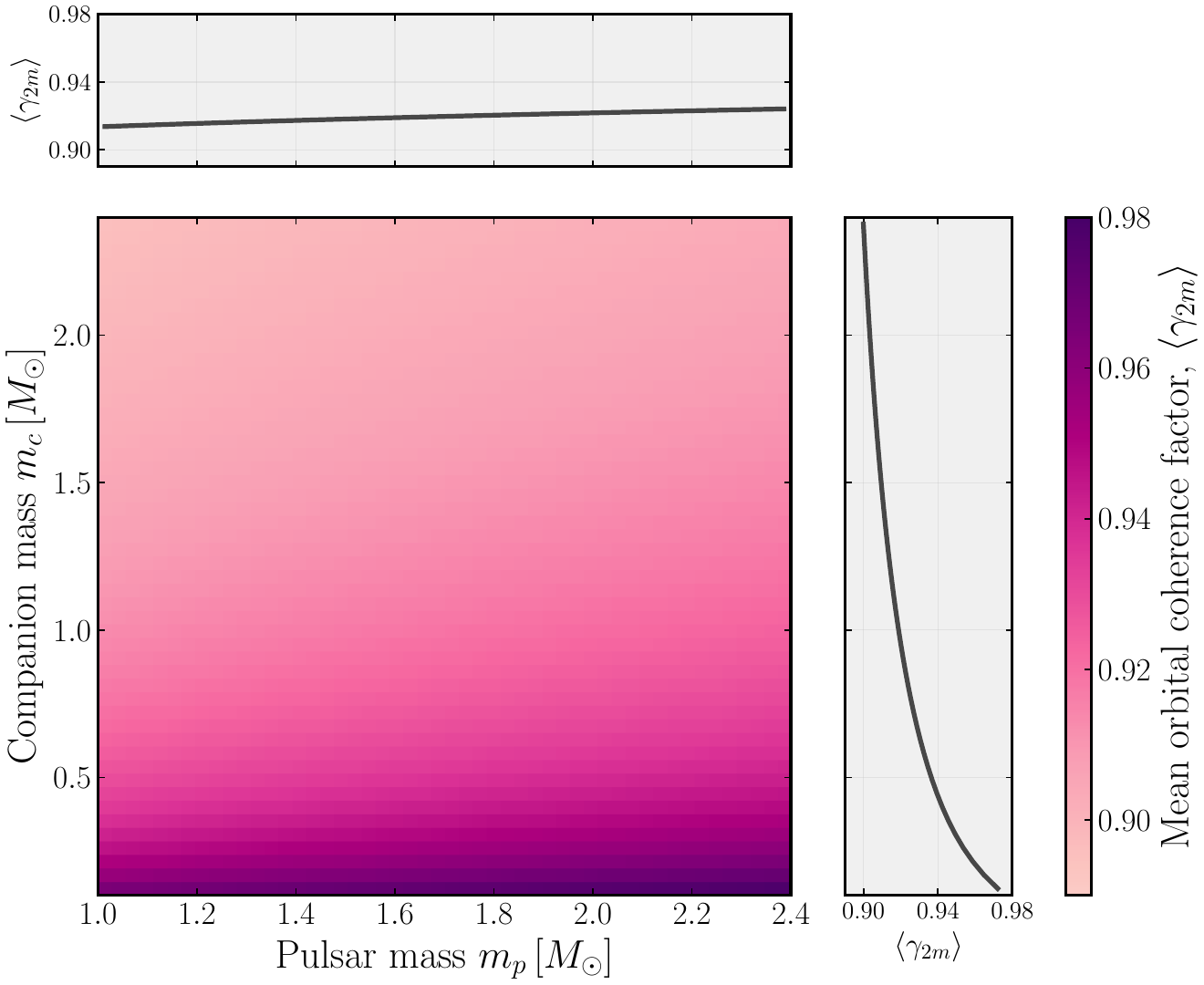}
\caption{Orbital coherence factor $\gamma_{2m}$ as a function of the pulsar mass $m_p$ and companion mass $m_c$, averaged over orbital parameters in the ranges $P_b\in[10^{-3},10^3]\,{\rm day}$, $e\in[0,0.9]$, $\iota\in[0^\circ,90^\circ]$, and $\omega_p\in[0^\circ,360^\circ]$, with $m=2$, $t_{\rm int}=134\,{\rm s}$, and $P_s=40.9\,{\rm ms}$. The upper panel shows $\gamma_{2m}$ as a function of $m_p$, averaged over $m_c$, while the right panel shows it as a function of $m_c$, averaged over $m_p$. On average, \textit{lower companion and higher pulsar masses improve detectability} as $\gamma_{2m}$ is closer to unity. The dependence on pulsar mass is weaker than that on companion mass.}
\label{fig:detectionfraction}
\end{figure}

Before proceeding to the full simulations, we seek intuition through the scaling relations between $\gamma_{2m}$\footnote{The acceleration search is the one that has yielded our dataset, and subsequent simulations are also based on $\gamma_{2m}$.} and key binary or search parameters.
Mismodeling is due to jerk and higher order terms. 
In the limit of a small jerk motion, we assume that $\alpha_v$ has fitted the velocity and $\alpha_a$ the acceleration along the line of sight and we ignore higher order terms.
Defining $\psi_{\rm res}\equiv m\omega_s j_{l,0}/(6 c)$, from Eq.~\eqref{eq:gamma2m} we have
\begin{equation}
\gamma_{2m}\sim  \frac{1}{t_{\rm int}} \Biggl\lvert  \int_{0}^{t_{\rm int}} {\rm exp} \left( i\psi_{\rm res} t^3  \right) dt  \Biggr\rvert\,.
\end{equation} 
For a small jerk (and small residual phase), we Taylor-expand the phase, integrate, and Taylor-expand the amplitude to obtain
\begin{align}
\gamma_{2m} 
&\sim 1- \frac{9}{224}\psi_{\rm res}^2t_{\rm int}^6 \Rightarrow \nonumber \\
&1-\gamma_{2m}^2\sim \psi_{\rm res}^2t_{\rm int}^6
\sim \omega_s^2 j_{l,0}^2 t_{\rm int}^6\,.
\end{align} 
Introducing a function ${\mathcal{G}}(f_0,e,\omega_p)$ that holds the eccentricity, true anomaly, and longitude of periastron dependence of Eq.~\eqref{eq:jerk-def}, we substitute the jerk motion to obtain the fractional SNR loss (one minus the coherence factor squared) as a function of the system parameters:
\begin{equation}
1-\gamma_{2m}^2\sim \omega_s^2 t_{\rm int}^6 P_b^{-14/3} \frac{m_c^2}{(m_p + m_c)^{4/3}} \sin^2 \iota \,\left[{\mathcal{G}}(f_0,e,\omega_p)\right]^2\,.
\end{equation}
This equation is only valid in the limit of a low jerk but reveals general trends:
\begin{enumerate}
\item $1-\gamma_{2m}^2 \sim\omega_s^2$; the signal loss is higher for pulsars with a faster spin that accumulate more phase during the observation time.
\item $1-\gamma_{2m}^2 \sim t_{\rm int}^6$; longer observation times mean more SNR loss, as a larger fraction of the orbit is observed and the pulsar is subject to a larger jerk.
\item $1-\gamma_{2m}^2 \sim P_b^{-14/3}$; tighter orbits with shorter periods have a larger motion variation and signal loss.
\item $1-\gamma_{2m}^2 \sim \sin^2 \iota$; orbits in the plane of the sky (face-on) for which $\iota=0$ have no line-of-sight motion and no signal loss.
\item $1-\gamma_{2m}^2 \sim m_c^{2}/(m_c+m_p)^{4/3}$; for comparable-mass systems with $m_p\approx m_c$, $1-\gamma_{2m}^2 \sim m_p^{2/3}$; for asymmetric systems with $m_p\gg m_c$, $1-\gamma_{2m}^2 \sim m_c^{2}/m_p^{4/3}$; in either case, the dependence on the binary masses is weak, shown in Fig.~\ref{fig:detectionfraction}. 
\end{enumerate}

\subsection{Mass measurements from pulsar timing}
\label{sec:measure}

\begin{figure*}[t]
\centering
\includegraphics[width=\textwidth]{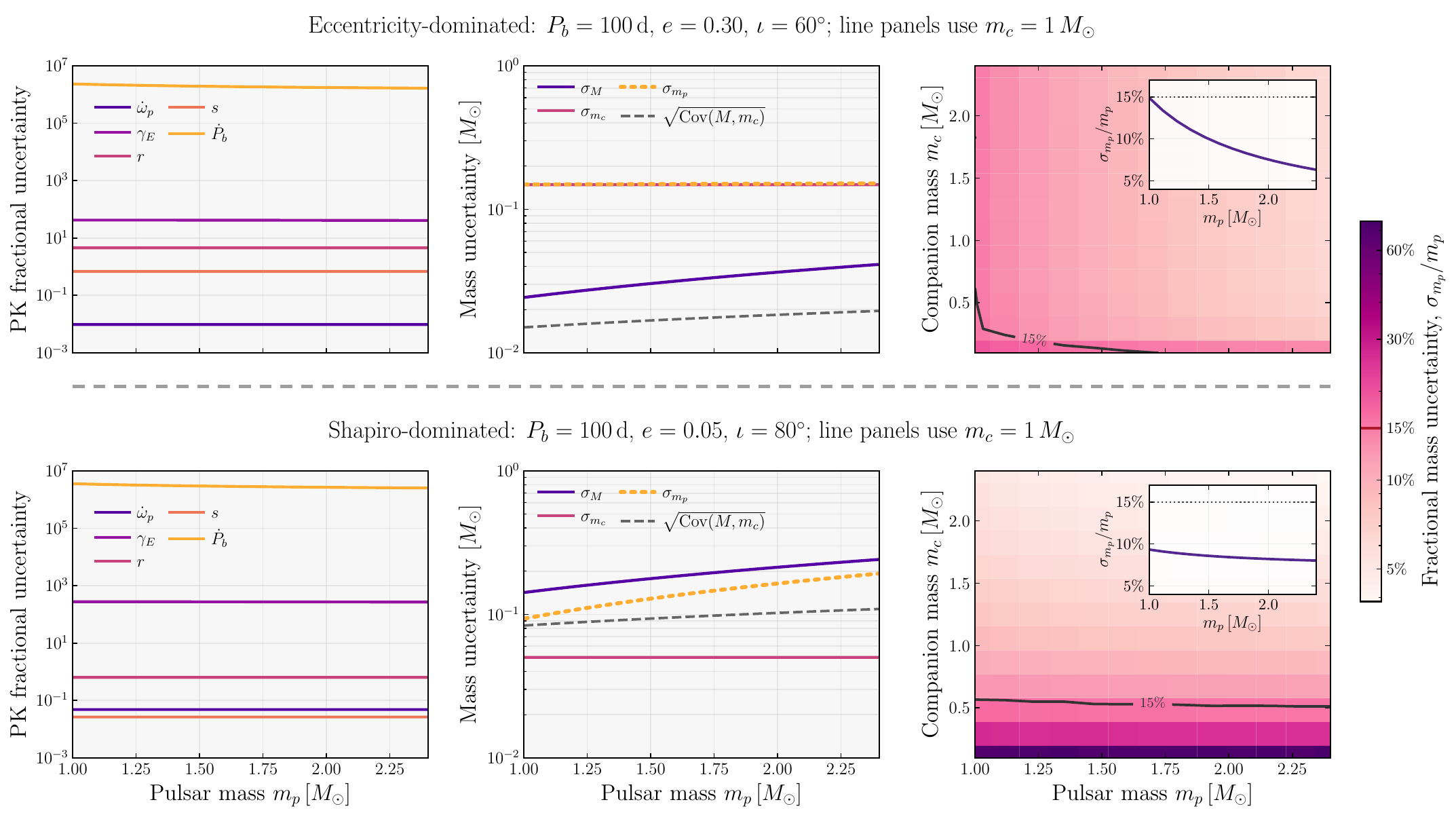}
\caption{
Measurement precision for the pulsar and companion masses for two representative systems: an eccentricity-dominated one with $e=0.30$ and $\iota=60^\circ$ (top row), and a Shapiro-dominated one with $e=0.05$ and $\iota=80^\circ$ (bottom row).
Both systems have $P_b=100\,$d and $m_c=1\,M_{\odot}$ for the left two columns and the right column inset.
Left: fractional uncertainty on the post-Keplerian parameters as a function of the pulsar mass.
Middle: corresponding  absolute uncertainty on the total mass, companion mass, and pulsar mass, together with the covariance between the total and companion masses. 
Right: fractional uncertainty on the pulsar mass as a function of both the pulsar and the companion mass for this configuration. 
The black solid line marks the $15\%$ measurability threshold adopted by the pulsar mass catalogs. 
The insets show the fractional uncertainty on the pulsar mass for $m_c=1\,M_\odot$.
In the eccentricity-dominated configuration, the pulsar mass fractional uncertainty decreases with pulsar mass approximately as $m_p^{-1}$, while in the Shapiro-dominated configuration the dependence is weaker.}
\label{fig:PKmeasure}
\end{figure*}

Verified pulsars receive followup radio observations, Panel 5 of Fig.~\ref{fig:selection_workflow_sketch}.
The radio data consist of pulse TOAs that are fit with a binary timing model. 
At Keplerian order, the projected orbital motion that modulates the TOAs depends on the orbital period $P_b$, eccentricity $e$, projected semi-major axis $a_p'$, longitude of periastron $\omega_p$, and epoch of periastron passage $T_0$ \cite{Lorimer2008}.   
Among them, $P_b$ and $a_p'$ are measured to high precision, the former from the periodicity of the TOA orbital modulation and the latter from the amplitude of the projected light-travel-time delay across the pulsar orbit. 
Combined, they yield the binary mass function:
\begin{equation}
f
=
\left( \frac{2\pi}{P_b} \right)^2 \frac{(a_p')^3}{G}
=
\frac{m_c^3 \sin^3 \iota}{(m_p + m_c)^2}\,.
\label{eq:massfunction}
\end{equation}
The first equality defines $f$ in terms of the measured Keplerian parameters, while the second equality links it to the component masses and inclination. 
Though $f$ is measured with typically negligible uncertainty, it cannot by itself break the degeneracy between the parameters of interest: pulsar mass, companion mass, inclination.

If the pulsar is in a sufficiently compact binary, radio timing can also constrain post-Keplerian parameters that describe relativistic effects on the orbital motion \cite{DamourTaylor1992}. 
In General Relativity, the post-Keplerian parameters are functions of the pulsar and companion masses and the leading-order Keplerian parameters. 
Therefore, post-Keplerian parameters can break the degeneracy of the binary mass function: two independent post-Keplerian measurements yield the component masses, while additional post-Keplerian measurements provide tests of relativistic gravity \cite{Stairs2003}.
The standard post-Keplerian parameters included in a binary timing model are the periastron advance $\dot{\omega}_p$, Einstein delay $\gamma_E$, Shapiro-delay parameters $r$ and $s$, and the orbital period derivative $\dot{P}_b$. 

The Shapiro delay is the additional propagation time incurred on pulses as they pass through the gravitational potential of the companion.
It is most readily measured in nearly edge-on systems, $\sin\iota\simeq 1$, where the pulses travel the closest to the companion along the line of sight. 
For low-eccentricity binaries in the weak-field limit, the delay in the pulse arrival time is
\begin{equation}
\Delta_{\rm SD}=-2r\ln (1-s\sin \Phi)\,,
\end{equation}
where $r$ is the range with units of time, $s$ is the dimensionless shape, and $\Phi$ is the orbital phase measured from the ascending node \cite{Pennucci2015}. 
With $T_\odot=GM_\odot c^{-3}$, the range parameter in General Relativity,
\begin{equation}
r = m_c T_\odot\,,
\label{eq:shapiro-range}
\end{equation}
sets the overall scale of the delay and directly constrains the companion mass: a more massive companion produces a deeper gravitational potential and therefore a larger propagation delay. 
The shape,
\begin{equation}
s = \sin \iota
=
\frac{a_p' M^{2/3}}{c\,m_c T_\odot^{1/3}}
\left(\frac{2\pi}{P_b}\right)^{2/3}\,,
\label{eq:shapiro-shape}
\end{equation}
controls the orbital-phase dependence of the delay and thus the shape of the Shapiro-delay signal across the orbit.
A larger $s=\sin\iota$ (larger shape) produces a sharper and more pronounced delay near superior conjunction, when the pulse passes closest to the companion along the line of sight.
In Eq.~\eqref{eq:shapiro-shape}, the second equality follows from the binary mass function and $M=m_p+m_c$ is the total mass. 
Measuring the Shapiro delay yields the companion mass and the inclination, which when combined with the binary mass function determine the pulsar mass.

The periastron advance $\dot{\omega}_p$ is the relativistic precession rate of the orbit within the orbital plane. 
In General Relativity, it is given by
\begin{equation}
\dot{\omega}_p = \frac{3}{1 - e^2} \left(T_\odot M \right)^{2/3} \left(\frac{2\pi}{P_b}\right)^{5/3}\,.
\label{eq:periastroadvance}
\end{equation}
Thus, a measurement of $\dot{\omega}_p$ combined with the Keplerian parameters $P_b$ and $e$ determines the total mass. 
Periastron advance is more prominent for eccentric binaries, where the point of closest approach can be identified and its advance tracked. 
In circular binaries the orbital separation is constant, so there is no well-localized point of closest approach from which to measure an advance.

The Einstein delay is the periodic variation in pulse arrival times caused by gravitational redshift and special relativistic time dilation over the pulsar's orbit. 
In General Relativity, its amplitude is given by 
\begin{equation}
\gamma_E =
e \left(\frac{P_b}{2\pi}\right)^{1/3}
T_\odot^{2/3}
\frac{m_c(m_p+2m_c)}{M^{4/3}}\,.
\label{eq:einstein-delay}
\end{equation}
The factor of $e$ reflects the fact that this is a phase-dependent effect: in an eccentric orbit, the pulsar's speed and gravitational potential vary around the orbit, so the relation between the pulsar's proper time and the binary barycentric time changes periodically. 
For a circular orbit, the pulsar's speed and potential are constant around the orbit, so there is no periodic variation and $\gamma_E \to 0$. 
Measurement of $\gamma_E$ provides a joint mass constraint.

The orbital period decay is the shrinking of the orbit as the binary loses energy to gravitational radiation, given in General Relativity by
\begin{equation}
\dot{P}_b = -\frac{192\pi}{5}
\left(\frac{2\pi}{P_b}\right)^{5/3}
\frac{1+\frac{73}{24}e^2+\frac{37}{96}e^4}{\left(1-e^2\right)^{7/2}}\,
T_\odot^{5/3}\,\frac{m_p m_c}{M^{1/3}}\,.
\label{eq:pbdot}
\end{equation}
The decay is stronger for more compact orbits, as well as eccentric orbits with closer periastron passages.
Measuring the orbital decay provides another joint constraint on the component masses, namely the chirp mass.

The binary timing model that is fit to pulse TOAs includes the full Keplerian and relativistic post-Keplerian effects either self-consistently within General Relativity or as independent parameters \cite{Fonseca2016}. 
The relative importance of the various effects depends on the binary geometry and observing baseline. 
Analytic scalings for the relative uncertainties of post-Keplerian parameters are given in Table~II of Ref.~\cite{DamourTaylor1992}. 
Secular effects such as $\dot{\omega}_p$ and $\dot{P}_b$ benefit from long timing baselines, because they accumulate coherently over many orbits. 
Eccentric systems are required for $\dot{\omega}_p$ and $\gamma_E$.  
Shapiro delay is primarily controlled by the line-of-sight geometry, with nearly edge-on systems being ideal as the pulses propagate close to the companion before reaching the observer.

Figure~\ref{fig:PKmeasure} illustrates component mass measurability using a timing model with all Keplerian and post-Keplerian parameters, see Sec.~\ref{sec:measurability-meas} for details.
The fit returns the uncertainty and covariance for Keplerian, post-Keplerian, and derived parameters.
We explore the dependence of measurability on the binary configuration for two representative cases: (i) an \textit{eccentricity-dominated} system and (ii) a \textit{Shapiro-dominated} system. 
The first two columns track how the constraints propagate from the post-Keplerian parameters to the system masses as a function of the pulsar mass.
The left column shows the fractional uncertainty on the post-Keplerian parameters, while the middle column shows the uncertainty on the pulsar, companion, and total mass as well as the covariance between the companion and total mass. 
The right column shows the fractional uncertainty on the pulsar mass as a function of the pulsar and companion mass.

The eccentricity-dominated system (top row) has a large eccentricity of $e=0.30$ and an inclination angle of $\iota=60^\circ$, suppressing Shapiro delay and making the relativistic periastron advance $\dot{\omega}_p$ the best-measured post-Keplerian parameter.
Its fractional uncertainty is ${\sim}2$ orders of magnitude lower than the second-best measured Shapiro shape $s$ (left).
The Shapiro-dominated system (bottom row) instead has $e=0.05$ and is viewed nearly edge-on at $\iota=80^\circ$, making $s$ the most precisely measured post-Keplerian parameter, with $\dot{\omega}_p$ a close second (left).
The fractional uncertainty of all post-Keplerian parameters is approximately independent of $m_p$.

All information from post-Keplerian parameters combines to constrain the masses (middle).
The uncertainty scalings with $m_p$ can be analytically understood by considering which post-Keplerian parameters are best constrained for each configuration. 
Overall, the total and component mass uncertainties, $\sigma_M$ and $\sigma_{m_c}$ respectively, scale similarly with $m_p$ for both systems.
But their relative sizes and covariance result in distinct scalings for the pulsar mass uncertainty $\sigma_{m_p}$.

A measurement of the periastron advance primarily constrains the total mass, as $\dot{\omega}_p\propto M^{2/3}$, and thus 
\begin{equation}
\sigma_M
\simeq
\frac{3}{2}
\frac{M}{\dot{\omega}_p}
\sigma_{\dot{\omega}_p}\,.
\end{equation}
The uncertainty in $M$ scales as $\sigma_M\propto M=m_p+m_c$ and consequently increases with $m_p$ (middle).

The companion mass is constrained by combining $s$, the mass function $f$, and $M$ inferred from $\dot{\omega}_p$:
\begin{equation}
m_c=\frac{f^{1/3}M^{2/3}}{s}\,.
\end{equation}
Neglecting the uncertainty in $f$, we have
\begin{equation}
\left(\frac{\sigma_{m_c}}{m_c}\right)^2
\simeq
\left(\frac{2}{3}\frac{\sigma_M}{M}\right)^2
+
\left(\frac{\sigma_s}{s}\right)^2\,.
\end{equation}
Since both terms are approximately independent of $m_p$, $\sigma_{m_c}$ is approximately constant with $m_p$ (middle). 

\paragraph{Eccentricity-dominated configuration.}
Since Shapiro delay is weak, $m_c$ is relatively poorly constrained. 
The companion mass uncertainty is substantially larger than the total mass uncertainty, while the covariance is small (middle). 
Consequently,
\begin{equation}
\sigma_{m_p}\approx\sigma_{m_c}\approx\mathrm{const.}\,,
\end{equation}
and hence (right)
\begin{equation}
\frac{\sigma_{m_p}}{m_p}\propto\frac{1}{m_p}\,.
\end{equation}
This results in a strong decrease of the pulsar mass fractional uncertainty with $m_p$. 
The two-dimensional map also shows that the fractional uncertainty increases for lower companion masses.

\paragraph{Shapiro-dominated configuration.}
Shapiro delay now provides a stronger constraint on $m_c$. 
The well-measured inclination and mass function also produce a strong positive correlation between $M$ and $m_c$ (middle). 
The pulsar mass uncertainty is
\begin{align}
\sigma_{m_p}
&=
\sqrt{
\sigma_M^2+\sigma_{m_c}^2
-2\,\mathrm{Cov}(M,m_c)
} \nonumber\\
&=
\sqrt{
\sigma_M^2+\sigma_{m_c}^2
-2\rho\,\sigma_M\sigma_{m_c}
}\,,
\end{align}
where $\rho$ is the correlation coefficient. 
Since $\rho\simeq1$ in this configuration,
\begin{equation}
\sigma_{m_p}\simeq\sigma_M-\sigma_{m_c}\,.
\end{equation}
Because $\sigma_M\propto M=m_p+m_c$ while $\sigma_{m_c}$ is approximately constant, $\sigma_{m_p}$ inherits a linear scaling with $m_p$ (middle). 
Dividing by $m_p$ therefore produces only a weak decrease in $\sigma_{m_p}/m_p$ with $m_p$ (right), while again the fractional uncertainty increases with lower companion masses.

\section{Selection function framework}
\label{sec:selectioneffects}


\begin{table*}[t]
\caption{Binary and pulsar parameter inputs to the selection function. We list the parameter symbol, its definition, its distribution or fixed value for the synthetic survey, and which pipeline stage (detectability or measurability) it enters.}
\label{tab:input-binary}
\centering
\begin{tabular}{@{}p{1.7cm}p{6cm}p{7cm}p{2.2cm}@{}}
\hline
Parameter & Description & Distribution or value & Pipeline stage \\
\hline

\rowcolor{lightgray}
\multicolumn{4}{@{}l}{\textbf{Binary parameters
$\boldsymbol{\theta}_{\rm bin}$}} \\

$m_p$ & Pulsar mass &
$\mathrm{Uniform}(1.0,2.9)\,M_\odot$ &
Both \\

$m_c$ & Companion mass &
$\mathrm{Uniform}(0.1,2.4)\,M_\odot$ &
Both \\

$\cos\iota$ & Cosine of orbital inclination &
$\mathrm{Uniform}(0,1)$; isotropic orientation
$0^\circ\leq\iota\leq90^\circ$ &
Both \\

$e$ & Orbital eccentricity &
$\mathrm{Uniform}(0,0.9)$&
Both \\

$P_b$ & Orbital period &
$\mathrm{LogUniform}(10^{-3},10^3)\,\mathrm{d}$ &
Both \\

$\omega_p$ & Longitude of periastron &
$\mathrm{Uniform}(0^\circ,360^\circ)$ &
Both \\

$\phi_0$ & Fraction of orbit since first TOA periastron&
$\mathrm{Uniform}(0,1)$ &
Measurability \\

$T_0$ & Epoch of periastron passage &
$T_0=t_{\rm start}-\phi_0P_b$ &
Measurability \\

\hline

\rowcolor{lightgray}
\multicolumn{4}{@{}l}{\textbf{Pulsar-emission parameters
$\boldsymbol{\theta}_{\rm psr}$}} \\

$P_s$ & Pulsar spin period &
$\mathrm{LogUniform}(10^{-3},5)\,\mathrm{s}$ &
Both \\

$w/P_s$ & Pulse duty cycle &
$19.79\%$ &
Both \\

$S$ & Pulsar flux density &
$98.21\,\mu{\rm Jy}$ &
Both \\
\hline
\end{tabular}
\end{table*}


\begin{table*}[t]
\caption{Telescope, discovery, and follow-up timing parameter inputs to the selection function. We list the parameter symbol, its definition, its fixed value for the synthetic survey, and which pipeline stage (detectability or measurability) it enters.
As the discovery survey and targeted timing observations use distinct telescope beam configurations, the corresponding parameters are denoted as $\boldsymbol{\theta}_{\rm tel}^{\rm det}$ and $\boldsymbol{\theta}_{\rm tel}^{\rm meas}$, respectively.}
\label{tab:input-survey}
\centering
\begin{tabular}{@{}p{2cm}p{6.5cm}p{6.5cm}p{2.3cm}@{}}
\hline
Parameter & Description & Value & Pipeline stage \\
\hline

\rowcolor{lightgray}
\multicolumn{4}{@{}l}{\textbf{Stage 1: Detectability telescope parameters
$\boldsymbol{\theta}_{\rm tel}^{\rm det}$}} \\

$G_{\rm det}$ & Telescope gain &
$8.2\,\mathrm{K\,Jy^{-1}}$ &
Detectability \\

$B_{\rm det}$ & Receiver bandwidth &
$100\,\mathrm{MHz}$ &
Detectability \\

$N_{p,{\rm det}}$ & Number of summed polarizations &
$2$ &
Detectability \\

$T_{{\rm sys},{\rm det}}$ & System temperature &
$24\,\mathrm{K}$ &
Detectability \\

\hline

\rowcolor{lightgray}
\multicolumn{4}{@{}l}{\textbf{Stage 1: Discovery observation parameters
$\boldsymbol{\theta}_{\rm obs}^{\rm det}$}} \\

$t_{\rm int}$ & Discovery integration time &
$134\,\mathrm{s}$ &
Detectability \\

$m$ & Fourier harmonic &
$2$ &
Detectability \\

\hline

\rowcolor{lightgray}
\multicolumn{4}{@{}l}{\textbf{Stage 2: Measurability telescope parameters
$\boldsymbol{\theta}_{\rm tel}^{\rm meas}$}} \\

$G_{\rm tim}$ & Follow-up telescope gain &
$10.4\,\mathrm{K\,Jy^{-1}}$ &
Measurability \\

$B_{\rm tim}$ & Follow-up receiver bandwidth &
$100\,\mathrm{MHz}$ &
Measurability \\

$N_{p,{\rm tim}}$ & Number of summed polarizations &
$2$ &
Measurability \\

$T_{{\rm sys},{\rm tim}}$ & Follow-up system temperature &
$24\,\mathrm{K}$ &
Measurability \\

\hline

\rowcolor{lightgray}
\multicolumn{4}{@{}l}{\textbf{Stage 2: Follow-up observation parameters
$\boldsymbol{\theta}_{\rm obs}^{\rm meas}$}} \\

$t_{\rm epoch}$ & Integration time for one epoch &
$7200\,\mathrm{s}$ &
Measurability \\

$N_{\rm epoch}$ & Number of uniformly spaced epochs &
$500$ &
Measurability \\

$t_{\rm baseline}$ & Total timing baseline &
$9.9095471\,\mathrm{yr}=3619.4621\,\mathrm{d}$ &
Measurability \\

$\Delta t_{\rm cadence}$ & Effective uniform cadence &
$t_{\rm baseline}/(N_{\rm epoch}-1)=7.25343\,\mathrm{d}$ &
Measurability \\

$\sigma_{\rm rn}$ & Uncertainty assigned to each epoch &
$w/{\rm SNR}_{\rm epoch}=18.0599\,\mu\mathrm{s}$ &
Measurability \\

\hline
\end{tabular}
\end{table*}

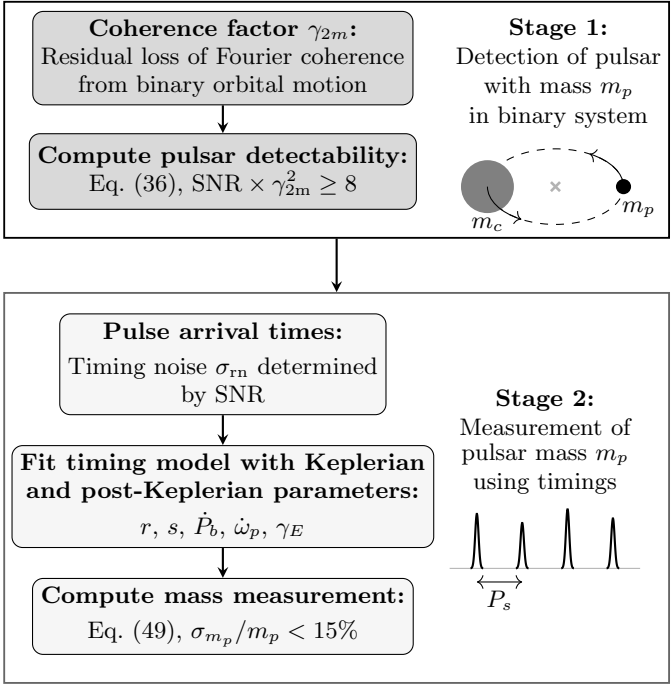
\begin{figure}
\hspace*{-0.35cm}
\begin{tikzpicture}[node distance=2.5cm]

    \node (start) [startstop] {\textbf{Coherence factor $\gamma_{2m}$:} \\ Residual loss of Fourier coherence \\ from binary orbital motion};

    \node (start2) [startstop, below of=start, yshift=+1cm] {\textbf{Compute pulsar detectability:} \\ Eq.~\eqref{eq:SNR-det}, $\rm{SNR} \times \gamma_{2m}^2 \geq 8$};

    \node (pk) [process, below of=start2, yshift=-0.05cm, align=center] {
        \textbf{Pulse arrival times:}\\[0.3em]
         Timing noise $\sigma_{\rm rn}$ determined \\  by SNR
    };
    
    \node (pk2) [process, below of=pk, yshift=+0.75cm, align=center] {
        \textbf{Fit timing model with Keplerian} \\ \textbf{and post-Keplerian parameters:}\\[0.3em]
         $r$, $s$, $\dot P_b$, $\dot\omega_p$, $\gamma_E$
    };

    \node (pk3) [process, align=center, anchor=north] at ($(pk2.south)+(pk2.north)-(pk.south)$) {
        \textbf{Compute mass measurement:}\\[0.3em]
         Eq.~\eqref{eq:meas-threshold}, $\sigma_{m_p}/m_p < 15\%$
    };

    \draw [arrow] (start.south) -- (start2.north);
    \draw [arrow] (pk.south) -- (pk2.north);
    \draw [arrow] (pk2.south) -- (pk3.north);

\node[right=12pt, yshift=14pt] at ($(start.north east)!0.5!(start2.south east)$)
{\shortstack{\textbf{Stage 1:}\\ Detection of pulsar\\ with mass  $m_p$ \\ in binary system}};

\node[right=19pt, yshift=16pt] at ($(pk.north east)!0.5!(pk3.south east)$)
{\shortstack{\textbf{Stage 2:}\\ Measurement of \\ pulsar mass $m_p$ \\ using timings}};

\begin{scope}[shift={(4.4,-1.7)}]

    \draw[gray!60, line width=1pt] (-0.06,-0.06) -- (0.06,0.06);
    \draw[gray!60, line width=1pt] (-0.06,0.06) -- (0.06,-0.06);

    \draw[dashed] (0,0) ellipse (0.9cm and 0.5cm);

    \fill[black] (0.9,0) circle (0.1cm);
    \node at (1.1,-0.3) {\textbf{$m_p$}};

    \fill[gray] (-0.9,0) circle (0.35cm);
    \node at (-0.9,-0.5) {\textbf{$m_c$}};

    \draw[->] (0.9,0) arc (0:60:0.9cm and 0.5cm);
    \draw[->] (-0.9,0) arc (180:240:0.9cm and 0.5cm);

\end{scope}

\begin{scope}[shift={(3,-6.75)}, scale=0.6]

\draw[gray!60] (0,0) -- (4.2,0);

\foreach \c/\a in {
  0.6/1.2,
  1.6/1.0,
  2.6/1.3,
  3.6/1.1
}{
  \draw[thick]
    plot[domain=\c-0.14:\c+0.14,samples=50]
    (\x,{\a*exp(-350*(\x-\c)^2)});
}

\draw[<->, thin] (0.6,-0.3) -- (1.6,-0.3);
\node at (1.1,-0.7) {$P_s$};

\end{scope}

\coordinate (stepone-left) at ($(start.west)+(-0.09cm,0)$);
\coordinate (stepone-right) at ($(start.east)+(3.1cm,0)$);
\coordinate (stepone-top) at ($(start.north)+(0,0.05cm)$);
\coordinate (stepone-bottom) at ($(start2.south)+(0,-0.3cm)$);

\coordinate (steptwo-top) at ($(pk.north)+(0,0.2cm)$);
\coordinate (steptwo-bottom) at ($(pk3.south)+(0,-0.3cm)$);
\coordinate (steptwo-left) at ($(stepone-left |- pk)$);
\coordinate (steptwo-right) at ($(stepone-right |- pk)$);

\node[
  draw=black,
  line width=0.7pt,
  inner xsep=8pt,
  inner ysep=2pt,
  fit=(stepone-left) (stepone-right) (stepone-top) (stepone-bottom)
] {};

\node[
  draw=black!60,
  line width=0.7pt,
  inner xsep=8pt,
  inner ysep=2pt,
  fit=(steptwo-left) (steptwo-right) (steptwo-top) (steptwo-bottom)
] {};

\coordinate (boxmid) at ($(stepone-left)!0.5!(stepone-right)$);
\draw[arrow]
    ($(boxmid |- stepone-bottom)+(0,-0.08cm)$) --
    ($(boxmid |- steptwo-top)+(0,0.08cm)$);

\coordinate (followup-mid) at ($(boxmid |- stepone-bottom)!0.5!(boxmid |- steptwo-top)$);

\end{tikzpicture}
\caption{
Illustration of the selection function calculation. 
We simulate binary systems with parameters listed in Tables~\ref{tab:input-binary} and~\ref{tab:input-survey} and determine whether they would enter the observed catalog in two stages.
\textbf{Step 1, detectability, Sec.~\ref{sec:detectability-meas}:} The binary SNR is adjusted by the coherence factor $\gamma_{2m}^{2}$ which accounts for the residual loss of Fourier coherence caused by binary orbital motion after correcting for constant acceleration, Eq.~\eqref{eq:gamma2m}.
Systems with $\rm{SNR}_{\rm det}$ above $8$ are deemed detected.
\textbf{Step 2, measurability, Sec.~\ref{sec:measurability-meas}:} Pulse TOAs are recorded with timing precision set by the SNR. A timing model with Keplerian and post-Keplerian effects is fit to the TOAs to yield a mass constraint. Systems with $\sigma_{m_p}/m_p<15\%$ are deemed measured and would enter observed catalogs.
}
\label{fig:detpipeline}
\end{figure}

We quantify the selection effects identified in Sec.~\ref{sec:observations} by performing a synthetic survey of a broad, physically-reasonable range of binary pulsars. 
The full process is illustrated in Fig.~\ref{fig:detpipeline}; detectability is discussed in Sec.~\ref{sec:detectability-meas} and measurability in Sec.~\ref{sec:measurability-meas}.
The full selection function that combines both steps is presented in Sec.~\ref{sec:selection-func}; it takes as input binary and  pulsar parameters as well as survey parameters and outputs whether the corresponding system enters the observed catalog.

The binary and pulsar parameters are listed in Table~\ref{tab:input-binary}.
The binary parameters are 
\begin{equation}
\boldsymbol{\theta}_{\rm bin}
=
\{m_p,\,m_c,\,e,\,\cos\iota,\,P_b,\,\omega_p,\,\phi_0\}\,,
\end{equation}
where $m_p$ and $m_c$ are the pulsar and companion masses, $e$ is the orbital eccentricity, $\iota$ is the orbital inclination, $P_b$ is the orbital period, $\omega_p$ is the longitude of periastron, and $\phi_0$ is the initial mean orbital phase.
From these we derive the epoch of periastron passage $T_0$.
The pulsar parameters are
\begin{equation}
\boldsymbol{\theta}_{\rm psr}
=
\{P_s,\,w/P_s,\,S\}\,,
\end{equation}
where $P_s$ is the pulsar spin period, $w$ is the observed pulse width, and $S$ is the pulsar flux density.
The simulated distribution for each parameter is listed in the table (this choice does not impact population inference, Sec.~\ref{sec:hierarchical-models}).

The survey (telescope and observation) parameters are listed in Table~\ref{tab:input-survey}. 
As they are unique to each stage, we define them in the corresponding subsections.
During simulation, all survey parameters are fixed to their observing value, also listed in the table.

\subsection{Stage 1: Detectability}
\label{sec:detectability-meas}

Detectability is based on the radiometer SNR of Eq.~\eqref{eq:SNR}, reproduced here with its parameter dependence:
\begin{align}
{\rm SNR}_{\rm det}
&=
\xi(G_{\rm det},B_{\rm det},N_{p,{\rm det}},T_{{\rm sys},{\rm det}})
\,S\,f_1(t_{\rm int},P_s,w)
\nonumber\\
&\quad\times
\gamma_{2m}^{2}(m_p,m_c,e,\iota,P_b,\omega_p,P_s,t_{\rm int})\,.
\label{eq:SNR-det}
\end{align}
The SNR is adjusted by the acceleration-search coherence factor as acceleration searches are widely implemented in modern surveys and were used to discover the systems considered here.\footnote{Jerk searches are recent and computationally expensive \cite{Andersen2018,Pol2021}.}
We classify a system as detected if ${\rm SNR}_{\rm det}\geq8$. 
While motivated from pulsar surveys~\cite{Lazarus:2015iua}, this threshold is an idealized criterion rather than a complete description of candidate selection, which also involves human verification and repeated observations, Sec.~\ref{sec:data}.

While most binary and pulsar parameters are sampled, the pulsar flux and pulse duty cycle are fixed to values provided in Table~\ref{tab:input-binary}.
This restriction reflects the practical problem that these parameters are in reality frequency- and propagation-dependent and remain poorly measured.
However, they clearly affect selection effects, for example if $S\rightarrow\infty$ all systems would be detectable with no selection effects.
We choose them such that the mass dependence of the combined selection is \emph{maximized}, see Appendix~\ref{app:scale-meas} for details.
Reassuringly, the resulting values are consistent with pulsar catalogs \cite{Manchester:2004bp,psrcat} and faint systems discovered with PALFA~\cite{Scholz2015}; selection effects are a weak function of $S$ and $w/P_s$ for physically plausible values.
Our population results are thus conservative estimates regarding the impact of observational selection effects on the pulsar mass.

The discovery SNR also depends on telescope and observation parameters.
The telescope parameters are
\begin{equation}
\boldsymbol{\theta}_{\rm tel}^{\rm det}
=
\{G_{\rm det},\,B_{\rm det},\,N_{p,{\rm det}},\,T_{{\rm sys},{\rm det}}\}\,,
\end{equation}
where $G_{\rm det}$ is the telescope gain, $B_{\rm det}$ is the observing bandwidth, $N_{p,{\rm det}}$ is the number of summed polarizations, and $T_{{\rm sys},{\rm det}}$ is the system temperature. 
The discovery observation parameters are
\begin{equation}
\boldsymbol{\theta}_{\rm obs}^{\rm det}
=
\{t_{\rm int},m\}\,,
\end{equation}
where $t_{\rm int}$ is the duration of the initial survey observation used to search for and detect the pulsar, and $m$ is the search Fourier harmonic.
We fix the telescope parameters to the Arecibo PALFA values~\cite{Cordes2006}, listed in Table~\ref{tab:input-survey}.\footnote{The ALFA central beam has a gain of ${\sim}10.4\,\mathrm{K\,Jy^{-1}}$, whereas the six outer beams have gains ${\sim}8.2\,\mathrm{K\,Jy^{-1}}$ \cite{Cordes2006}. We use the outer-beam value for the discovery calculation because a blind-survey source need not lie in the central beam, and the central-beam value for targeted timing because a known pulsar can be placed at the most sensitive beam position.\label{footnote:beamgain}}
Similarly, the discovery integration time is set to $t_{\rm int}=134\,\mathrm{s}$ and we adopt the $m=2$ detection harmonic for an orthogonal-rotator approximation~\cite{Pol2021}.
In reality, the pulsars in our dataset have been detected with heterogeneous timing campaigns. 
These choices are a representative configuration and, moreover, their impact is absorbed by the process that leads to the pulsar flux and duty cycle determination, Appendix~\ref{app:scale-meas}.

The detectability dependence on the binary parameters enters through the coherence factor $\gamma_{2m}$, defined in Eq.~\eqref{eq:gamma2m}.
Though in principle $\gamma_{2m}$ should depend on $\phi_0$, namely where in the orbit the pulsar is,\footnote{The observation duration is $t_{\rm int}=134\,\mathrm{s}$ and we simulate orbital periods up to $10^3$\,d.} for computational feasibility we rely on the phase-averaged neural-network emulator for the coherence factor  of \citet{Pol2021}.
The underlying calculation averages the coherence factor over the initial orbital phase, weighted by the time spent in the corresponding part of the orbit \cite{Bagchi:2013wga}.

Finally, the neural-network emulator was trained for $m_c\geq0.2\, M_\odot$ and $m_p\leq2.4\,M_{\odot}$, whereas our population includes companions with support down to $0.1\,M_\odot$ and pulsars up to $2.9\,M_\odot$. 
Although the emulator takes $m_p$, $m_c$ and $\iota$ as separate inputs, these quantities enter the equations exclusively through the pulsar's projected semi-major axis, Eq.~\eqref{eq:app}. 
We exploit this physical degeneracy (rather than extrapolating the neural network below its training range) to gain access to the range $0.1\leq m_c/M_\odot<0.2$ and $2.4\leq m_p/M_\odot<2.9$.
For masses in this range, we adjust the emulator inputs to $m_c\rightarrow0.2\,M_\odot$ and $m_p\rightarrow2.4\,M_\odot$ and adjust the inclination such that the projected semi-major axis remains the same.
This equivalence is exact, as we have confirmed against the full Eq.~\eqref{eq:gamma2m}.

\subsection{Stage 2: Measurability}
\label{sec:measurability-meas}

Systems satisfying the detectability criterion proceed to the timing stage. 
This stage includes $N_{\rm epoch}$ observing epochs, each with duration $t_{\rm epoch}$\footnote{The quantity $t_{\rm epoch}$ is analogous to the $t_{\rm int}$ integration time during discovery, but they refer to distinct campaigns and have distinct numerical values.} over a total time span of $t_{\rm baseline}$.
The observing epochs are distributed uniformly over the timing baseline
with effective cadence
\begin{equation}
\Delta t_{\rm cadence}
=
\frac{t_{\rm baseline}}{N_{\rm epoch}-1}\,.
\end{equation}
The values for the follow-up timing-campaign parameters
\begin{equation}
\boldsymbol{\theta}_{\rm obs}^{\rm meas}
=
\left\{
t_{\rm epoch},\,
N_{\rm epoch},\,
t_{\rm baseline}
\right\}\,,
\end{equation}
are given in Table~\ref{tab:input-survey}.
Each of the $500$ epochs is a two-hour follow-up observation, spread over $10$\,yr, for an effective cadence of ${\sim}7\,\mathrm{days}$. 
A multi-year baseline and hundreds of epochs are representative of timing campaigns used for precision mass measurements, e.g., Ref.~\cite{Martinez2015}. 
Real timing schedules are of course irregular, but uniform spacing isolates the dependence on the binary parameters and avoids introducing additional variation through a randomly chosen schedule. 
The telescope parameters for these observations are
\begin{equation}
\boldsymbol{\theta}_{\rm tel}^{\rm meas}
=
\{G_{\rm tim},\,B_{\rm tim},\,N_{p,{\rm tim}},\,T_{{\rm sys},{\rm tim}}\}\,,
\end{equation}
with values again given in Table~\ref{tab:input-survey}.

Unlike in Sec.~\ref{sec:detectability-meas} where we average over the binary phase, here we assign the pulsar randomly within its orbit.
This is achieved via 
\begin{equation}
\phi_0=(t_{\rm start}-T_0)/P_b\,,
\end{equation}
which corresponds to the fraction of an orbital period elapsed since periastron at the first observation at $t_{\rm start}$.
Per Table~\ref{tab:input-binary}, $\phi_0$ is sampled uniformly and the epoch of periastron is then $T_0=t_{\rm start}-\phi_0P_b$.

Each observing epoch contains approximately
\begin{equation}
N_{\rm pulse}^{\rm epoch}
\simeq
\frac{t_{\rm epoch}}{P_s}
\end{equation}
individual pulses, which are folded to form one average broadband profile. 
Since we fold based on a full binary-orbit timing model, the pulse signal adds coherently while the radiometer noise averages down. 
We associate one folded-profile SNR with each observing epoch, again based on Eq.~\eqref{eq:SNR}:
\begin{align}
{\rm SNR}_{\rm epoch}
&=
\xi(G_{\rm tim},B_{\rm tim},N_{p,{\rm tim}},T_{{\rm sys},{\rm tim}})
\nonumber\\
&\quad\times
S\,f_1(t_{\rm epoch},P_s,w)\,.
\label{eq:snr-meas}
\end{align}
Compared to the discovery SNR of Eq.~\eqref{eq:SNR-det}, we do not include the coherence factor $\gamma_{2m}$, as the analysis uses a full binary-orbit timing model rather than the constant-acceleration approximation.

Based on the epoch's SNR, we calculate the corresponding radiometer-noise timing uncertainty $\sigma_{\rm rn}$. 
Following \citet{Liu2011}, this is
\begin{equation}
\sigma_{\rm rn}
=
\frac{1}{\beta\,{\rm SNR}_{\rm pulse}}
\sqrt{\frac{\Delta}{N_{\rm pulse}^{\rm epoch}}},
\end{equation}
where ${\rm SNR}_{\rm pulse}$ is the single-pulse SNR, $\beta$ describes the sharpness of the pulse template, and $\Delta$ is the noise decorrelation timescale. 
Coherently folding the pulses implies ${\rm SNR}_{\rm epoch}={\rm SNR}_{\rm pulse}\sqrt{N_{\rm pulse}^{\rm epoch}}$, and thus
\begin{equation}
\sigma_{\rm rn}
=
\frac{\sqrt{\Delta}}
{\beta\,{\rm SNR}_{\rm epoch}}
\equiv
\frac{w_{\rm tim}}{{\rm SNR}_{\rm epoch}}\,,
\end{equation}
where $w_{\rm tim}\equiv\sqrt{\Delta}/\beta$.
The timing uncertainty therefore scales inversely with the square root of the epoch duration, $\sigma_{\rm rn}\propto1/\sqrt{t_{\rm epoch}}$.
The term $w_{\rm tim}$ has dimensions of time and depends on the shape of the pulse profile \citep{Liu2011}. 
Since the TOA uncertainty is approximately the characteristic width of the pulse profile divided by the signal-to-noise ratio of the observation \cite{VigelandVallisneri2014}, we identify it with the pulse width $w$, reaching the final expression for the timing uncertainty
\begin{equation}
\sigma_{\rm rn}
\approx
\frac{w}{{\rm SNR}_{\rm epoch}}\,.
\end{equation}
We assign this uncertainty to every epoch and neglect pulse jitter, red timing noise, dispersion-measure variations, and instrumental systematics. 
We also do not add random displacements to the TOAs, so the calculation is deterministic.
As discussed in Sec.~\ref{sec:detectability-meas}, measurability depends on the assumed TOA uncertainty.
We select $S$ and $w/P_s$ to maximize the pulsar-mass dependence of the combined detectability and measurability selection, Appendix~\ref{app:scale-meas}.

The timing model is fit to the pulse TOAs with a per-epoch timing uncertainty $\sigma_{\rm rn}$ using the pulsar-timing software 
\textsc{PINT} \cite{pint}.
The main models available are the ``DD'' model that provides a theory-independent result by fitting the post-Keplerian parameters independently, and ``DDGR'' that imposes their General-Relativity form.
In this work, the DD model is only used for the left column of Fig.~\ref{fig:PKmeasure} to show the relevant measurement accuracy of the post-Keplerian effects.
All other analyses, including the other Fig.~\ref{fig:PKmeasure} columns and the synthetic survey, adopt the DDGR model.  
A suite of \textsc{PINT}-specific parameters is reported in Appendix~\ref{app:timing-fit-configuration}.

\begin{figure*}[t]
\centering
\includegraphics[width=0.94\textwidth]{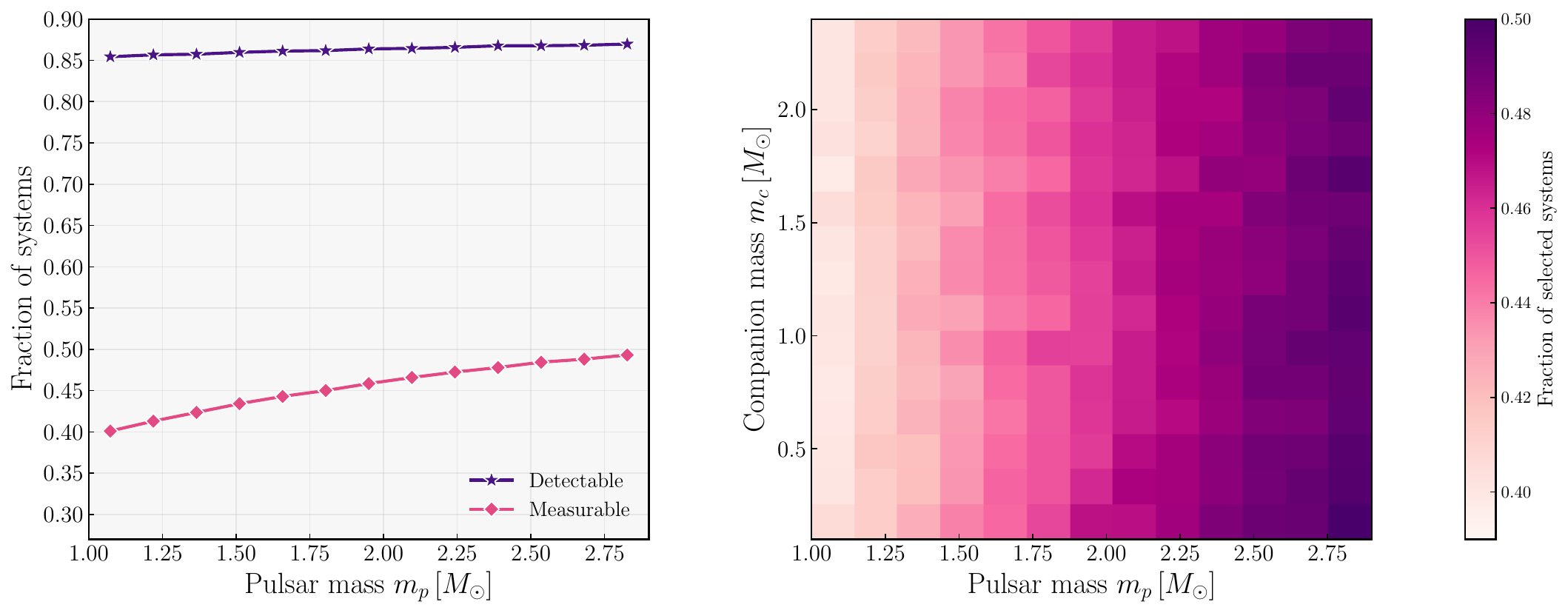}
\caption{
Summary of the synthetic catalog used to compute the selection function. 
The catalog contains $10^7$ simulated binary systems generated using the setup described in Sec.~\ref{sec:selectioneffects}. Left: Fractions of detectable and measurable systems as a function of the pulsar mass $m_p$, marginalizing over all other parameters. Detectability depends only weakly on $m_p$, while measurability has a stronger but still modest dependence. Right: Fractions of systems that are both detectable and measurable as functions of pulsar mass $m_p$ and companion mass $m_c$, marginalizing over all other parameters. 
}
\label{fig:selectionfunction}
\end{figure*}

The fit is based on \texttt{WLSFitter}, the weighted least-squares fitter implemented in \textsc{PINT}.  
The calculation returns the full parameter covariance matrix, from which we extract the 1-$\sigma$ pulsar-mass measurement uncertainty:
\begin{align}
\sigma_{m_p}^{2}
&=
\sigma_M^2+\sigma_{m_c}^2
-2\,\mathrm{Cov}(M,m_c).
\end{align}
A system is counted as measurable if
\begin{equation}
\frac{\sigma_{m_p}}{m_p}<0.15\,,
\label{eq:meas-threshold}
\end{equation}
matching the construction criterion of the catalog \cite{FreireNSMasses}.

\subsection{Selection function}
\label{sec:selection-func}

The selection function that includes both the detectability and measurability stages is depicted in Fig.~\ref{fig:selectionfunction}.
Out of $10^7$ simulated systems, $86\%$ are detectable and $45\%$ are additionally measurable.
Out of the detected systems, $53\%$ yield a measurable pulsar mass. 
By construction, no system is measurable but not detectable.
These numbers show that our simulations probe a wide range of outcomes, however they should not be interpreted as predictions of \emph{how many} systems will realistically be detected by a survey.
The utility of the selection function for our purposes is its dependence on the binary parameters, rather than absolute numbers.

The left panel shows that detectability depends only weakly on pulsar mass, increasing slightly toward higher masses, consistent with Fig.~\ref{fig:detectionfraction} and the analytical scalings of Sec.~\ref{sec:detect}. 
Overall, when marginalizing over all other parameters, a larger pulsar mass produces a smaller orbit about the barycenter, which reduces the line-of-sight motion and the associated loss of signal coherence. 
The dependence of measurability on pulsar mass is moderately stronger, but still weak, consistent with Fig.~\ref{fig:PKmeasure}: ${\sim}40\%$ of the simulated systems have measurable masses at $m_p=1\,M_{\odot}$, rising to ${\sim}46\%$ at $m_p=2\,M_{\odot}$, and ${\sim}49\%$ at $m_p=2.9\,M_{\odot}$.
That dependence is the outcome of a population average of factors that vary over binary configurations, as shown in the systems of Fig.~\ref{fig:PKmeasure}.
The right panel shows the fraction of selected systems over both companion and pulsar masses. 
There is a correlation between the two, as systems with high pulsar and low companion mass are more easily detectable and measurable than systems with low pulsar and high companion mass.

\begin{figure*}
\hspace{-3mm}
\includegraphics[width=0.94\textwidth]{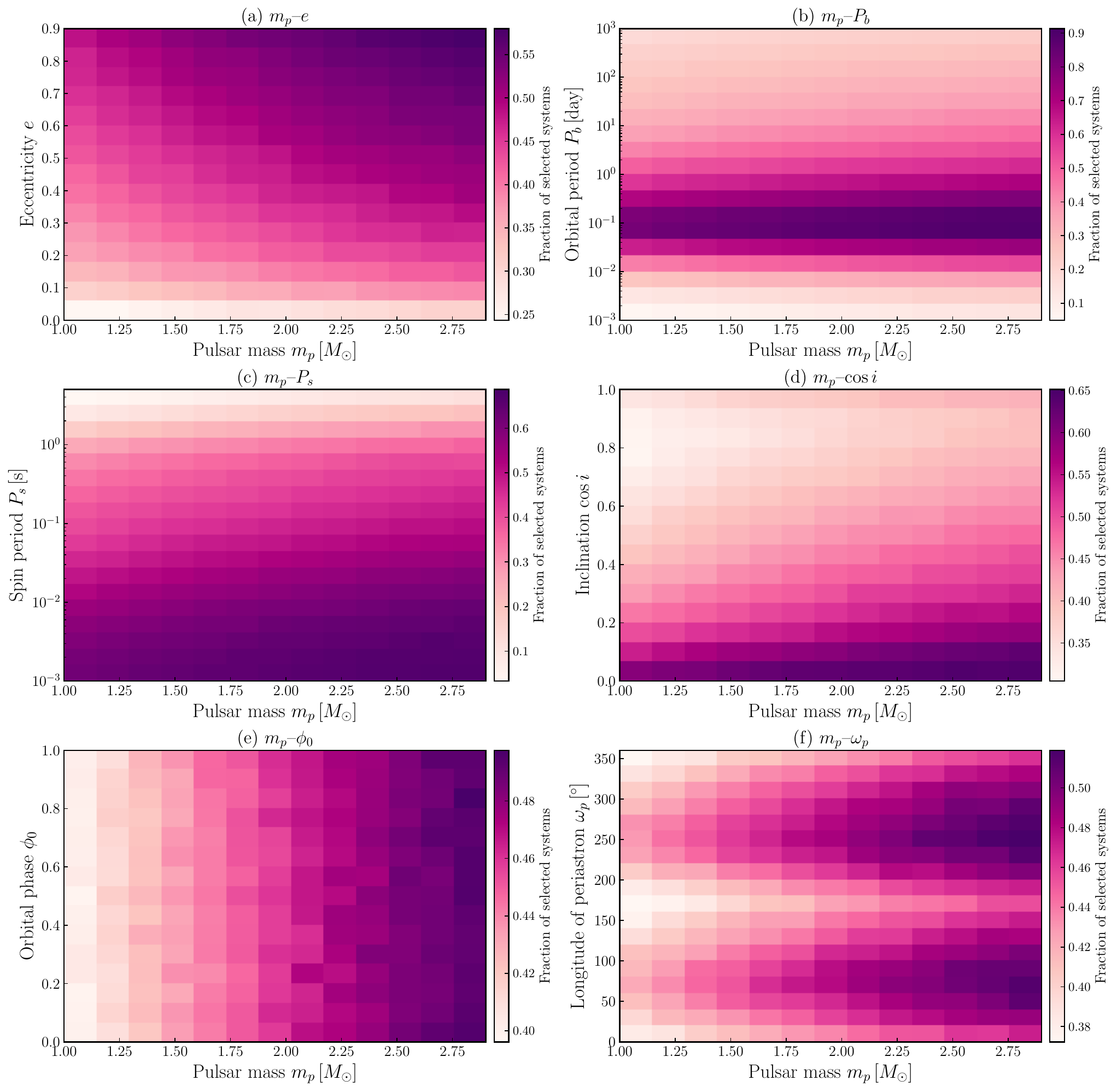}
\caption{
Same as the right panel of Fig.~\ref{fig:selectionfunction} but for other system parameters: fraction of synthetic systems that are both detectable and measurable as a function of the pulsar mass $m_p$ and the eccentricity $e$ in panel (a), the orbital period $P_b$ in panel (b), the pulsar spin period $P_s$ in panel (c), the cosine of the inclination in panel (d), the initial orbit fraction in panel (e), and the longitude of periastron in panel (f). 
In each panel, non-plotted parameters are marginalized over.
The scale of the colormap changes between panels. 
}
\label{fig:selection-other-params}
\end{figure*}

Beyond the binary masses, Fig.~\ref{fig:selection-other-params} shows the fraction of selected systems as a function of different parameters, with all non-plotted parameters marginalized over in each panel.
The selection function varies more prominently as a function of these parameters.
Systems with higher eccentricities (with a fraction of selected systems ranging from ${\sim}0.25$ to ${\sim}0.55$), orbital periods of a few hours (${\sim}0.1$ to ${\sim}0.9$), lower spin periods (${\sim}0.1$ to ${\sim}0.6$), and higher inclination angles (${\sim}0.35$ to ${\sim}0.65$) are more likely to end up in an observational catalog. 
The only additional parameter (beyond the companion mass of Fig.~\ref{fig:selectionfunction}) that is noticeably correlated with the pulsar mass is the eccentricity, with a preference for systems with higher pulsar mass and higher eccentricity.  

\section{Full dataset}
\label{sec:posteriorreconstruction}

\begin{table*}[t]
\caption{Pulsars in our observational dataset. For each pulsar we list the post-Keplerian parameters measured, the reported mass measurement~\cite{FreireNSMasses}, and the reconstructed pulsar mass, companion mass, and inclination. Reconstructed values are posterior medians with 16th--84th percentile intervals. Here $h_3$ and $\varsigma$ are an orthometric re-parametrization of the Shapiro-delay parameters $r$ and $s$~\cite{Freire:2010sv}.}
\label{tab:consolidated-mass-reconstructions}
\centering
\begingroup
\fontsize{6.7}{7.15}\selectfont
\setlength{\tabcolsep}{2.2pt}
\renewcommand{\arraystretch}{0.91}
\resizebox{0.99\textwidth}{!}{%
\begin{tabular}{@{}p{1.80cm}p{1.90cm}p{3.25cm}@{\hspace{1pt}}p{2.45cm}|p{2.42cm}p{2.42cm}p{2.21cm}@{}}
\hline
Pulsar & System type & \shortstack[l]{PK parameters used\\in reconstruction} & \shortstack[l]{Reported $m_p$\\$[M_\odot]$} & \shortstack[l]{Reconstructed\\$m_p\,[M_\odot]$} & \shortstack[l]{Reconstructed\\$m_c\,[M_\odot]$} & \shortstack[l]{Reconstructed\\$\iota\,[{}^\circ]$} \\
\hline
J0218+4232 & PSR--WD & $r,\,s$\,\cite{2024ApJ...966...26T} & $1.49^{+0.23}_{-0.20}$ & $1.50^{+0.23}_{-0.21}$ & $0.180^{+0.017}_{-0.017}$ & $84.8^{+1.1}_{-1.0}$ \\
J0437-4715 & PSR--WD & $r,\,s$\,\cite{2024ApJ...971L..18R} & $1.418(44)$ & $1.415^{+0.040}_{-0.040}$ & $0.2210^{+0.0040}_{-0.0039}$ & $42.494^{+0.016}_{-0.016}$ \\
J0453+1559 & PSR--NS & $\dot{\omega}_p,\,h_3,\,\varsigma^{\ast}$\,\cite{2015ApJ...812..143M} & $1.559(5)$ & $1.5632^{+0.0075}_{-0.0089}$ & $1.1684^{+0.0087}_{-0.0072}$ & $76.2^{+1.5}_{-1.6}$ \\
J0509+3801 & PSR--NS & $\dot{\omega}_p,\,\gamma_E,\,\dot{P}_b$\,\cite{2024ApJ...962..167M} & $1.399(6)$ & $1.3954^{+0.0071}_{-0.0071}$ & $1.4151^{+0.0071}_{-0.0071}$ & $34.31^{+0.20}_{-0.20}$ \\
J0514-4002A & PSR--WD & $\dot{\omega}_p,\,\gamma_E$\,\cite{2503.05466} & $1.39(3)$ & $1.375^{+0.021}_{-0.021}$ & $1.098^{+0.021}_{-0.021}$ & $61.1^{+2.1}_{-1.9}$ \\
J0621+1002$^{\ddagger}$ & PSR--WD & $\dot{\omega}_p,\,s$\,\cite{2012PhDT.........4K} & $1.53^{+0.10}_{-0.20}$ & -- & -- & -- \\
J0641+0448 & PSR--NS & $\dot{\omega}_p,\,r,\,s$\,\cite{2603.10788} & $1.319^{+0.021}_{-0.035}$ & $1.308^{+0.026}_{-0.029}$ & $1.280^{+0.026}_{-0.021}$ & $77.3^{+4.8}_{-4.1}$ \\
J0740+6620 & PSR--WD & $r,\,s$\,\cite{2021ApJ...915L..12F} & $2.08(7)$ & $2.086^{+0.073}_{-0.067}$ & $0.2536^{+0.0057}_{-0.0053}$ & $87.54^{+0.18}_{-0.17}$ \\
J0751+1807 & PSR--WD & $\dot{P}_b,\,h_3,\,\varsigma$\,\cite{2016MNRAS.458.3341D} & $1.64(15)$ & $1.75^{+0.11}_{-0.12}$ & $0.1583^{+0.0075}_{-0.0075}$ & $72.7^{+3.3}_{-3.9}$ \\
J0955-6150 & PSR--WD & $\dot{\omega}_p,\,h_3,\,\varsigma$\,\cite{2203.00607} & $1.71(3)$ & $1.705^{+0.018}_{-0.017}$ & $0.2536^{+0.0018}_{-0.0018}$ & $83.07^{+0.98}_{-1.01}$ \\
J1012-4235 & PSR--WD & $h_3,\,\varsigma$\,\cite{2311.13563} & $1.44^{+0.13}_{-0.12}$ & $1.44^{+0.11}_{-0.10}$ & $0.276^{+0.013}_{-0.013}$ & $87.94^{+0.29}_{-0.29}$ \\
J1125-6014 & PSR--WD & $h_3,\,\varsigma$\,\cite{2023MNRAS.520.1789S} & $1.68^{+0.17}_{-0.15}$ & $1.68^{+0.12}_{-0.12}$ & $0.328^{+0.016}_{-0.015}$ & $77.61^{+0.76}_{-0.77}$ \\
J1141-6545 & PSR--WD & $\dot{\omega}_p,\,\gamma_E,\,\dot{P}_b$\,\cite{2008PhRvD..77l4017B} & $1.27(1)$ & $1.275^{+0.011}_{-0.011}$ & $1.014^{+0.011}_{-0.011}$ & $73.9^{+2.4}_{-2.0}$ \\
J1227-6208 & PSR--WD & $\dot{\omega}_p,\,h_3,\,\varsigma$\,\cite{2407.13593} & $1.54(15)$ & $1.43^{+0.13}_{-0.13}$ & $1.344^{+0.066}_{-0.065}$ & $78.8^{+1.2}_{-1.2}$ \\
J1518+4904$^{\ddagger}$ & PSR--NS & $\dot{\omega}_p,\,s$\,\cite{2024ApJ...966...26T} & $1.470^{+0.030}_{-0.034}$ & -- & -- & -- \\
J1528-3146 & PSR--WD & $\dot{\omega}_p,\,h_3,\,\varsigma$\,\cite{2304.06578} & $1.61^{+0.14}_{-0.13}$ & $1.55^{+0.14}_{-0.14}$ & $1.249^{+0.080}_{-0.078}$ & $59.7^{+1.9}_{-2.0}$ \\
B1534+12 & PSR--NS & $\dot{\omega}_p,\,\gamma_E,\,r,\,s^{\ast}$\,\cite{2014ApJ...787...82F} & $1.3330(2)$ & $1.33274^{+0.00024}_{-0.00024}$ & $1.34576^{+0.00024}_{-0.00024}$ & $77.096^{+0.045}_{-0.045}$ \\
J1543-5149 & PSR--WD & $h_3,\,\varsigma$\,\cite{2025MNRAS.537.2462C} & $1.349^{+0.043}_{-0.061}$ & $1.350^{+0.047}_{-0.046}$ & $0.2233^{+0.0049}_{-0.0049}$ & $88.925^{+0.075}_{-0.076}$ \\
J1614-2230 & PSR--WD & $h_3,\,\varsigma$\,\cite{2023MNRAS.520.1789S} & $1.94(3)$ & $1.944^{+0.026}_{-0.026}$ & $0.4959^{+0.0040}_{-0.0040}$ & $89.181^{+0.017}_{-0.017}$ \\
J1713+0747 & PSR--WD & $h_3,\,\varsigma$\,\cite{2018ApJS..235...37A} & $1.35(7)$ & $1.28^{+0.14}_{-0.13}$ & $0.282^{+0.020}_{-0.019}$ & $72.27^{+0.74}_{-0.75}$ \\
J1748-2446ap & PSR--WD & $\dot{\omega}_p,\,\gamma_E$\,\cite{2403.17799} & $1.700^{+0.015}_{-0.045}$ & $1.697^{+0.016}_{-0.053}$ & $0.294^{+0.053}_{-0.015}$ & $71^{+13}_{-18}$ \\
J1756-2251 & PSR--NS & $\dot{\omega}_p,\,\gamma_E,\,\dot{P}_b,\,s$\,\cite{2014MNRAS.443.2183F} & $1.341(7)$ & $1.3314^{+0.0073}_{-0.0072}$ & $1.2385^{+0.0072}_{-0.0072}$ & $66.16^{+0.77}_{-0.74}$ \\
J1757-1854 & PSR--NS & $\dot{\omega}_p,\,\gamma_E,\,h_3,\,\varsigma^{\ast}$\,\cite{2026arXiv260623926S} & $1.3384(2)$ & $1.33917^{+0.00054}_{-0.00056}$ & $1.39365^{+0.00056}_{-0.00054}$ & $84.35^{+0.23}_{-0.22}$ \\
J1802-2124 & PSR--WD & $r,\,s$\,\cite{1002.0514} & $1.24(11)$ & $1.22^{+0.12}_{-0.11}$ & $0.780^{+0.040}_{-0.039}$ & $79.84^{+0.64}_{-0.62}$ \\
J1807-2500B & PSR--massive companion & $\dot{\omega}_p,\,r,\,s^{\ast}$\,\cite{1112.2612} & $1.3655(21)$ & $1.36734^{+0.00056}_{-0.00056}$ & $1.20457^{+0.00033}_{-0.00033}$ & $85.67^{+0.15}_{-0.15}$ \\
J1829+2456 & PSR--NS & $\dot{\omega}_p,\,\varsigma$\,\cite{2021MNRAS.500.4620H} & $1.306(7)$ & $1.30622^{+0.00084}_{-0.00080}$ & $1.29930^{+0.00080}_{-0.00084}$ & $75.76^{+0.14}_{-0.14}$ \\
B1855+09 & PSR--WD & $h_3,\,\varsigma$\,\cite{2018ApJS..235...37A} & $1.37^{+0.13}_{-0.10}$ & $1.344^{+0.087}_{-0.084}$ & $0.2409^{+0.0098}_{-0.0096}$ & $88.02^{+0.29}_{-0.30}$ \\
J1856-0039 & PSR--NS & $\dot{\omega}_p,\,\gamma_E,\,\dot{P}_b$\,\cite{2607.27333} & $1.304(22)$ & $1.293^{+0.022}_{-0.022}$ & $1.196^{+0.022}_{-0.022}$ & $46.2^{+1.1}_{-1.1}$ \\
J1903+0327 & PSR--MS star & $h_3,\,\varsigma^{\ast}$\,\cite{2011MNRAS.412.2763F} & $1.667(7)^{\ast\ast}$ & $1.641^{+0.068}_{-0.065}$ & $1.020^{+0.025}_{-0.024}$ & $77.53^{+0.41}_{-0.42}$ \\
J1906+0746 & PSR--massive companion & $\dot{\omega}_p,\,\gamma_E^{\ast}$\,\cite{2602.05947} & $1.316(5)$ & $1.3142^{+0.0046}_{-0.0045}$ & $1.2989^{+0.0046}_{-0.0046}$ & $44.68^{+0.20}_{-0.20}$ \\
J1909-3744 & PSR--WD & $h_3,\,\varsigma$\,\cite{2023MNRAS.520.1789S} & $1.45(3)$ & $1.439^{+0.022}_{-0.022}$ & $0.2038^{+0.0020}_{-0.0020}$ & $86.707^{+0.096}_{-0.096}$ \\
J1910-5958A & PSR--WD & $r,\,s$\,\cite{2301.04055} & $1.55(7)$ & $1.548^{+0.072}_{-0.070}$ & $0.2020^{+0.0060}_{-0.0059}$ & $88.922^{+0.012}_{-0.012}$ \\
J1913+1102 & PSR--NS & $\dot{\omega}_p,\,\gamma_E,\,\dot{P}_b,\,h_3$\,\cite{2026arXiv260619276M} & $1.599(8)$ & $1.5995^{+0.0081}_{-0.0081}$ & $1.2899^{+0.0081}_{-0.0081}$ & $54.04^{+0.51}_{-0.49}$ \\
B1913+16 & PSR--NS & $\dot{\omega}_p,\,\gamma_E,\,\dot{P}_b,\,h_3,\,\varsigma$\,\cite{2016ApJ...829...55W} & $1.438(1)$ & $1.43774^{+0.00097}_{-0.00097}$ & $1.39045^{+0.00097}_{-0.00097}$ & $47.109^{+0.043}_{-0.043}$ \\
J1918-0642 & PSR--WD & $h_3,\,\varsigma$\,\cite{2018ApJS..235...37A} & $1.29^{+0.10}_{-0.09}$ & $1.290^{+0.067}_{-0.064}$ & $0.2308^{+0.0076}_{-0.0074}$ & $84.67^{+0.43}_{-0.44}$ \\
J1933-6211$^{\dagger}$ & PSR--WD & $r,\,s$\,\cite{2304.09060} & $1.4^{+0.3}_{-0.2}$ & $1.47^{+0.29}_{-0.27}$ & $0.430^{+0.050}_{-0.049}$ & $54.98^{+1.00}_{-0.98}$ \\
J1943+2210 & PSR--WD & $r,\,s$\,\cite{2025RAA....25a4002Y} & $1.84^{+0.11}_{-0.09}$ & $1.85^{+0.12}_{-0.10}$ & $1.036^{+0.037}_{-0.033}$ & $88.83^{+0.15}_{-0.13}$ \\
J1946+2052 & PSR--NS & $\dot{\omega}_p,\,\gamma_E,\,\dot{P}_b,\,h_3,\,\varsigma$\,\cite{2510.12506} & $1.2842(21)$ & $1.2842^{+0.0020}_{-0.0020}$ & $1.2478^{+0.0020}_{-0.0020}$ & $73.75^{+0.33}_{-0.32}$ \\
J1946+3417 & PSR--WD & $\dot{\omega}_p,\,h_3,\,\varsigma$\,\cite{2017MNRAS.465.1711B} & $1.828(22)$ & $1.829^{+0.019}_{-0.019}$ & $0.2659^{+0.0022}_{-0.0022}$ & $76.2^{+1.1}_{-1.1}$ \\
J1949+3106 & PSR--WD & $\dot{\omega}_p,\,h_3,\,\varsigma$\,\cite{2019ApJ...881..165Z} & $1.34^{+0.17}_{-0.15}$ & $1.32^{+0.12}_{-0.11}$ & $0.802^{+0.041}_{-0.039}$ & $79.92^{+0.78}_{-0.79}$ \\
J1950+2414 & PSR--WD & $\dot{\omega}_p,\,h_3$\,\cite{2019ApJ...881..165Z} & $1.496(23)$ & $1.501^{+0.023}_{-0.023}$ & $0.2783^{+0.0058}_{-0.0044}$ & $76.9^{+3.3}_{-3.9}$ \\
J2023+2853 & PSR--WD & $r,\,s$\,\cite{2025RAA....25a4002Y} & $1.28^{+0.06}_{-0.06}$ & $1.276^{+0.055}_{-0.054}$ & $0.850^{+0.020}_{-0.020}$ & $83.67^{+0.21}_{-0.20}$ \\
J2043+1711 & PSR--WD & $h_3,\,\varsigma$\,\cite{2018ApJS..235...37A} & $1.38^{+0.12}_{-0.13}$ & $1.368^{+0.093}_{-0.088}$ & $0.1718^{+0.0075}_{-0.0073}$ & $82.96^{+0.57}_{-0.58}$ \\
J2045+3633 & PSR--WD & $\dot{\omega}_p,\,h_3,\,\varsigma$\,\cite{2020MNRAS.499.4082M} & $1.251(21)$ & $1.256^{+0.027}_{-0.027}$ & $0.880^{+0.020}_{-0.018}$ & $63.3^{+2.0}_{-2.2}$ \\
J2053+4650 & PSR--WD & $h_3,\,\varsigma$\,\cite{2017MNRAS.470.4421B} & $1.40^{+0.21}_{-0.18}$ & $1.37^{+0.17}_{-0.15}$ & $0.847^{+0.056}_{-0.054}$ & $85.11^{+0.86}_{-0.87}$ \\
B2127+11C & PSR--massive companion & $\dot{\omega}_p,\,\gamma_E,\,\dot{P}_b$\,\cite{2006ApJ...644L.113J} & $1.358(10)$ & $1.3586^{+0.0085}_{-0.0084}$ & $1.3542^{+0.0084}_{-0.0085}$ & $50.13^{+0.43}_{-0.42}$ \\
J2222-0137 & PSR--WD & $\dot{\omega}_p,\,\dot{P}_b,\,h_3,\,\varsigma$\,\cite{2107.09474} & $1.831(10)$ & $1.834^{+0.012}_{-0.012}$ & $1.3207^{+0.0049}_{-0.0049}$ & $85.241^{+0.068}_{-0.068}$ \\
J2234+0611 & PSR--WD & $\dot{\omega}_p,\,h_3$\,\cite{2019ApJ...870...74S} & $1.353^{+0.014}_{-0.017}$ & $1.354^{+0.015}_{-0.021}$ & $0.298^{+0.020}_{-0.015}$ & $41.5^{+2.8}_{-3.2}$ \\
\hline
\end{tabular}
}%
\vspace{0.35em}
\begin{minipage}{0.985\textwidth}
\fontsize{6.4}{7.0}\selectfont
$^{\ast}$The post-Keplerian list is not identical to the one in Freire~\cite{FreireNSMasses}; individual cases are discussed in Sec.~\ref{sec:fullpulsarsample}. $^{\ast\ast}$The reported mass is the source paper's 99.7\% interval, rather than a 68.3\% interval; we divide its half-width by three to obtain the $1\sigma$ uncertainty. $^{\dagger}$Although the reported pulsar mass satisfies the 15\% fractional uncertainty criterion, the reconstructed mass does not. $^{\ddagger}$No component-mass posterior is reconstructed; see Sec.~\ref{sec:fullpulsarsample}. 
\end{minipage}
\endgroup
\end{table*}

While our main target is the pulsar mass, its posterior for a given system is correlated with the companion mass and inclination through the post-Keplerian parameters.
In other words, the 1-dimensional constraints reported in catalogs may not be representative of the full multidimensional posterior for each observed system.
Before proceeding with the population analysis, therefore, we reconstruct approximate joint posteriors for the masses and inclination from the reported post-Keplerian measurements.
We describe the method in Sec.~\ref{sec:pkposteriorreconstruction} and present our final dataset in Sec.~\ref{sec:fullpulsarsample} and Table~\ref{tab:consolidated-mass-reconstructions}. 

\subsection{Reconstruction of the correlated posteriors}
\label{sec:pkposteriorreconstruction}

Denoting collectively $\theta_{\rm PK}=\{m_p,m_{c},\iota\}$, the multidimensional posterior for each pulsar is $p(\theta_{\rm PK}\mid d)$,
where $d$ are the data (timing constraints).
The set of measured post-Keplerian parameters is
\begin{equation}
\boldsymbol{y}
=
(y_{1},\ldots,y_{K})\,,
\end{equation}
with measured values $y_{k}^{\rm obs}$ and uncertainties $\sigma_{k}$. 
Since the post-Keplerian parameters are individually constrained from distinct timing phenomena, we assume that their likelihood is a product of independent Gaussians for each parameter:
\begin{equation}
p(d\mid\theta_{\rm PK})
\propto
\prod_{k=1}^{K}
\exp\left[
-\frac{
\left(
y_{k}^{\rm GR}(\theta_{\rm PK})
-y_{k}^{\rm obs}
\right)^2
}{
2\sigma_{k}^2
}
\right]\,,
\label{eq:pk-reconstruction-likelihood}
\end{equation}
where $y_{k}^{\rm GR}$ is the General-Relativity prediction. 
This form is not exact, but the full likelihood and covariance between post-Keplerian measurements are generally unavailable; it is a more defensible approximation than an independent-Gaussian form for $\theta_{\rm PK}$.
For each pulsar, we adopt Eq.~\eqref{eq:pk-reconstruction-likelihood} with the reported post-Keplerian measurements together with a perfectly-measured binary mass function $f^{\rm obs}$ to reconstruct the $\theta_{\rm PK}$ posterior.
We also impose a uniform prior \emph{on the post-Keplerian parameters} which induces a nontrivial prior on $\theta_{\rm PK}$ via the Jacobian of the transformation. Overall, for the case of $2$ post-Keplerian parameters
\begin{equation}
p(\theta_{\rm PK}\mid d)
\propto
p(y_{a},y_{b}\mid d)\delta(f-f^{\rm obs})
\left|
\det
\frac{
\partial(y_{a}^{\rm GR},y_{b}^{\rm GR},f)
}{
\partial(m_p,m_{c},\iota)
}
\right|.
\label{eq:two-pk-reconstruction}
\end{equation}
An analogous calculation applies if additional post-Keplerian parameters are available.

This reconstruction step allows us to only include mass information that comes from standard post-Keplerian timing, per Sec.~\ref{sec:measurability-meas}, in our analysis.

\begin{figure*}[t]
\hspace{-2em}
\includegraphics[width=0.95\textwidth]{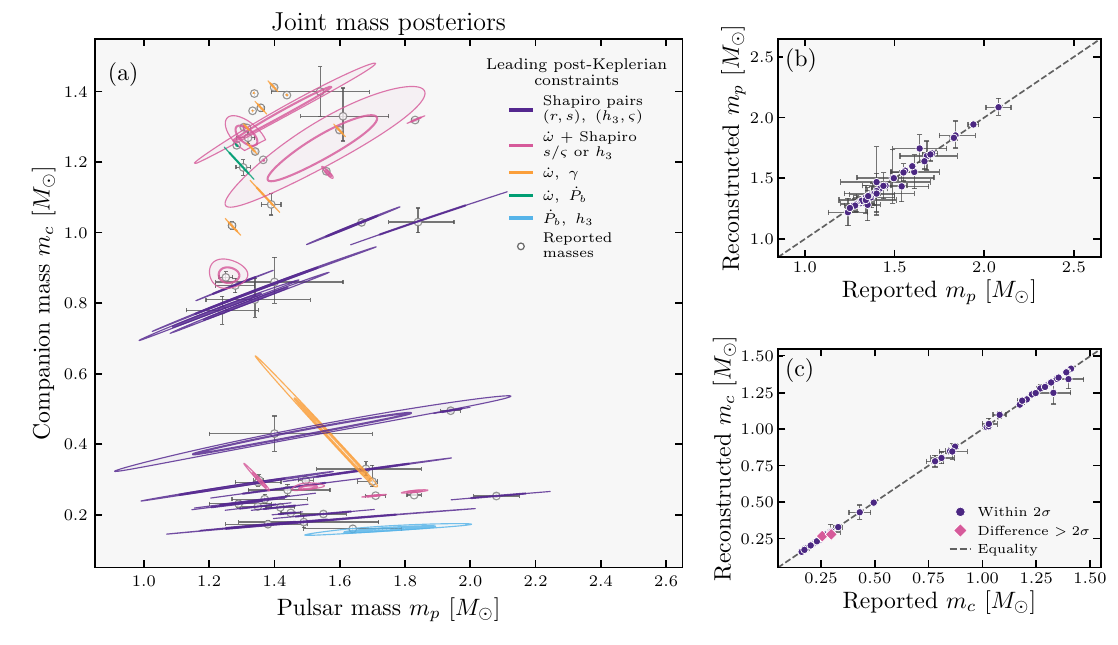}
\caption{Reported and reconstructed component masses for the pulsars of Table~\ref{tab:consolidated-mass-reconstructions}. Panel (a) shows contours enclosing 50\% and 90\% of each reconstructed joint $(m_p,m_c)$ posterior, colored by the leading constraints (the two post-Keplerian measurements with the smallest fractional uncertainties). Grey circles show the reported component masses, with horizontal and vertical bars indicating their quoted 1-$\sigma$ uncertainties \cite{FreireNSMasses}. Panels (b) and (c) compare the reconstructed and reported pulsar and companion masses, respectively. Horizontal bars show the reported uncertainties, while vertical bars show the reconstructed 16th--84th percentile intervals after marginalizing the multidimensional reconstructed posterior. Dashed lines indicate equality and pink diamonds mark differences exceeding 2-$\sigma$. }
\label{fig:reported-reconstructed-masses}
\end{figure*}

\subsection{Final pulsar sample}
\label{sec:fullpulsarsample}

Table~\ref{tab:consolidated-mass-reconstructions} lists the final dataset, with the measured post-Keplerian parameters, the reported 1-dimensional masses from the Freire catalog \cite{FreireNSMasses}, and our reconstructed $\theta_{\rm PK}$ posteriors.
The reconstructed-vs-reported 2-dimensional mass posterior for the component masses is shown in Fig.~\ref{fig:reported-reconstructed-masses}.
The pulsar and companion masses are strongly correlated, and the shape of the correlation depends on which post-Keplerian parameters are measured (left).
When marginalized to 1 dimension, the reconstructed posteriors agree well with the reported 1-dimensional component masses (right).

\paragraph{Reported sample.}
The pulsar list was assembled from the pulsar mass catalog maintained by Freire \cite{FreireNSMasses}. 
As of August 2026, the catalog contains 54 systems with pulsar mass measurements, corresponding to 55 pulsars (PSR~J0737--3039 contains two pulsars). 
We retain the 48 systems whose individual pulsar masses are inferred from binary radio timing analogous to Sec.~\ref{sec:measurability-meas} and satisfy the $15\%$ precision criterion (for asymmetric measurements, the criterion is applied to the smaller of the reported upper and lower $1\sigma$ uncertainties).

Six systems, with seven pulsars, are excluded: 
\begin{itemize}
    \item PSR~J0337+1715 requires a specialized dynamical model of its hierarchical triple system \cite{Ransom2014}. 
    \item  PSRs~J0348+0432, J1012+5307 and J1738+0333 require optical measurements of the white-dwarf companions \cite{Antoniadis2012,Antoniadis2013}. 
    \item B2303+46 combines an inferred white dwarf mass with a timing measurement of the total mass \cite{vanKerkwijkKulkarni1999}.
    \item PSR~J0737--3039A/B require the double-pulsar mass ratio and a specialized timing model \cite{Kramer2021}. 
\end{itemize}

For PSR~J1906+0746, the catalog lists both a standard timing solution and an alternative solution incorporating information from relativistic geodetic precession \cite{Desvignes2019}. 
We report only the former result inferred from $\dot{\omega}_p$, $\gamma_E$ and $\dot P_b$.

\paragraph{Reported post-Keplerian constraints.}
For secular post-Keplerian parameters, we use the intrinsic relativistic contribution after removing the kinematic corrections reported in the source paper. 
We use the observed value only when the source reports that these corrections are negligible relative to the measurement uncertainty.

We convert $\dot{\omega}_p$ into a total mass constraint using the leading-order relation of Eq.~\eqref{eq:periastroadvance}. 
We do not include post-Newtonian corrections in order to keep the reconstruction consistent across the dataset. 
For PSRs~J1757--1854 \citep{2026arXiv260623926S} and J1946+2052 \citep{2510.12506}, $\dot{\omega}_p$ is sufficiently precisely measured that higher-order terms and spin-orbit contributions may be relevant. 
Their reconstructed posteriors are therefore approximations.

For most systems, we use the exact post-Keplerian measurements listed in the catalog~\cite{FreireNSMasses}. 
We make the following alterations:
\begin{itemize}
    \item For PSR~J0453+1559, we additionally include the published $\varsigma$ measurement \citep{2015ApJ...812..143M}.

    \item For PSR~B1534+12, we omit $\dot P_b$ because its intrinsic value requires a substantial distance-dependent kinematic correction \citep{2014ApJ...787...82F}.

    \item For PSR~J1757--1854, we use the published $\dot{\omega}_p$, $\gamma_E$, $h_3$, and $\varsigma$\footnote{The quantities $h_3$ and $\varsigma$ are an orthometric re-parametrization of the Shapiro-delay parameters $r$ and $s$~\cite{Freire:2010sv}.} measurements. We omit post-Keplerian parameters not considered in Sec.~\ref{sec:measurability-meas}. We also omit $\dot P_b$, because its intrinsic value has a poorly constrained kinematic correction \citep{2026arXiv260623926S}.

    \item For PSR~J1903+0327, we omit $\dot{\omega}_p$ because its kinematic contribution depends on the unknown orbital orientation and is larger than the measurement uncertainty. We reconstruct the component masses from $h_3$ and $\varsigma$
    \citep{2011MNRAS.412.2763F}.

    \item For PSR~J1906+0746, we omit $\dot P_b$ because the uncertain kinematic correction dominates the uncertainty of the intrinsic orbital decay \citep{2602.05947}.
\end{itemize}

\paragraph{Reconstructed masses.}
We obtain $\theta_{\rm PK}$ posteriors for 46 of the 48 systems. The exceptions are:
\begin{itemize}
    \item For PSR~J0621+1002, the Shapiro delay non-detection constrains the masses and inclination, but the source does not report a measured value and uncertainty for $s$.

\item For PSR~J1518+4904, an inclination is reported, but it is not derived from the Shapiro delay alone: the analysis also uses the constraint from $\dot{\omega}$ together with the mass function. Converting this inclination into $s$ and then combining it with $\dot{\omega}$ would therefore use the $\dot{\omega}$ information twice, so we do not reconstruct a mass posterior for this system.
\end{itemize}
Finally, although the reported pulsar mass for PSR~J1933--6211 satisfies the 15\% fractional uncertainty criterion, its reconstructed mass uncertainty does not.
We retain this pulsar and in what follows analyze the 46 systems with reconstructed mass posteriors.

\section{Binary pulsar population models}
\label{sec:hierarchical-models}

With the dataset and simulated survey in hand, we infer the population distribution for the binary parameters and the pulsar spin using the hierarchical inference framework detailed in Appendix~\ref{app:derivseleffects}.
The framework includes the following elements:
\begin{enumerate}
    \item For each of the $N_{\rm p}$ pulsars, we have timing data $d_i$.
    \item With data $d_i$ we constrain for each pulsar parameters $\theta_i$, via the likelihood $p(d_i\mid\theta_i)$. The individual-system parameters include the sampled parameters of Table~\ref{tab:input-binary}: $\theta=\{m_p,m_c,\iota,e,P_b,\omega_p,\phi_0,P_s\}$.
    \item Individual-system parameters have an astrophysical population distribution, $p(\theta\mid\Lambda)$ parametrized by hyperparameters $\Lambda$ that we infer.
\end{enumerate}

The posterior for $\Lambda$ given all the data $d=\{d_i\}$ combines the likelihood derived in Appendix~\ref{app:derivseleffects} with a prior for the hyperparameters $p(\Lambda)$,
\begin{align}
p(\Lambda\mid d)
&\propto
p(\Lambda)\,
\left[
\frac{1}{\xi(\Lambda)}
\right]^{N_p}
\prod_{i=1}^{N_p}
\mathcal{L}_i(d_i\mid\Lambda)\,,
\\
\mathcal{L}_i(d_i\mid\Lambda)&=\int \diff \theta_i\, p\left(d_i \mid \theta_i\right) p\left(\theta_i \mid \Lambda\right)\,,
\\
\xi(\Lambda)&=\int \diff d\, \diff \theta\, p(\text {det} \mid {d}) p({d}, {\theta} \mid \Lambda)\,.
\end{align}
Here $\mathcal{L}_i(d_i\mid\Lambda)$ is the contribution of each individual system to the population likelihood, in practice computed with Eq.~\eqref{eq:hier-numerator-samples} and
$\xi(\Lambda)$ is the selection function, evaluated through the simulated systems of Sec.~\ref{sec:selectioneffects} as given in Eq.~\eqref{eq:hier-selection-samples}
\begin{equation}
\xi(\Lambda)
=
\frac{1}{N_{\rm sim}}
\sum_{k\in{\rm select}}
\frac{
p(\theta_k\mid\Lambda)
}{
p(\theta_k\mid{\rm sim})
}\,.
\end{equation}
Out of $N_{\rm sim}$ simulated systems, the sum runs only over the ``selected'' ones that pass the detectability and measurability thresholds, which are then reweighted from the ``simulation'' distribution of Table~\ref{tab:input-binary} to the population distribution.\footnote{The Monte-Carlo evaluation of this integral is validated in Appendix~\ref{app:validationofselfunc}.}
This equation confirms that the chosen simulation distribution does not impact the population inference as it is reweighted away.
It further clarifies that population inference among different parameters is linked through the selection function, e.g. the population distribution of the pulsar spin impacts its mass inference.
This is why, even if we were only interested in the pulsar mass, in the population analysis we would need to consider all relevant parameters.

\begin{widetext}
We model the population as
\begin{align}
p\left(\theta \mid \Lambda\right) &= 
{\cal{N}}^{(2)}_{[1.0,M_{\rm pop}]}(m_p\mid \mu_{1,m_p},\sigma_{1,m_p},\mu_{2,m_p},\sigma_{2,m_p},f_{m_p})\,
{\cal{N}}^{(2)}_{[0.1,2.4]}(m_c\mid \mu_{1,m_c},\sigma_{1,m_c},\mu_{2,m_c},\sigma_{2,m_c},f_{m_c})\nonumber \\
&\quad \times
 {\cal{N}}_{[-3,3]}(\log_{10} P_b\mid \mu_{P_b},\sigma_{P_b})\,
 {\cal{N}}_{[-3,\log_{10} 5]}(\log_{10} P_s\mid \mu_{P_s},\sigma_{P_s})\,
 {\cal{N}}^{(2)}_{[0,0.9]}(e\mid 0,\sigma_{1,e},\mu_{2,e},\sigma_{2,e},f_{e})\nonumber \\
&\quad\times {\cal{U}}_{[0,1]}(\cos \iota)\, 
 {\cal{U}}_{[0,2\pi]}\,(\omega_p)\,
 {\cal{U}}_{[0,1]}(\phi_0) \,,
 \label{eq:pop-model-func}
\end{align}
where ${\cal{U}}_{[A,B]}(X)$ is a uniform distribution for $X\in[A,B]$, ${\cal{N}}_{[A,B]}(X\mid \mu_{X},\sigma_{X})$ is a Gaussian distribution for parameter $X\in[A,B]$ with mean $\mu_X$ and standard deviation $\sigma_X$, and ${\cal{N}}^{(2)}_{[A,B]}(X\mid \mu_{1},\sigma_{1},\mu_{2},\sigma_{2},f)$ is a 2-component Gaussian mixture for variable $X\in[A,B]$:
\begin{equation}
\label{eq:nsdist}
{\cal{N}}^{(2)}_{[A,B]}(X\mid \mu_{1},\sigma_{1},\mu_{2},\sigma_{2},f)
=f\,\mathcal{N}_{[A,B]}(X\mid\mu_1,\sigma_1)
+
(1-f)\,\mathcal{N}_{[A,B]}(X\mid\mu_2,\sigma_2)\,.
\end{equation}
\end{widetext}

The functional form of these distributions is motivated by physical considerations and inspection of the measured dataset in Appendix~\ref{app:all-posteriors}.
All orientation parameters, $(\cos\iota,\omega_p)$, are distributed isotropically, while the orbital phase $\phi_0$ is distributed uniformly; their population distributions are fixed.
The (base 10) logarithmic orbital and spin periods are modeled with a Gaussian whose mean and standard deviation are inferred.
For the component masses we adopt the standard Gaussian mixture, with the pulsar mass distribution truncated at a maximum $M_{\rm pop}$.
We assume the same functional form for the eccentricity within its physical bounds, but also center the first Gaussian at $\mu_1=0$.
The adopted model assumes that parameters are independently distributed, i.e., there is no term that couples two or more parameters. 
This is again motivated by Appendix~\ref{app:all-posteriors}; exploring more complete models is left for future work. 
Hyperparameter priors are uniform with bounds listed in Appendix~\ref{app:hyperparam-prior}.

\begin{figure*}[!t]
\includegraphics[width=\textwidth]{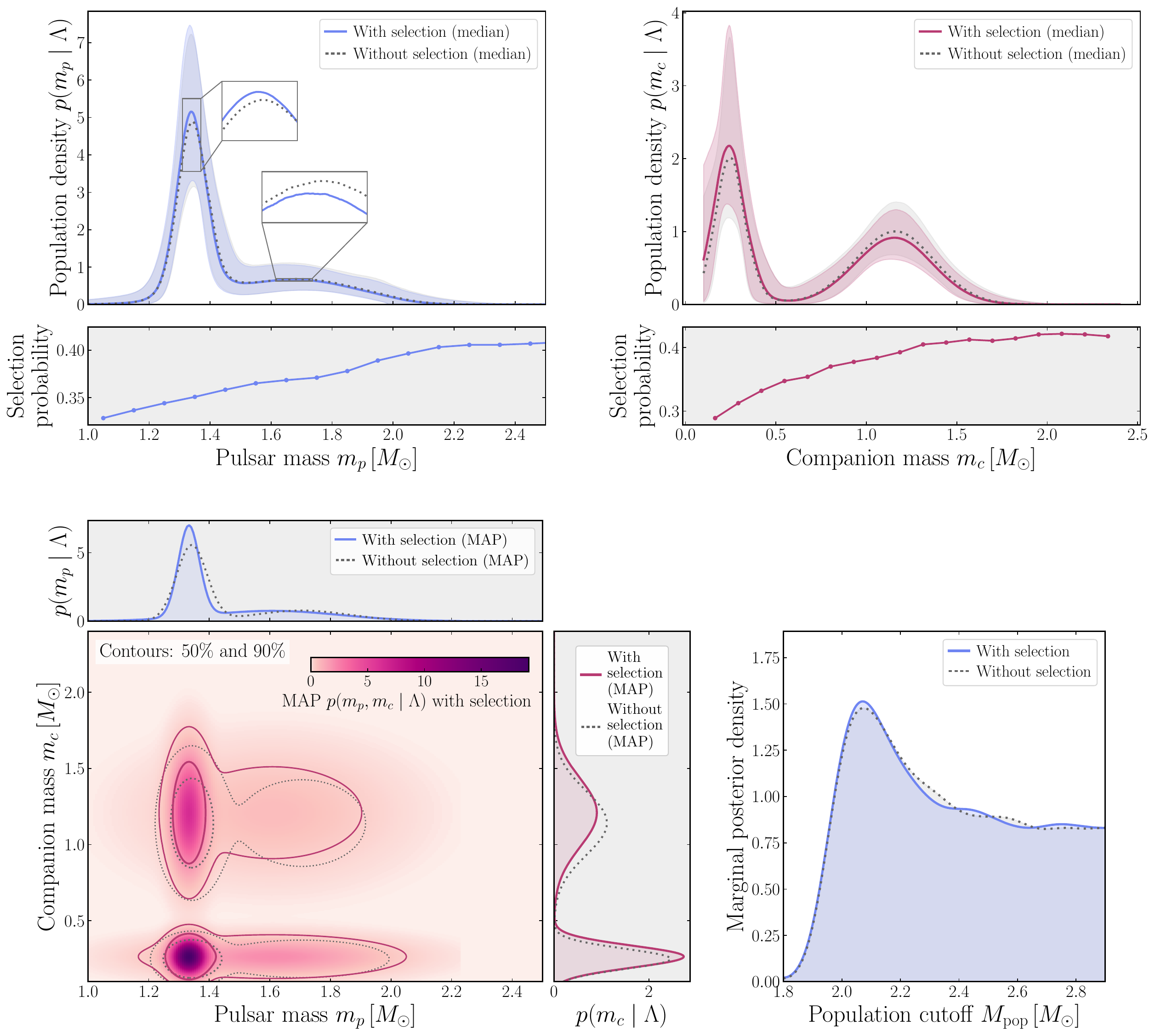}
\caption{Population distributions for the pulsar (top left) and companion (top right) mass. Solid colored curves include selection effects, while dotted gray curves neglect them. The curves show the median population density across the hyperposterior; the shaded bands show the corresponding central $90\%$ credible interval. The insets magnify the two  peaks in the pulsar mass distribution. The lower part of the top panels show the selection probabilities marginalized over the inferred distributions of the other binary parameters. Points denote binned simulated selection catalog values. The bottom-left panel shows the maximum \emph{a posteriori} (MAP) joint density $p(m_p,m_c\mid\Lambda)$ and its marginals. The colored density and solid magenta contours include selection effects, while the dotted gray contours neglect them; the contours enclose $50\%$ and $90\%$ of each joint density. The bottom-right panel shows the marginalized posterior for $M_{\rm pop}$, with and without selection effects.}
\label{fig:analysisI}
\end{figure*}

\begin{figure}
\hspace{-2em}
\centering
\includegraphics[width=0.41\textwidth]{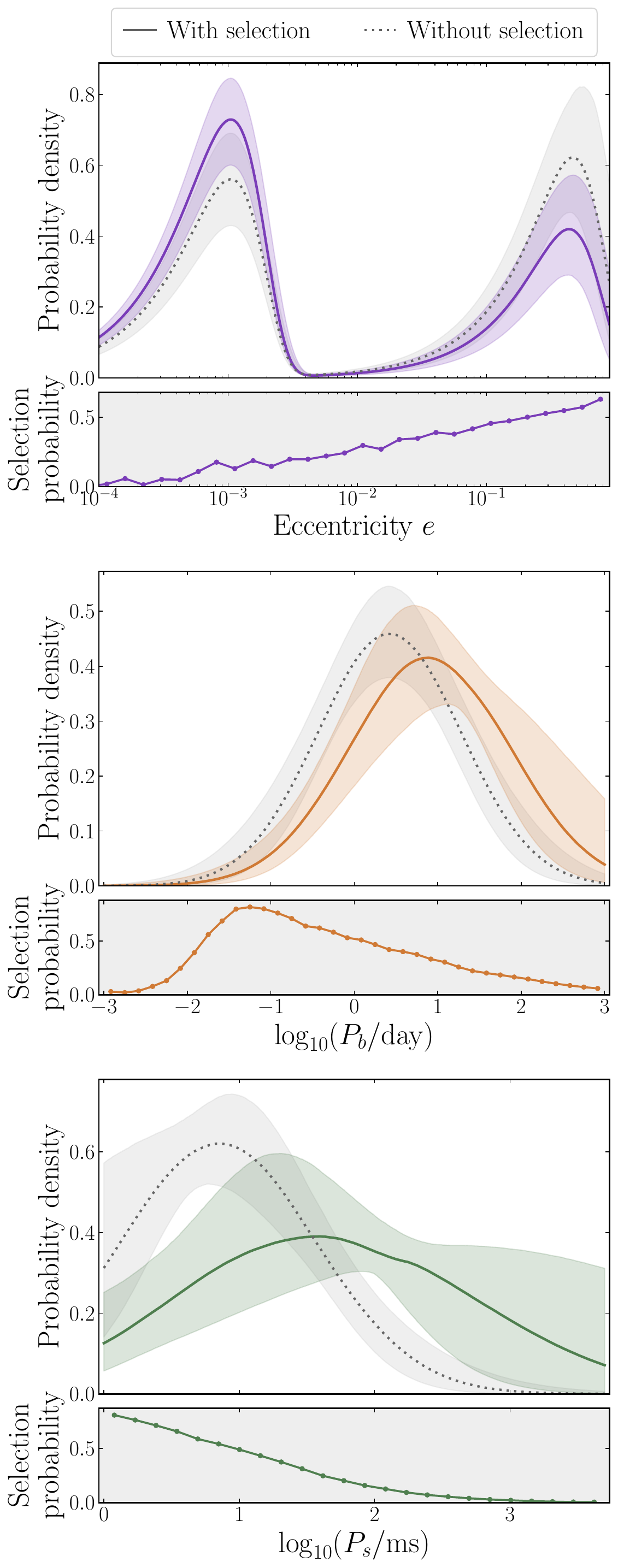}
\caption{Inferred eccentricity (top), orbital period (middle) and pulsar spin period (bottom) populations. Solid colored curves include selection effects, while dotted gray curves neglect them; shaded regions show the central $90\%$ credible intervals. Curves shows the pointwise population median evaluated over the hyperposterior; they therefore need not be Gaussian, although every individual population model is. The lower panels show the selection probability for each parameter, marginalized over the inferred populations of the other parameters. To display the bimodal structure, the eccentricity distribution is shown as a density with respect to $\log_{10}e$ on a logarithmic axis, although the population model is defined and fitted in linear $e$.}
\label{fig:otherpopulations}
\end{figure}

\section{Results}
\label{sec:results}

Figures~\ref{fig:analysisI} and \ref{fig:otherpopulations} present the results of the multidimensional population analysis. 
Posterior measurements for each hyperparameter are listed in Appendix~\ref{app:hyperparam-prior}, with and without selection effects.
For comparison to past results, we also present a 1-dimensional analysis in Appendix~\ref{app:1d}, which infers only the pulsar mass population and averages the selection function over the fixed simulation distributions of the remaining parameters.
In contrast, the multidimensional analysis jointly infers the population distributions of $m_p$, $m_c$, $e$, $\log_{10}P_b$, and $\log_{10}P_s$ according to the model of Eq.~\eqref{eq:pop-model-func}. 
The selection efficiency $\xi(\Lambda)$ is evaluated by reweighting the simulated systems according to all five inferred population distributions, per Eq.~\eqref{eq:hier-selection-samples}. 
Consequently, the effective selection as a function of any one parameter depends on the inferred, rather than uniform, distributions of the other parameters. 
The analysis also retains the correlations between $m_p$, $m_c$, and $\iota$ in the individual system posteriors.

Consistent with existing results~\cite{Alsing2018,Farr2020,Golomb2025,Anik2025}, the inferred pulsar mass population has two components: a narrow component with mean $\mu_{1,m_p}=1.340^{+0.015}_{-0.014}\,M_\odot$ and a broader component with mean $\mu_{2,m_p}=1.67^{+0.09}_{-0.09}\,M_\odot$, Fig.~\ref{fig:analysisI} top left panel. 
Accounting for selection effects slightly increases the relative weight of the low-mass peak and suppresses the higher-mass component, because systems with more massive pulsars are more likely to enter the dataset. 
But the change is modest and remains well within the statistical uncertainty. 
The upper mass cutoff changes from $M_{\rm pop}=2.33^{+0.38}_{-0.27}\,M_\odot$ without selection effects to $2.32^{+0.39}_{-0.27}\,M_\odot$ with selection effects, again well within the statistical uncertainty, Fig.~\ref{fig:analysisI} bottom right panel.

The effective pulsar-mass selection function is also shown in Fig.~\ref{fig:analysisI} (lower part of the top left panel). 
It differs from the 1-dimensional case of Appendix~\ref{app:1d} and left panel of Fig.~\ref{fig:selectionfunction} because now it is marginalized over the \emph{inferred} population rather than the \emph{simulated} uniform distributions of the other parameters. 
The multidimensional effective selection is overall lower, though the ratio of selection probability between the two pulsar-mass peaks is approximately the same; the ratio $S(1.67\,M_\odot)/S(1.34\,M_\odot)$ is $\sim1.05$ in both cases.

The companion mass population is strongly bimodal, with means $\mu_{1,m_c}=0.238^{+0.023}_{-0.040}\,M_\odot$ and $\mu_{2,m_c}=1.151^{+0.043}_{-0.049}\,M_\odot$, Fig.~\ref{fig:analysisI} top right panel. 
The lower-mass peak consists primarily of low-mass white dwarf companions, while the higher-mass peak contains both massive white dwarfs and neutron stars.
Population weighting has a more pronounced effect on the companion-mass effective selection function, reversing the dependence of Fig.~\ref{fig:selectionfunction}. 
After reweighting to the inferred population distributions, higher-mass companions are more likely to enter the dataset. 
Correcting for this selection effect therefore increases the inferred relative weight of the low-mass peak and decreases the relative weight of the broader component near $1.15\,M_\odot$. 
The resulting correction due to selection effects remains small compared with the statistical uncertainty.

The bottom left panel of Fig.~\ref{fig:analysisI} shows the corresponding joint pulsar and companion mass population for the maximum \emph{a posteriori} hyperposterior sample.
Including selection effects produces only
modest changes in the joint density contours.

Figure~\ref{fig:otherpopulations} shows the inferred eccentricity, orbital period, and pulsar spin period populations together with their selection functions. 
The selection correction has a more visible effect on these parameters than on the component masses, as anticipated from Fig.~\ref{fig:selectionfunction}. 
The orbital-period selection probability peaks at short periods and decreases across much of the region containing the inferred population; accounting for this shifts the inferred orbital period distribution towards \emph{longer periods}. 
The population mean of $\log_{10}(P_b/\rm{day})$ shifts from $0.42^{+0.12}_{-0.13}$ without selection effects to $0.91^{+0.26}_{-0.19}$ when we correct for them.
The spin-period selection probability decreases strongly with increasing $\log_{10}(P_s/\rm{s})$, so the selection-corrected population is likewise shifted towards \emph{longer spin periods}. 
The population mean of $\log_{10}(P_s/\rm{s})$ shifts from $-2.16^{+0.15}_{-0.24}$ without selection effects to $-1.40^{+0.63}_{-0.32}$ when we correct for them. This is because mass measurability improves for rapidly spinning pulsars in our framework: at fixed duty cycle, shorter spin periods produce narrower pulses and hence smaller TOA uncertainties.
These results demonstrate that jointly inferring these distributions is important, since their strong selection dependence affects both their own population inference and the effective selection functions of the component masses.

The eccentricity population contains a sharply concentrated component near $e=0$ together with a broader component extending to larger eccentricities; the logarithmic scale in Fig.~\ref{fig:otherpopulations} was chosen to elucidate both components. 
Because the selection probability increases with eccentricity, accounting for selection increases the inferred relative weight of the low eccentricity component and suppresses the broader component. 
The inferred mixture fraction assigns $65^{+7}_{-7}\%$ of systems to the near-circular component and $35^{+7}_{-7}\%$ to the broader component.
That breakdown would have been closer to 50\%--50\% had we ignored selection effects.

The width of the component centered at zero is not resolved: its posterior accumulates at the lower prior bound $\sigma_{1,e}=10^{-3}$. 
This railing indicates that the data favor a component narrower than the range allowed by the adopted population model. 
We impose this lower bound to keep the eccentricity population within the resolution supported by the simulated catalog.

Posterior constraints for further hyperparameters are reported in Table~\ref{tab:population_hyperpriors}. 
The full hyperposterior for the model of Eq.~\eqref{eq:pop-model-func} is available through this work's data release \cite{Drummond2026Data}.

\section{Discussion and conclusions}
\label{sec:discussion}

For the radio pulsar sample considered here, namely pulsars with timing-based mass measurements, accounting for observational selection produces \textit{only modest changes in the mass population distributions} but \textit{impacts the eccentricity, orbital period, and spin period distributions more than their statistical uncertainty}.  
The population distributions we infer here are the astrophysical pulsar distributions \emph{at the time of detection} and should serve as the starting point for subsequent astrophysical interpretations, e.g.~\cite{Grichener:2026nph,Karim:2026bva,Disberg:2026gfy}.
Indeed our aim was to correct only for observational selection effects, while retaining the impact of less certain astrophysical processes that impact the present-day distributions.

Additionally, since our results (and especially the maximum population mass of Fig.~\ref{fig:analysisI}) are based on the full catalog of pulsars, they evade issues related to statistical \emph{regression to the mean}.
Selecting ``extreme" pulsar measurements, such as the highest- or lowest-mass system, for detailed followup or interpretation selects extreme noise fluctuations.
Subsequent data collection is then likely to result in a ``less extreme" measurement, e.g., the $\sim 1{-}\sigma$ reduction in the mass of J0740+6620 between the original~\cite{NANOGrav:2019jur} and subsequent~\cite{Fonseca2021} analysis.
Hierarchical inference probes the full set of (random) noise fluctuations and thus is not susceptible to extreme fluctuations, instead pulling the most ``extreme" measurements closer to the rest of the population~\cite{Farr2020}.

Our results are subject to a few qualifications.
First, the selection function is designed to approximate, rather than fully replicate, the full detection and measurement process, similar in spirit to Ref.~\cite{Magee:2023muf}.
Second, we ignore human decisions to follow up interesting systems, as no correction is required when all systems are retained regardless of their follow-up status, see Sec.~\ref{sec:data}.
Third, population inference is based on a model, chosen to provide simple descriptions of the data of Fig.~\ref{fig:mpXposteriors}.
The factorized population model we adopt is supported by the absence of clear correlations among the system parameters. 
We also adopt an isotropic inclination distribution because binary orientations are expected to be randomly distributed. 
Finally, the fixed observing and telescope parameters were selected to represent realistic survey conditions, but we fixed the pulsar flux and pulse width. 
As such, we cannot characterize \emph{how many} systems are detectable, we only target the parameter dependence.

Besides incorporating selection effects, the other novelty of this work is in our treatment of the multi-dimensional posterior for each binary system and in the population.
Previous pulsar mass population analyses have treated the likelihood for each system as a one-dimensional distribution in pulsar mass~\cite{Alsing2018,Farr2020,Golomb2025,Anik2025,Chattopadhyay:2026vfv}, without retaining its correlations with other binary parameters. 
This might partially be due to the fact that constraints are typically presented in the literature as a series of marginalized 1-dimensional measurements for each parameter, rather than a multi-dimensional distribution for all parameters.
This process ignores strong correlations, e.g., Fig.~\ref{fig:reported-reconstructed-masses}.
Here instead, we reconstruct the correlations among $m_p$, $m_c$, and $\iota$ for each binary system based on the reported post-Keplerian parameters. 
We further infer the companion mass, eccentricity, orbital period, and spin period populations \emph{together} with the pulsar mass. 
This more complete treatment does not qualitatively change our conclusions for the pulsar mass distribution. 
We do find, however, that selection effects have a greater impact on parameters that directly influence detectability and timing measurability, such as the orbital period, eccentricity and spin period. 
For example, we conclude that the bimodal eccentricity distribution is not an artifact of observational selection effects.

Finally, the inferred pulsar mass distribution is based on fewer pulsars than previous studies that combine radio pulsar timing with other observational channels.  For example, Refs.~\cite{Alsing2018,Farr2020} combined radio pulsar measurements with X-ray and optical constraints and included systems without a precise individual pulsar mass measurement, using the available constraints on the binary mass function or total mass instead. 
Gravitational-wave observations were considered in Refs.~\cite{Anik2025,Golomb2025}. 
By contrast, we include only pulsars for which radio timing provides a precise measurement of their mass. 
Our sample also includes more recent measurements from the Freire catalog \cite{FreireNSMasses} and applies stricter criteria to auxiliary constraints: i.e., we exclude mass constraints derived from white dwarf surface-gravity modeling or from optical radial velocity measurements.
This is a necessity driven by our main goal: in order to study observational selection effects, we need to restrict to a \emph{specific observational avenue}.
Similar studies are required to quantify the impact of different observational avenues on those more extensive datasets.

\acknowledgments
We thank Eliot Finch for useful discussions and Scott Ransom for comments on the draft.
KC and LVD acknowledge support from the Department of Energy under award number DE-SC0023101 and by a grant from the Simons Foundation (MP-SCMPS-00001470).  LVD is supported by the Sherman Fairchild Postdoctoral Fellowship at the California Institute of Technology.
This work made use of PINT~\cite{Luo2021} for pulsar timing calculations and Bilby~\cite{Ashton2019} with the \texttt{dynesty} sampler~\cite{Speagle2020} for Bayesian inference. We thank Nihan Pol and Terrence Pierre Jacques for sharing the code accompanying Ref.\ \cite{Pol2021}; we use their neural network to evaluate the detectability component of our selection function. We thank Jacob Golomb for sharing the inference code used in Ref.\ \cite{Golomb2025}. We also used JAX, NumPy, SciPy, Astropy and Matplotlib.
Calculations in this work were conducted using the Resnick High Performance Computing Center, a facility supported by Resnick Sustainability Institute at the California Institute of Technology. 

We used AI agents (Claude and Codex 5.6 Sol) to refine figures, check data tables against the literature, and proofread the text. We verified the AI-assisted output manually: minor figure refinements were reviewed directly in the code, and the data table was validated manually before again being checked using AI.

The hyperparameter posterior samples, multidimensional selection function and reconstructed posterior samples for the individual pulsars are publicly available in the GitHub repository
\href{https://github.com/lisadrummond/RadioPulsarSelectionEffects}
{\texttt{RadioPulsarSelectionEffects}}, archived on Zenodo
\cite{Drummond2026Data}.

\appendix

\section{Flux and pulse duty cycle choice}
\label{app:scale-meas}

\begin{figure}[t]
\centering
\hspace{-4mm}
\includegraphics[width=0.5\textwidth]{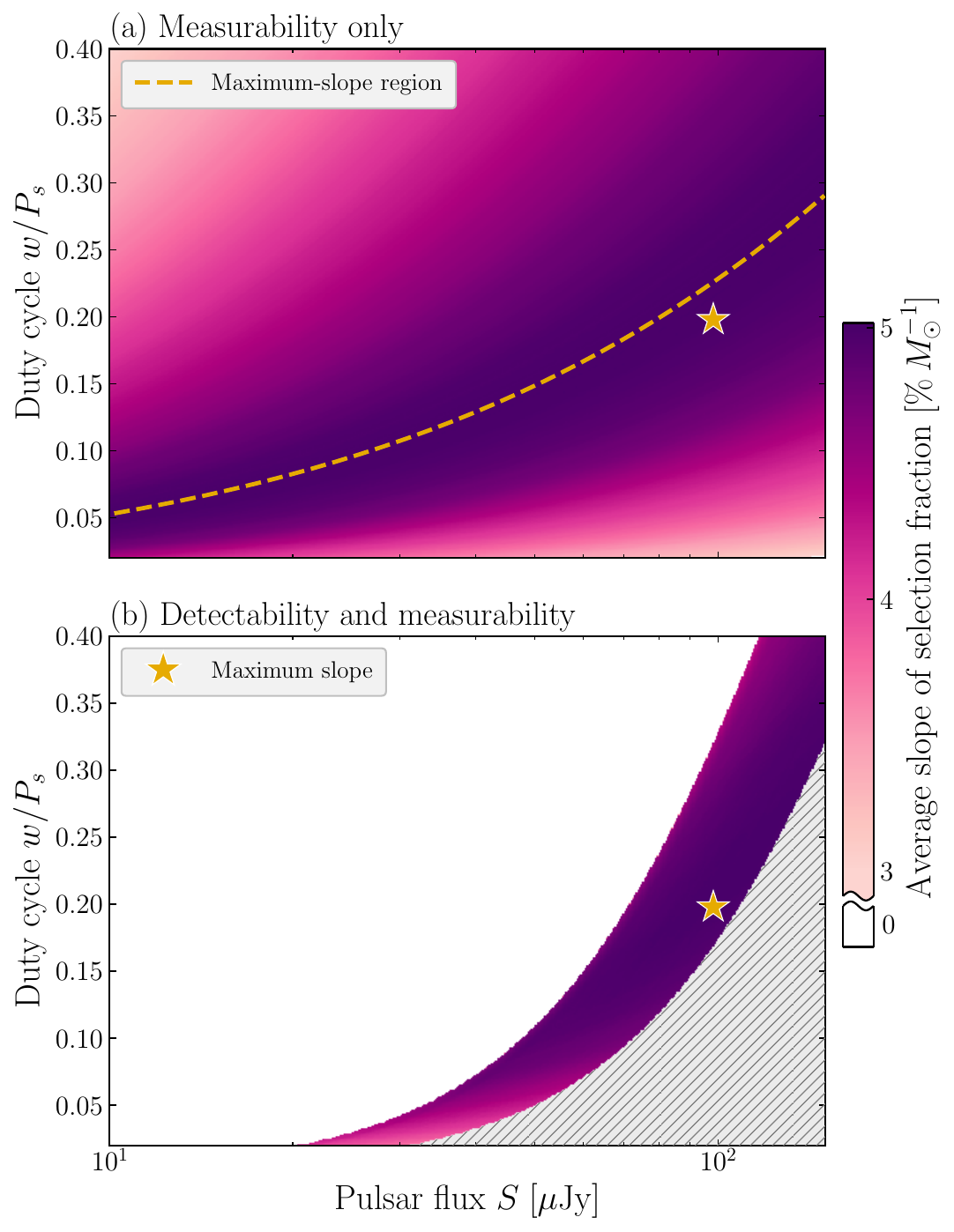}
\caption{ Mass dependence of measurability (top) and combined detectability and measurability (bottom) as functions of the pulsar flux $S$ and pulse duty cycle $w/P_s$, after marginalizing over the other simulated parameters. The color scale shows the average slope of the selection fraction, defined as the difference between the fractions at the highest and lowest pulsar masses divided by the full mass interval, $2.9-1.0=1.9\,M_\odot$. The dashed curve marks the combinations that maximize this slope for measurability alone. The star denotes the adopted combination, $S=98.21\,\mu{\rm Jy}$ and $w/P_s=0.1979$, which maximizes the combined detectability-and-measurability slope; the same point is shown in both panels for comparison. In the white region on the left of the bottom panel, no systems are both detectable and measurable, so the selection fraction vanishes. In the hatched region on the right, the rescaling makes additional systems detectable that were below the threshold in the original numerical run, hence we cannot compute a selection function here.}
\label{fig:worsecase}
\end{figure}

We select the pulsar flux and duty cycle for the synthetic survey of Table~\ref{tab:input-binary} by maximizing the mass-dependence of the selection effects.
Since both the detectability SNR and the measurability criterion $\sigma_{m_p}/m_p$ scale analytically with $S$ and $w/P_s$, in practice we select fiducial values of $S=0.05$\,mJy and $w/P_s=5\%$, carry out the synthetic survey and re-scale the results to explore the range $10^{-3} \leq S/\mu{\rm Jy} \leq 700$ and $10^{-3} \leq w/P_s \leq 0.9$.
We assess the mass dependence of selection effects through the slope of the percentage of selected systems with $m_p$ after marginalizing over all other parameters, e.g. the left panel of Fig.~\ref{fig:selectionfunction}.
The slope is defined as the difference between the percentage of selected systems at $m_p=1\,M_{\odot}$ and $m_p=2.9\,M_{\odot}$, divided by the mass difference.
The results are shown in Fig.~\ref{fig:worsecase} for both the measurability and the combined detectability and measurability effects.
Upon detection, the mass dependence of the selection function depends on $S$ and $w/P_s$ only weakly as shown by the scale of the colormap.
The values that maximize the slope are $S=98.21\,\mu{\rm Jy}$ and $w/P_s=19.79\%$, which (reassuringly) are comparable to dim pulsars detected in the PALFA survey whose settings we adopt~\cite{Scholz2015}.

\section{The \textsc{PINT} timing configuration}
\label{app:timing-fit-configuration}

In Table~\ref{tab:timing-fit-configuration} we provide additional settings and parameters for the \textsc{PINT} timing configuration used to assess measurability in Sec.~\ref{sec:measurability-meas}.
We fit the spin frequency $F_0=1/P_s$, its first time derivative $F_1=\dot{F}_0$, the projected semi-major axis $a_p\sin\iota/c$, the remaining Keplerian parameters $P_b$, $e$, $\omega_p$, $T_0$, and the two mass parameters, $M$ and $m_c$. 
These jointly determine the rotational phase, binary orbit, and relativistic timing signatures. 
The astrometric and dispersion parameters are held fixed to their simulated values, so the calculation is conditional on the adopted sky position, proper motion, parallax and dispersion measure.


\begin{table*}[t]
\caption{Synthetic TOA and timing fit configuration. Quantities already
listed in Tables~\ref{tab:input-binary} and \ref{tab:input-survey} are not
repeated. The astrometric and spin values are inherited from our fiducial parameter file.
\texttt{simDDGR\_current.par}.}
\label{tab:timing-fit-configuration}
\centering
\begin{tabular}{@{}p{9cm}p{7cm}@{}}
\hline
Parameter & Value \\
\hline

\rowcolor{lightgray}
\multicolumn{2}{@{}l}{\textbf{Synthetic observations and timing program}} \\

Observatory &
Arecibo \\

$t_{\rm start}$ &
MJD $55004.4475751$ \\

Epoch placement &
Uniform over the timing baseline \\

Random TOA perturbations &
None (\texttt{add\_noise=False}) \\

Fitting program &
\textsc{PINT} \texttt{WLSFitter} \\

Time units &
TDB \\

Clock correction &
TT(BIPM2021) \\

Solar-system ephemeris &
DE440 \\

Solar-wind electron density, $n_0$ &
$0\,\mathrm{cm}^{-3}$ \\

\hline

\rowcolor{lightgray}
\multicolumn{2}{@{}l}{\textbf{Spin and astrometric parameters}} \\

Right ascension, $\alpha$ (RAJ) &
$15^{\mathrm{h}}\,18^{\mathrm{m}}\,16.79782212^{\mathrm{s}}$ \\

Declination, $\delta$ (DECJ) &
$+49^{\circ}\,04^{\prime}\,34.0834413^{\prime\prime}$ \\

Proper motion in right ascension,
$\mu_\alpha\cos\delta$ (PMRA) &
$-0.7113\,\mathrm{mas\,yr^{-1}}$ \\

Proper motion in declination,
$\mu_\delta$ (PMDEC) &
$-8.9150\,\mathrm{mas\,yr^{-1}}$ \\

Parallax, $\varpi$ (PX) &
$1.8031\,\mathrm{mas}$ \\

Spin-frequency derivative, $F_1=\dot F_0$ &
$-1.621593472179\times10^{-17}\,\mathrm{Hz\,s^{-1}}$ \\

Dispersion measure, DM &
$11.611711\,\mathrm{pc\,cm^{-3}}$ \\

\hline

\rowcolor{lightgray}
\multicolumn{2}{@{}l}{\textbf{Binary timing model and additional fitted quantities}} \\

Binary timing model &
DDGR \\

Projected semi-major axis, $A_1=x=a_p\sin\iota/c$ &
Unique per each system \\

Total mass, $M=m_p+m_c$ &
Unique per each system \\

\hline
\end{tabular}
\end{table*}

\section{Multiparameter hierarchical inference with selection effects}
\label{app:derivseleffects}

In this appendix we describe the hierarchical inference formalism. 
In Sec.~\ref{sec:formalism} we derive the population likelihood including selection effects~\cite{10.1063/1.1835214,Mandel:2018mve} and detail its implementation~\cite{Callister:2024cdx}.
In Sec.~\ref{sec:multipleparams} we discuss multiparameter inference and the treatment of the orbital and pulsar parameters.

\subsection{Formalism}
\label{sec:formalism}

Given an ensemble of $N_{\textrm{p}}$ pulsar systems whose parameters $\theta_i$ have been measured with data $d_i$, where $i\in\{1\ldots N_p\}$, we target the population distribution of $\theta$.
We parametrize the population distribution with hyperparameters $\Lambda$ as $p(\theta|\Lambda)$.
The likelihood for all data $d=\{d_i\}$ given $\Lambda$ is
\begin{equation}
p\left(d \mid \Lambda\right) = \prod_{i=1}^{N_{\mathrm{p}}} p\left({d}_i \mid \text {det} , \Lambda\right)\,,
\end{equation}
where the ``det'' condition enforces that the data from pulsar $i$ are in our dataset, i.e., the system has been detected and the pulsar properties have been measured. 
Using the rules of conditional probabilities (a special case of which is Bayes' Theorem), we have
\begin{align}
p\left(d \mid \Lambda\right) &= \prod_{i=1}^{N_{\mathrm{p}}} p\left({d}_i \mid \text {det} , \Lambda\right)= \prod_{i=1}^{N_{\mathrm{p}}} \frac{p(\text {det} \mid {d_i}, \Lambda) p({d_i} \mid \Lambda)}{p(\text {det} \mid \Lambda)}\nonumber\\ 
&\propto \prod_{i=1}^{N_{\mathrm{p}}} \frac{ p({d_i} \mid \Lambda)}{p(\text {det} \mid \Lambda)}\,,
\label{eq:like-der}
\end{align}
since $p(\text {det} \mid {d_i}, \Lambda)$ is constant upon conditioning on the data ${d_i}$.\footnote{
This simplification typically invokes the assumption that given some data, detectability is deterministic~\cite{Essick:2023upv}.
This is likely not true in detection campaigns where an element of human decision-making is involved.
\citet{Essick2026} generalized this statement to data that are not detected deterministically, as long as detectability does not depend on the true system parameters, or further data not included in the likelihood.}

The denominator of Eq.~\eqref{eq:like-der} is the detection efficiency, commonly denoted as $\xi(\Lambda)=p(\text {det} \mid \Lambda)$. 
It is the probability that a pulsar drawn from the population enters the dataset, integrated over the system parameters and data
\begin{equation}
\xi(\Lambda)=\int \diff  d\, \diff {\theta}\, p(\text {det} \mid {d}) p({d}, {\theta} \mid \Lambda)\,.
\end{equation}
Moreover,
\begin{equation}
    p({d_i} \mid \Lambda) = \int \diff \theta_i\, p\left(d_i \mid \theta_i\right) p\left(\theta_i \mid \Lambda\right)\,,
\end{equation}
hence putting it all together we obtain
\begin{align}
p\left(d \mid \Lambda\right) & =\frac{1}{\xi(\Lambda)^{N_{\mathrm{p}}}} \prod_{i=1}^{N_{\mathrm{p}}} \int \diff \theta_i\, p\left(d_i \mid \theta_i\right) p\left(\theta_i \mid \Lambda\right)\,.
\label{eq:hier-like}
\end{align}
This equation is the full hierarchical likelihood we aim to compute. 
In practice, we evaluate the various integrals via Monte Carlo sums.

Given $N_{\mathrm{s}}$ samples from the parameter posterior $\theta_i^j \sim p(\theta_i|d_i,\text{pr})$, where $j\in\{1\ldots N_{\rm s}\}$, with prior $p(\theta|\text{pr})$, the integral in the product of Eq.~\eqref{eq:hier-like} is 
\begin{align}
    \int \diff \theta_i\, p\left(d_i \mid \theta_i\right) p\left(\theta_i \mid \Lambda\right) &\propto \int \diff \theta_i\, \frac{p({\theta_i} \mid {d_i}, \text {pr})}{p({\theta_i} \mid \text {pr})} p\left(\theta_i \mid \Lambda\right) \nonumber\\
    &\sim \frac{1}{N_{\mathrm{s}}} \sum_j \frac{p\left({\theta}_i^j \mid \Lambda\right)}{p\left({\theta}_i^j \mid \text {pr}\right)}\,.
    \label{eq:hier-numerator-samples}
\end{align}
Effectively, each posterior sample is reweighted from the original analysis prior to the population distribution.

The detection efficiency in the denominator of Eq.~\eqref{eq:hier-like} is estimated with the simulated systems of Sec.~\ref{sec:selectioneffects}.
We select some distribution for the pulsar parameters $p ({\theta} \mid \text {sim})$ (Table~\ref{tab:input-binary}), draw systems, simulate data $p({d}, {\theta} \mid \text {sim})$ and determine whether both conditions are met: the pulsar is detectable and its parameters are measured. 
Assuming $N_{\mathrm{sim}}$ simulations out of which $N_{\mathrm{f}}$ pass our criteria, then for $k\in\{1\ldots N_{\rm sim}\}$
\begin{align}
\xi(\Lambda)&=\int \diff d\, \diff \theta\, p(\text {det} \mid {d}) p({d}, {\theta} \mid \Lambda) \nonumber\\
&=\int \diff d\, \diff \theta\, p({d}, {\theta} \mid \text {sim}) \frac{p(\text {det} \mid {d}) p({d}, {\theta} \mid \Lambda)}{p({d}, {\theta} \mid \text {sim})} \nonumber\\ 
 &\sim \frac{1}{N_{\text {sim}}} \sum_k^{N_{\text {sim}}} \frac{p\left(\text {det} \mid {d}_k\right) p\left({d}_k, {\theta}_k \mid \Lambda\right)}{p\left({d}_k, {\theta}_k \mid \text {sim}\right)}\,.
 \end{align}
In the last equation, we again treat $p\left(\text {det} \mid {d}_k\right)$ as a deterministic function that is either 0 or 1 depending on whether data $d_k$ are detectable or not. 
This reduces the sum over all simulated systems to a sum over only the detected ones, with $n\in\{1\ldots N_{\rm f}\}$, and 
 \begin{align}
 \xi(\Lambda)& = \frac{1}{N_{\text {sim}}} \sum_{n}^{N_{\mathrm{f}}} \frac{p\left({d}_n, {\theta}_n \mid \Lambda\right)}{p\left({d}_n, {\theta}_n \mid \text {sim}\right)}\nonumber \\ 
 &=\frac{1}{N_{\text {sim}}} \sum_{n}^{N_{\mathrm{f}}} \frac{p\left({d}_n \mid {\theta}_n\right) p\left({\theta}_n \mid \Lambda\right)}{p\left({d}_n \mid {\theta}_n\right) p\left({\theta}_n \mid \text {sim}\right)}\nonumber \\ 
 &=\frac{1}{N_{\text {sim}}} \sum_{n}^{N_{\mathrm{f}}} \frac{ p\left({\theta}_n \mid \Lambda\right)}{ p\left({\theta}_n \mid \text {sim}\right)}\,,
 \label{eq:hier-selection-samples}
\end{align}
where in the second line we have used the product rule and the fact that the data do not depend on $\Lambda$.
Effectively, each simulated system that has been detected is reweighted from the original simulation distribution to the population distribution.
This equation also demonstrates that the original simulation distribution of Table~\ref{tab:input-binary} does not impact the inference as it is reweighted away.

\subsection{Multiple parameters}
\label{sec:multipleparams}

In the generic formalism of Sec.~\ref{sec:formalism}, $\theta$ is a collection of all relevant binary and pulsar parameters that are sampled over in Table~\ref{tab:input-binary}.
We split them into groups:
\begin{enumerate}
\item Orientation parameters, $\theta_U=\{\omega_p, \phi_0\}$, are distributed uniformly in the population.  
\item Well-measured parameters, $\theta_\delta=\{e,P_b,P_s\}$, have effectively negligible measurement uncertainty.
\item Post-Keplerian parameters, $\theta_m=\{m_p,m_c,\iota\}$.
\end{enumerate}
Overall, $\theta =\{\theta_m,\theta_U,\theta_\delta\}$ and the corresponding hyperparameters are $\Lambda =\{\Lambda_m,\Lambda_U,\Lambda_\delta\}$.
We use the following simplifying conditions.
\begin{enumerate}
\item We adopt separable population models in Sec.~\ref{sec:hierarchical-models}:
\begin{align}
p&\left(\theta_m,\theta_U,\theta_\delta \mid \Lambda_m,\Lambda_U,\Lambda_\delta\right)\nonumber \\
&=p\left(\theta_m\mid \Lambda_m\right)p\left(\theta_U\mid \Lambda_U\right)p\left(\theta_\delta\mid \Lambda_\delta\right)\,.
\end{align}
\item We impose a fixed, uniform distribution for the orientation parameters, $p\left(\theta_U\mid \Lambda_U\right)= \rm const.$
\item We ignore the uncertainty in the well-measured parameters
\begin{equation}
p(\theta|d_i)=p(\theta_m,\theta_U|d_i)\delta(\theta_\delta-\theta_{\delta,i})\,,
\end{equation}
where $\theta_{\delta,i}$ is their reported value for pulsar $i$.
\item We assume separable priors, $p(\theta_m,\theta_U,\theta_\delta)=p(\theta_m)p(\theta_U)p(\theta_\delta)$. This requirement is satisfied given our parameter groups. While $p(\theta_U)$ and $p(\theta_\delta)$ are further separable, $p(\theta_m)$ is not: we have separable priors among the post-Keplerian parameters, but converting to the masses and inclination couples the corresponding priors, Sec.~\ref{sec:pkposteriorreconstruction}.
\item We assume that all priors of the original analysis were uniform $p\left(\theta_U\right){\propto} p\left(\theta_\delta\right)= \rm const.$ Again, it is the prior on the post-Keplerian parameters that is also constant, and not that of the reconstructed $\theta_m$.
\end{enumerate}

Starting with the numerator of Eq.~\eqref{eq:hier-like}, i.e., the contribution of a single system in the population, we have (suppressing the parameter $i$ subscript for clarity)
\begin{widetext}
\begin{align}
    p({d_i} \mid \Lambda) &= \int \diff \theta_m \diff \theta_U \diff \theta_\delta\, p\left(d_i \mid \theta_m,\theta_U,\theta_\delta\right) p\left(\theta_m,\theta_U,\theta_\delta \mid \Lambda_m,\Lambda_U,\Lambda_\delta\right)\nonumber \\
    &\propto \int \diff \theta_m \diff \theta_U \diff \theta_\delta\, p\left(\theta_m,\theta_U,\theta_\delta \mid d_i \right) \frac{p\left(\theta_m,\theta_U,\theta_\delta \mid \Lambda\right)}{p(\theta_m,\theta_U,\theta_\delta)}\nonumber \\
    &= \int \diff \theta_m \diff \theta_U \diff \theta_\delta\, p(\theta_m,\theta_U|d_i)\delta(\theta_\delta-\theta_{\delta,i}) \frac{p\left(\theta_m\mid \Lambda_m\right)p\left(\theta_U\mid \Lambda_U\right)p\left(\theta_\delta\mid \Lambda_\delta\right)}{p(\theta_m)p(\theta_U)p(\theta_\delta)}\nonumber \\
    &= \frac{p\left(\theta_{\delta,i}\mid \Lambda_\delta\right)}{p(\theta_{\delta,i})}\int \diff \theta_m \, p(\theta_m|d_i) \frac{p\left(\theta_m\mid \Lambda_m\right)}{p(\theta_m)}\,,
    \label{eq:hier-like-nuisance}
\end{align}
\end{widetext}
where in the second line we have used Bayes' theorem and dropped the normalization $p(d)$, in the third line we have substituted the simplifying conditions, and in the final line we have carried out the integrals over $\theta_U$ and $\theta_\delta$.
Since the population distribution for $\theta_U$ is fixed and equal to its prior, $\theta_U$ drops out of the equation.
The only individual-system posterior we need is $p(\theta_m|d_i)$, the posterior for the component masses and inclination that has been marginalized over all other parameters; this is reconstructed in Sec.~\ref{sec:posteriorreconstruction}.
The contributions of $\theta_\delta$ and $\theta_m$ have factorized and do not impact each other.
At this stage, if we were only interested in the pulsar mass alone or the binary parameters alone, we could ignore the contribution of $\theta_\delta$ or $\theta_m$ respectively, as overall normalizations.

Moving to the denominator of the population likelihood of Eq.~\eqref{eq:hier-like}, the selection term becomes
\begin{widetext}
\begin{align}
\xi(\Lambda)&=\int \diff d \diff \theta_m \diff \theta_U \diff \theta_\delta \,p(\text {det} \mid {d}) p({d}, \theta_m,\theta_U,\theta_\delta \mid \Lambda)\nonumber \\
&=\int \diff d \diff \theta_m \diff \theta_U \diff \theta_\delta \,p(\text {det} \mid {d}) p({d}| \theta_m,\theta_U,\theta_\delta) p\left(\theta_m\mid \Lambda_m\right)p\left(\theta_U\right)p\left(\theta_\delta\mid \Lambda_\delta\right)\,,
\end{align}
\end{widetext}
where we have again used the fact that the data do not depend on $\Lambda$ and substituted the simplifying conditions.
The parameters $\theta_U,\theta_\delta$ are now linked with $\theta_m$ through the data generation term $p({d}\mid \theta_m,\theta_U,\theta_\delta)$ which is not separable.
Indeed this is why all parameters are sampled over in Table~\ref{tab:input-binary}.
The simulation distribution for $\theta_U$ matches the population, but the one for $\theta_\delta$ generically does not. 
Unlike the case above for the numerator of Eq.~\eqref{eq:hier-like}, now if we were only interested in the pulsar mass alone or the binary parameters alone, we could \emph{not} assume the population of $\theta_\delta$ or $\theta_m$ respectively to be fixed, instead they would have to be inferred together.

To sum up: 
\begin{enumerate}
\item The uniform angles have a fixed distribution and do not impact the numerator of the hierarchical likelihood in Eq.~\eqref{eq:hier-like}. They still affect the selection term in the denominator, which is why we have sampled over them according to their uniform population distribution in Sec.~\ref{sec:selectioneffects}.
\item The perfectly-measured binary and spin parameters separate from the mass parameters in the likelihood numerator and do not impact each other's inference. They still couple in the selection term. If the simulation distribution of Sec.~\ref{sec:selectioneffects} is distinct from the true population distribution for each parameter, then this discrepancy will impact inference. This is why the population distribution of all parameters is inferred in Sec.~\ref{sec:results}.
\end{enumerate}

\section{Validation of the selection function}
\label{app:validationofselfunc}

\begin{figure}
\hspace{-3mm}
\includegraphics[width=0.48\textwidth]{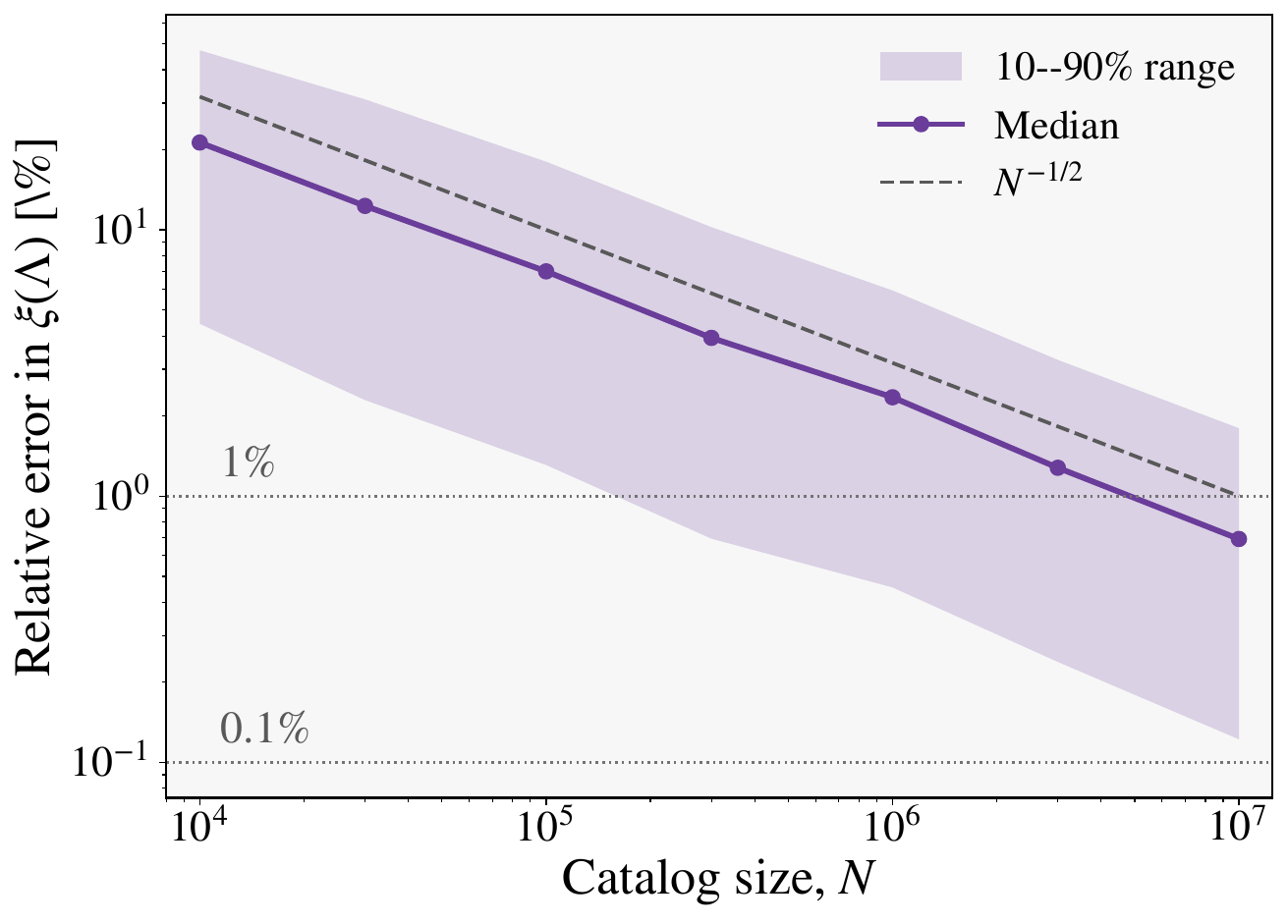}
\caption{
Convergence of the Monte Carlo estimate of the selection efficiency, $\xi(\Lambda)$. 
The synthetic catalog is divided into independent chunks of $10^4$ systems, which are resampled to construct catalogs with sizes between $N=10^4$ and $10^7$. For each catalog size, the relative difference from the selection efficiency obtained using the complete $10^7$-system catalog is evaluated for $400$ bootstrap realizations and $32$ samples from the population posterior. The solid purple curve shows the median relative error and the shaded region shows the $10$--$90\%$ range. The dashed line shows the expected $N^{-1/2}$ scaling. At $N=10^7$, the median Monte-Carlo error is ${\sim}0.69\%$.
}
\label{fig:selection-convergence}
\end{figure}

In this Appendix, we validate the Monte-Carlo evaluation of the selection function
\begin{equation}
\xi(\Lambda)
=
\frac{1}{N_{\rm sim}}
\sum_{k\in{\rm select}}
\frac{
p(\theta_k\mid\Lambda)
}{
p(\theta_k\mid{\rm sim})
}\,,
\end{equation}
with $N_{\rm sim}=10^7$ simulated systems. Specifically:
\begin{enumerate}
\item We divide the full synthetic catalog into $1000$ independent chunks, each with $10^4$ systems.

\item For catalog sizes between $N=10^4$ and $10^7$, we construct $400$ bootstrap catalogs by resampling the $1000$ chunks with replacement.

\item We select $32$ random samples from the population posterior of Sec.~\ref{sec:results}, $p(\Lambda\mid d)$. 

\item For every bootstrap catalog and population, we compare the resulting selection function, $\xi_N(\Lambda)$, with that of the complete catalog, $\xi_{10^7}(\Lambda)$, using
\begin{equation}
\epsilon_N(\Lambda)
=
\frac{
\left|\xi_N(\Lambda)-\xi_{10^7}(\Lambda)\right|
}{
\xi_{10^7}(\Lambda)
}\,.
\end{equation}
\end{enumerate}

Figure~\ref{fig:selection-convergence} shows the median relative error and the $10$--$90\%$ range across bootstrap realizations and population samples. 
The error decreases as ${\sim}N^{-1/2}$, as expected for Monte-Carlo sampling. 
At $N=N_{\rm sim}=10^7$, the median relative error is ${\sim}0.69\%$, with $90\%$ of the realizations below ${\sim}2\%$.  
We also assess the criterion $N_{\rm eff}>4N_{\rm obs}$~\cite{Farr2019Selection}, where $N_{\rm eff}$ is the effective number of weighted simulations contributing to $\xi(\Lambda)$, i.e.,
\begin{equation}
N_{\rm eff}\equiv\frac{\left(\sum_{k\in{\rm select}} w_k\right)^2}{\sum_{k\in{\rm select}} w_k^2},
\qquad
w_k\equiv\frac{p(\theta_k\mid\Lambda)}{p(\theta_k\mid{\rm sim})}\,,
\end{equation}
and $N_{\rm obs}$ is the number of observed pulsars. 
For the 32 population-posterior samples tested above, we find a minimum $N_{\rm eff}\sim3.6\times10^3$, comfortably exceeding $4N_{\rm obs}=184$.

\section{Individual posteriors for the full dataset}
\label{app:all-posteriors}

\begin{figure*}
\hspace{-3mm}
\includegraphics[width=0.8\textwidth]{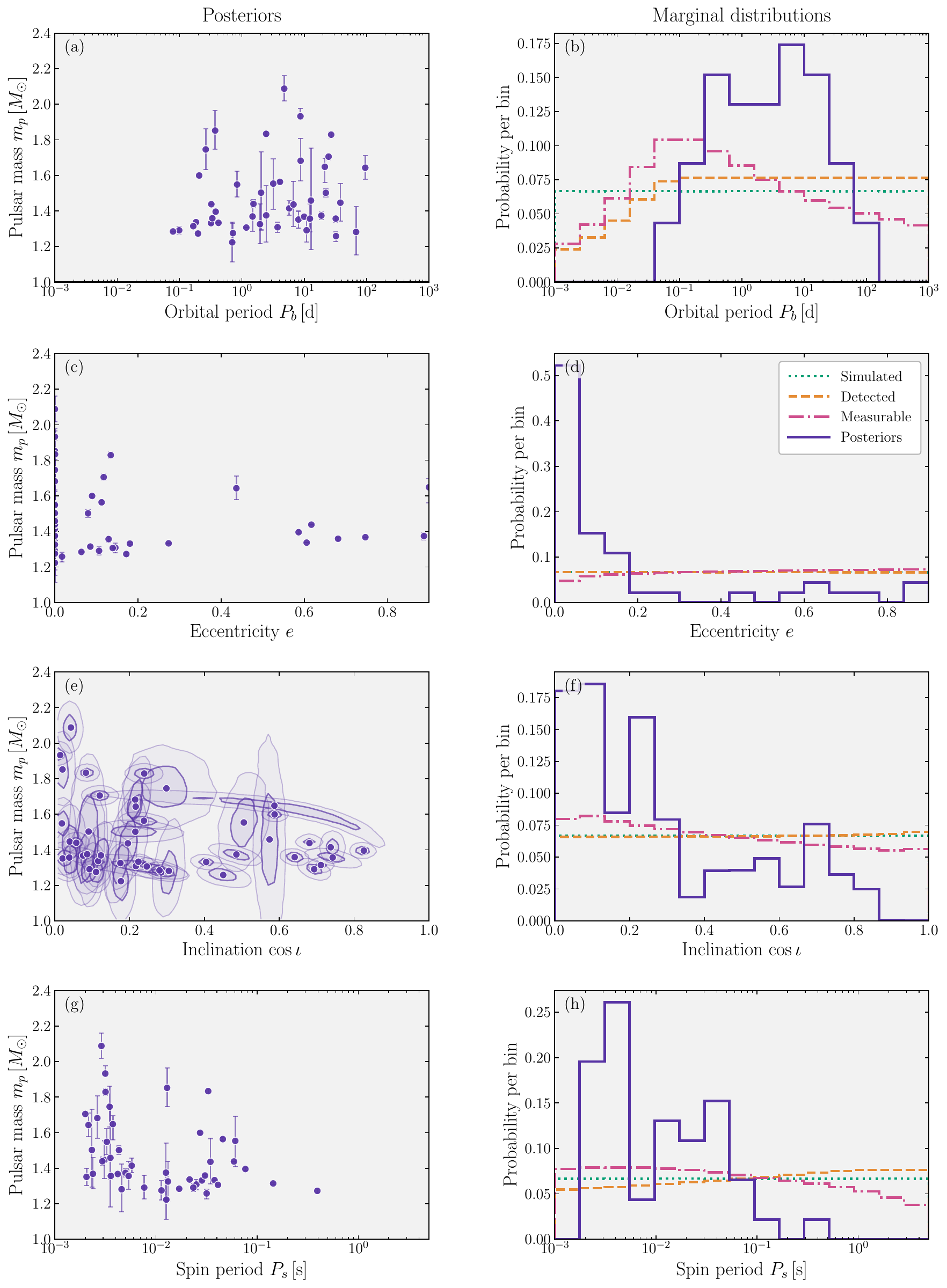}
\caption{
Posteriors and marginal distributions for the 46 analyzed systems. Left panels show the $m_p$ posterior medians and $\sim1\sigma$ intervals at the measured $P_b$, $e$, and $P_s$ values; panel (e) shows the 1- and 2-$\sigma$-joint $m_p$--$\cos\iota$ posterior contours. Right panels show the corresponding 1-D distributions, together with the simulated, detected, and measurable populations.}
\label{fig:mpXposteriors}
\end{figure*}

In this Appendix, we examine the posteriors for the parameters of each of the $46$ pulsars from Table~\ref{tab:consolidated-mass-reconstructions} that we analyze.
Figure~\ref{fig:mpXposteriors} shows the 2-dimensional posteriors between the pulsar mass and the orbital period, eccentricity, inclination, and spin period, as well as 1-dimensional histograms of the posterior medians.
The pulsar mass and inclination posteriors are reconstructed in Sec.~\ref{sec:posteriorreconstruction}, while the eccentricity, orbital period, and spin period have negligible uncertainty.
A plot of the pulsar mass and companion mass posteriors is available in Fig.~\ref{fig:reported-reconstructed-masses} and not reproduced here.
These plots motivate the population model of Eq.~\eqref{eq:pop-model-func} from their shape and absence of obvious population-level correlations among parameters.

\section{Hyperparameter priors and posteriors}
\label{app:hyperparam-prior}

Table~\ref{tab:population_hyperpriors} lists the priors for the hyperparameters of Sec.~\ref{sec:hierarchical-models} and their posteriors from the analysis of Sec.~\ref{sec:results}.

\begin{table*}[!ht]
\centering
\caption{Hyperpriors and posterior constraints for the joint population model, with and without selection effects. Posterior values are medians with 68\% credible intervals. Here, $\mathcal{U}(a,b)$ denotes a uniform distribution. The first eccentricity component mean is fixed at zero. The orbital and spin period populations are modeled as Gaussian distributions in $\log_{10}(P_b/\mathrm{day})$ and $\log_{10}(P_s/\mathrm{s})$, respectively.}
\label{tab:population_hyperpriors}
\small
\begin{tabular}{lllcc}
\toprule
& & & \multicolumn{2}{c}{Posterior} \\
\cmidrule(lr){4-5}
Population & Hyperparameter & Hyperprior
& Without selection & With selection \\
\midrule

Pulsar mass
& $\mu_{1,m_p}$
& $\mathcal{U}(1.0,\,2.2)\,M_\odot$
& $1.343^{+0.015}_{-0.014}\,M_\odot$
& $1.340^{+0.015}_{-0.014}\,M_\odot$ \\

& $\mu_{2,m_p}$
& $\mathcal{U}\!\left(\mu_{1,m_p},\,2.8\right)\,M_\odot$
& $1.68^{+0.10}_{-0.08}\,M_\odot$
& $1.67^{+0.09}_{-0.09}\,M_\odot$ \\

& $\sigma_{1,m_p}$
& $\mathcal{U}(0.03,\,0.45)\,M_\odot$
& $0.049^{+0.015}_{-0.011}\,M_\odot$
& $0.047^{+0.014}_{-0.010}\,M_\odot$ \\

& $\sigma_{2,m_p}$
& $\mathcal{U}(0.03,\,0.45)\,M_\odot$
& $0.22^{+0.09}_{-0.05}\,M_\odot$
& $0.22^{+0.09}_{-0.05}\,M_\odot$ \\

& $f_{m_p}$
& $\mathcal{U}(0,\,1)$
& $0.61^{+0.10}_{-0.11}$
& $0.62^{+0.09}_{-0.10}$ \\

& $M_{\mathrm{pop}}$
& $\mathcal{U}(1.8,\,2.9)\,M_\odot$
& $2.33^{+0.38}_{-0.27}\,M_\odot$
& $2.32^{+0.39}_{-0.27}\,M_\odot$ \\

\midrule

Companion mass
& $\mu_{1,m_c}$
& $\mathcal{U}(0.1,\,1.3)\,M_\odot$
& $0.243^{+0.023}_{-0.030}\,M_\odot$
& $0.238^{+0.023}_{-0.040}\,M_\odot$ \\

& $\mu_{2,m_c}$
& $\mathcal{U}\!\left(\mu_{1,m_c},\,2.4\right)\,M_\odot$
& $1.16^{+0.05}_{-0.05}\,M_\odot$
& $1.15^{+0.04}_{-0.05}\,M_\odot$ \\

& $\sigma_{1,m_c}$
& $\mathcal{U}(0.02,\,0.60)\,M_\odot$
& $0.087^{+0.042}_{-0.030}\,M_\odot$
& $0.092^{+0.042}_{-0.031}\,M_\odot$ \\

& $\sigma_{2,m_c}$
& $\mathcal{U}(0.02,\,0.60)\,M_\odot$
& $0.219^{+0.051}_{-0.036}\,M_\odot$
& $0.218^{+0.050}_{-0.036}\,M_\odot$ \\

& $f_{m_c}$
& $\mathcal{U}(0,\,1)$
& $0.43^{+0.07}_{-0.07}$
& $0.48^{+0.07}_{-0.07}$ \\

\midrule

Eccentricity
& $\sigma_{1,e}$
& $\mathcal{U}(0.001,\,0.35)$
& $0.001034^{+0.000059}_{-0.000025}$
& $0.001035^{+0.000058}_{-0.000026}$ \\

& $\mu_{2,e}$
& $\mathcal{U}(0.01,\,0.9)$
& $0.13^{+0.12}_{-0.08}$
& $0.09^{+0.09}_{-0.06}$ \\

& $\sigma_{2,e}$
& $\mathcal{U}(0.003,\,0.45)$
& $0.40^{+0.03}_{-0.05}$
& $0.39^{+0.04}_{-0.05}$ \\

& $f_e$
& $\mathcal{U}(0,\,1)$
& $0.50^{+0.07}_{-0.07}$
& $0.65^{+0.07}_{-0.07}$ \\

\midrule

Orbital period
& $\mu_{P_b}$
& $\mathcal{U}(-3,\,3)$
& $0.42^{+0.12}_{-0.13}$
& $0.91^{+0.26}_{-0.19}$ \\

& $\sigma_{P_b}$
& $\mathcal{U}(0.05,\,3.0)$
& $0.86^{+0.10}_{-0.09}$
& $0.96^{+0.17}_{-0.12}$ \\

\midrule

Spin period
& $\mu_{P_s}$
& $\mathcal{U}\!\left(-3,\,\log_{10}5\right)$
& $-2.16^{+0.15}_{-0.24}$
& $-1.40^{+0.63}_{-0.32}$ \\

& $\sigma_{P_s}$
& $\mathcal{U}(0.05,\,2.0)$
& $0.72^{+0.18}_{-0.11}$
& $1.12^{+0.54}_{-0.32}$ \\

\bottomrule
\end{tabular}
\end{table*}

\section{Analysis of the 1-dimensional reported pulsar masses}
\label{app:1d}

\begin{figure*}[t]
\centering
\includegraphics[width=0.84\textwidth]{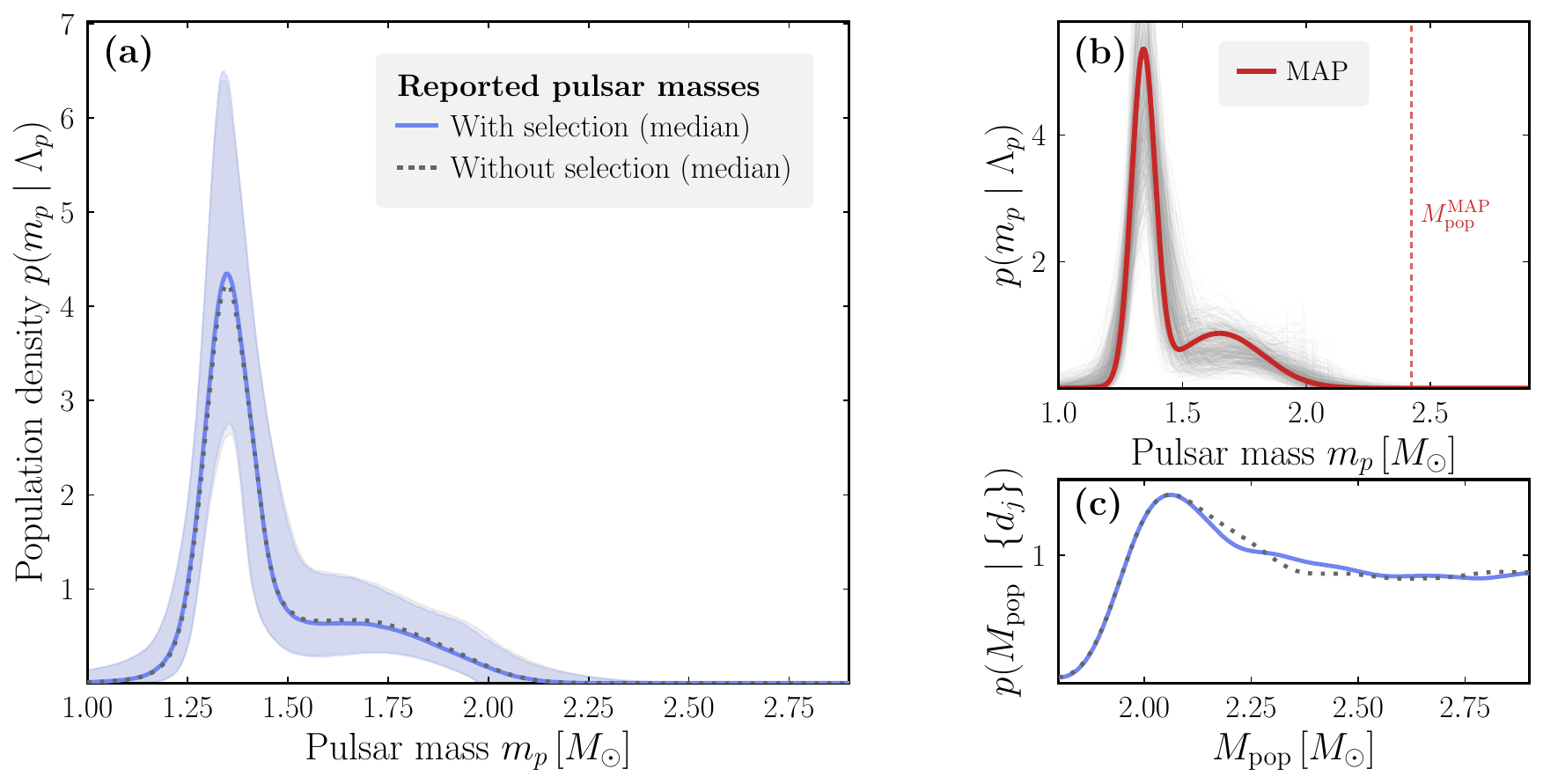}
\caption{Population distribution for the pulsar mass obtained using directly the published 1-dimensional mass constraints in Table \ref{tab:consolidated-mass-reconstructions}, without accounting for correlations with other parameters. 
We show results with and without accounting for selection effects in the pulsar mass.
Panel (a) shows the inferred population density $p(m_p\mid\Lambda_p)$. The solid blue curve includes the selection correction, whereas the dashed gray curve neglects it. Each curve is the pointwise posterior median; the shaded regions are the corresponding central $90\%$ credible intervals. Panel (b) shows 500 population posterior samples (gray curves) from the selection-corrected analysis, together with the maximum-a-posteriori population model in red; the vertical dashed line marks its value of $M_{\rm pop}$. Panel (c) shows the marginalized posterior for $M_{\rm pop}$, with and without selection effects, smoothed using a Gaussian kernel density estimate.}
\label{fig:pulsarpopulation}
\end{figure*}

For comparison to past analyses, in this Appendix we infer the pulsar mass population using the reported, marginalized mass constraints. 
Each of the 48 pulsars\footnote{Here, we include the $2$ pulsars whose posteriors we were unable to reconstruct.} is represented by its reported 1-dimensional mass measurement and uncertainty from Table \ref{tab:consolidated-mass-reconstructions}. 
We approximate each mass constraint using a one-dimensional split-Gaussian likelihood, which preserves asymmetric uncertainties and reduces to a Gaussian when the reported uncertainty is symmetric. 
Unlike the analysis of Sec.~\ref{sec:results}, this analysis makes a number of simplifying assumptions, see Appendix~\ref{app:derivseleffects} for details: (i) it does not account for correlations between $m_p$ and other parameters in the individual-system likelihoods, e.g., Fig.~\ref{fig:reported-reconstructed-masses}, (ii) it assumes a uniform prior in $m_p$ for each pulsar, and (iii) it assumes that the remaining, neglected orbital and pulsar parameters are distributed in the population according to the simulated distributions of Table~\ref{tab:input-binary}.

The pulsar mass population is modeled using the same two-component Gaussian mixture defined in Eq.~\eqref{eq:nsdist}, truncated to $1\,M_\odot\leq m_p\leq M_{\rm pop}$. 
The corresponding hyperparameters $\Lambda_p$ are the component means $\mu_{1,m_p}$ and $\mu_{2,m_p}$, component widths $\sigma_{1,m_p}$ and $\sigma_{2,m_p}$, mixture fraction $f_{m_p}$, and upper population cutoff $M_{\rm pop}$, with priors listed in Table~\ref{tab:population_hyperpriors}.
The population distributions of the remaining binary parameters are not inferred; for the purpose of estimating the selection function, they are fixed to the distributions used to generate the selection simulations. 
The hierarchical likelihood reduces to
$$
\mathcal{L}(\Lambda_p)\propto\frac{\displaystyle\prod_{i=1}^{N_{\rm p}}\int p(d_i\mid m_p)\,p(m_p\mid\Lambda_p)\,\mathrm{d}m_p}{\xi(\Lambda_p)^{N_{\rm p}}}\,,
$$
where the selection efficiency is
$$
\xi(\Lambda_p)=\int S_{\rm 1D}(m_p)\,p(m_p\mid\Lambda_p)\,\mathrm{d}m_p\,,
$$
and we have defined $S_{\rm 1D}(m_p)$ as the selection probability marginalized over the fixed distributions of the remaining binary parameters, i.e., the left panel of Fig.~\ref{fig:selectionfunction}. 
From that figure alone, it is clear that the selection function has a weak dependence on the pulsar mass \emph{alone}.

Figure~\ref{fig:pulsarpopulation} shows the inferred pulsar-mass population with and without selection correction. 
The median population distributions are similar over the mass range. 
As shown in Fig.~\ref{fig:selectionfunction}, higher-mass pulsars are slightly more likely to be selected.
Correcting for this effect marginally shifts the inferred population toward lower masses, although the shift remains well within the statistical uncertainty. 
The marginalized constraint on the upper population cutoff is broad. 
With selection effects, we obtain $M_{\rm pop}=2.32^{+0.39}_{-0.28}\,M_\odot$ whereas neglecting selection effects gives $M_{\rm pop}=2.31^{+0.41}_{-0.27}\,M_\odot$; these values are posterior medians with $68\%$ credible intervals. 
The overlap between the intervals shows that the selection correction has little effect on the inferred upper cutoff.

\bibliographystyle{apsrev4-1}
\bibliography{main}

\end{document}